\documentclass[prl,aps,twocolumn,longbibliography]{revtex4-1}
\usepackage[utf8]{inputenc}

\pdfoutput=1

\newcommand{\cs}{$\clubsuit\;$}

\usepackage{float}

\usepackage{fourier}

\usepackage{color}
\usepackage{graphicx}
\usepackage{amsmath,amssymb}

\usepackage{times}
\usepackage{upgreek}
\usepackage{psfrag} 
\usepackage{latexsym} 
\usepackage{amstext}
\usepackage{amsxtra} 

\usepackage{textcomp}
\usepackage{amsfonts}
\usepackage{graphicx}
\usepackage{bm}
\usepackage{color}
\usepackage{braket}
\usepackage{units}
\usepackage[svgnames]{xcolor}

\usepackage{tcolorbox}
\usepackage{ragged2e}

\AtBeginDocument{%
    \newwrite\bibnotes
    \def\bibnotesext{Notes.bib}
    \immediate\openout\bibnotes=\jobname\bibnotesext
    \immediate\write\bibnotes{@CONTROL{REVTEX41Control}}
    \immediate\write\bibnotes{@CONTROL{%
    apsrev41Control,author="08",editor="1",pages="1",title="0",year="1"}}
     \if@filesw
     \immediate\write\@auxout{\string\citation{apsrev41Control}}%
    \fi
}%

\definecolor{myColor}{rgb}{0.02,0.12,0.3}
\definecolor{myciteColor}{rgb}{0.39,0.7,0.89}
\usepackage[colorlinks=true,citecolor=myColor,linkcolor=myColor,urlcolor=myColor]{hyperref}

\def\be{\begin{equation}}
\def\ee{\end{equation}}

\begin{document}

\title{Quantum Gases in Optical Boxes}

\begin{abstract}
Advances in light shaping for optical trapping of neutral particles have led to the development of box traps for ultracold atoms and molecules. These traps have allowed the creation of homogeneous quantum gases and opened new possibilities for studies of many-body physics. They simplify the interpretation of experimental results, provide more direct connections with theory, and in some cases allow qualitatively new, hitherto impossible experiments. Here we review progress in this emerging field.
\end{abstract}

\author{Nir Navon}
\affiliation{Department of Physics, Yale University, New Haven, Connecticut 06520, USA}
\author{Robert P. Smith}
\affiliation{Clarendon Laboratory, University of Oxford, Parks Road, Oxford OX1 3PU, United Kingdom}
\author{Zoran Hadzibabic}
\email[]{zh10001@cam.ac.uk}
\affiliation{Cavendish Laboratory, University of Cambridge, J. J. Thomson Avenue, Cambridge CB3 0HE, United Kingdom}

\maketitle

Since the earliest days of ultracold atomic gases, the success in using them to study many-body physics~\cite{Dalfovo:1999,Giorgini:2008,Bloch:2008} has owed a lot to the possibility to trap the atoms in versatile potentials, including low-dimensional traps~\cite{Gorlitz:2001b}, double wells~\cite{Andrews:1997a}, and optical lattices~\cite{Anderson:1998,Greiner:2001b, Bloch:2005}. The electromagnetic trapping potentials are often also dynamically tuneable, which has allowed experiments ranging from studies of elementary excitations~\cite{Jin:1996a,Mewes:1996b,Andrews:1997} to reversible crossing of phase transitions~\cite{Stamper-Kurn:1998c,Greiner:2002a}.

More recent advances in the shaping of optical potentials have opened many new possibilities. One major development is the increasingly popular use of the uniform (flat-bottom) optical-box traps~\cite{Gaunt:2013, Chomaz:2015,Mukherjee:2017,Hueck:2018,Tajik:2019,Bause:2021}, as opposed to the traditionally used (optical or magnetic) harmonic ones. 
Making a quantum gas homogeneous almost invariably makes the interpretation of the experiments and the comparisons with theory more direct and easier. Moreover, in some cases it allows qualitatively different experiments and discoveries of new physics.

In this brief review we outline the developments that have allowed the creation of box traps and highlight some of the scientific successes in this new and growing field.

Before starting, we also draw the reader's attention to contemporary reviews on two related emerging fields - (i) the creation of `atomtronics' circuits for coherent matter waves, such as ring traps that support persistent currents~\cite{Amico:2020}, and  (ii) the trapping of individual atoms or molecules in arrays of optical tweezers~\cite{Ni:2021}.
All three fields take advantage of the advances in light shaping, and there are also scientific connections; as a prominent example, the dynamics of phase transitions in a homogeneous system have been studied in ring traps~\cite{Corman:2014}, 2D and 3D box traps~\cite{Chomaz:2015,Navon:2015}, and a 1D tweezers-array~\cite{Keesling:2019}.

\begin{figure*}
\centering
\includegraphics[width=\textwidth]{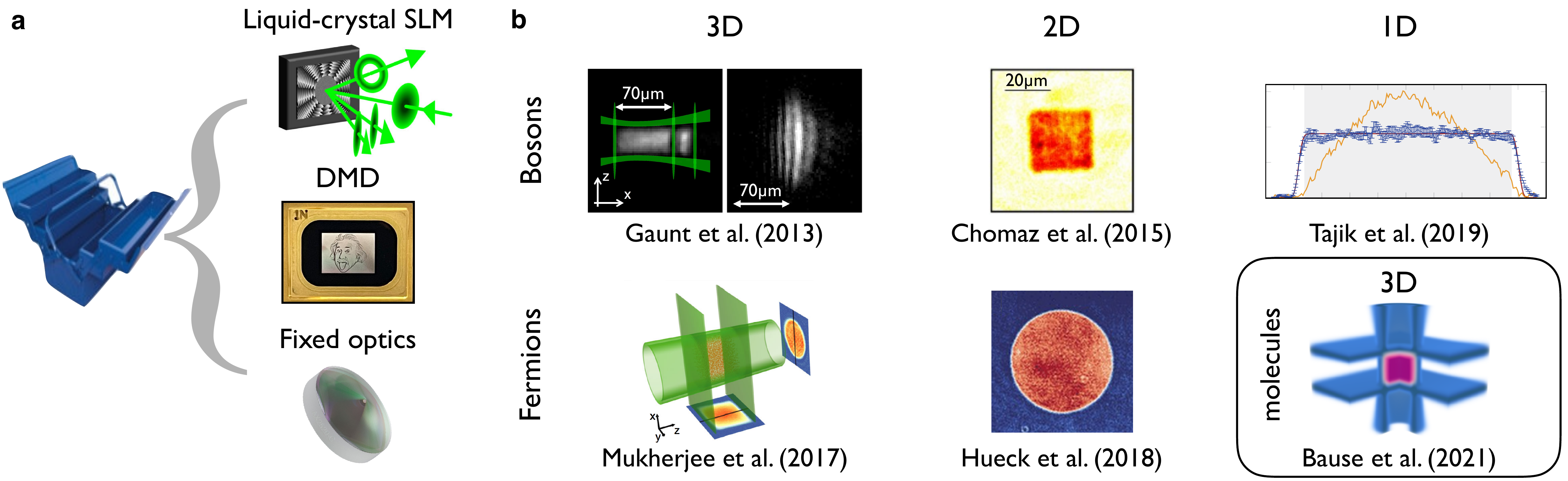}
\caption{\textbf{Optical box traps.} (a) Box tools. The first 
3D box trap~\cite{Gaunt:2013} was generated using a phase-controlling spatial light modulator (SLM) and a blue-detuned laser to create the box walls that repel and confine the atoms. As illustrated in the cartoon, a phase pattern imprinted by the SLM onto a single Gaussian beam creates three separate beams, a tube and two sheets; the three beams are then `folded' to form a cylindrical box trap (such as shown for a 3D Fermi gas in (b)). Other experiments have used Digital Micromirror Devices (DMDs) that spatially modulate the amplitude of a laser beam. While these versatile spatial modulators were instrumental for the development of first box traps, in many cases one can also create them using more traditional tools, such as specialised fixed (non-tuneable) optics.  (b) Box-trapped gases have been realised for Bose and Fermi atoms in various dimensionalities, as well as for (fermionic) ground-state diatomic molecules in 3D~\cite{Gaunt:2013,Chomaz:2015,Mukherjee:2017,Hueck:2018,Tajik:2019,Bause:2021}. One can also create other geometries, such as the two-trap configuration shown in the left panel for the 3D Bose gas; in this first experiment, this was used to demonstrate the coherence of the gas through matter-wave interference (after release and free expansion) seen in the right panel. For the 1D Bose gas the figure directly compares the line density distribution in a box trap (blue) and a harmonic one (orange).
\label{fig:box_traps}}
\end{figure*}


{\setcounter{figure}{0}
\renewcommand{\figurename}{Box}
\begin{figure}
\begin{tcolorbox}[colback=blue!5,colframe=blue!75!black,title=\textbf{Box 1: Characterizing box traps}]

\centering
\includegraphics[width=\columnwidth]{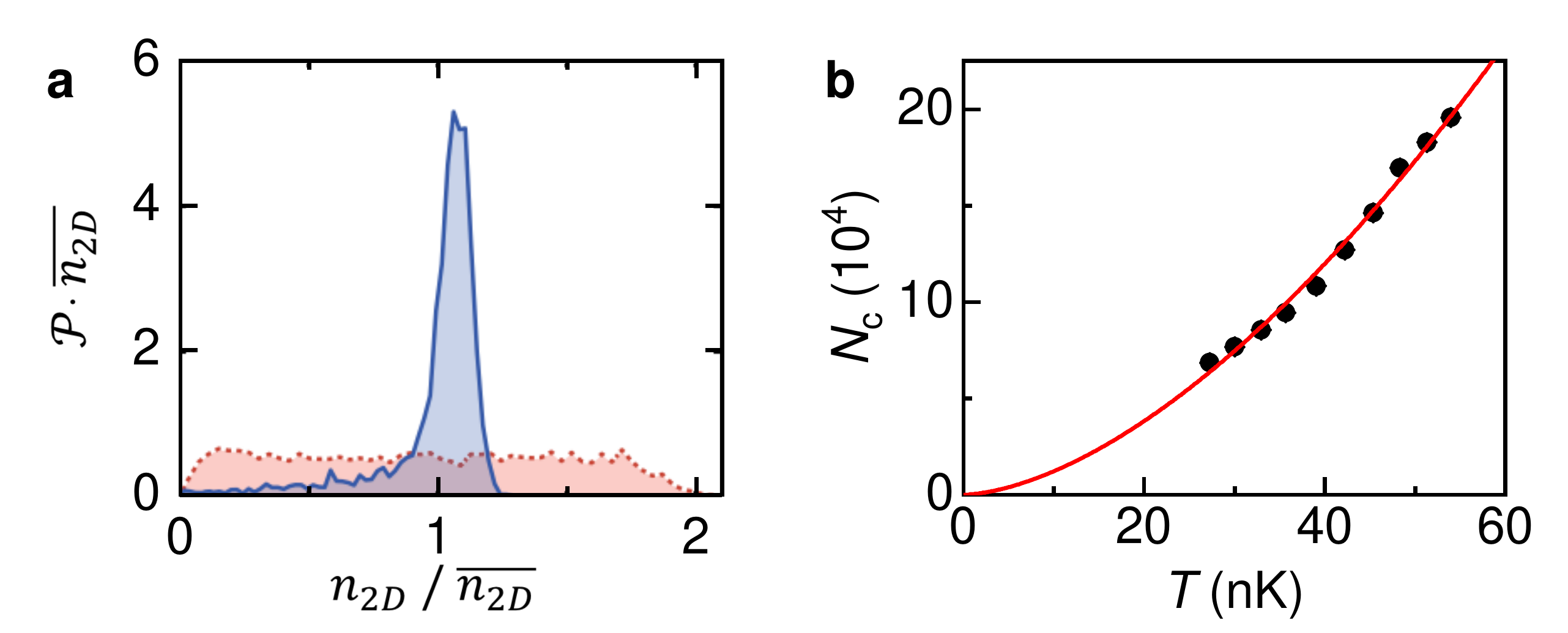}
\justify
(a) One simple measure of the gas homogeneity is the distribution of the real-space densities. 
Here we show the distribution of column densities, $n_{\rm 2D}$, extracted from {\it in situ} images of 3D clouds~\cite{Mukherjee:2017}.  In a box trap (blue) the probability distribution $\mathcal{P}(n_{\rm 2D})$ is narrow, strongly peaked near the average value $\overline{n_{\rm 2D}}$. This means, for example, that most atoms experience essentially the same mean-field potential. For comparison, the corresponding distribution in a non-degenerate harmonically trapped gas (red) is very broad; the expected distribution is uniform between $0$ and $2 \overline{n_{\rm 2D}}$.  

\noindent (b) A complementary way to characterise box traps is to look at the single-particle density of states. This is seen, for example, in the dependence of the critical atom number for Bose-Einstein condensation, $N_{\rm c}$, on the temperature $T$~\cite{Schmidutz:2014}. In a 3D harmonic trap $N_{\rm c} \propto T^3$, while in a perfect 3D box $N_{\rm c} \propto T^{3/2}$. Here the experimental data is captured by $N_{\rm c} \propto T^{\alpha}$ with $\alpha = 1.65$ (red line). A common way~\cite{Gaunt:2013} to characterise (imperfect) box traps is to model them by an isotropic power-law potential $V({\bf r}) \propto r^p$ with $p\gg 1$. Then $\alpha = 3/2 + 3/p$, and for a Fermi gas one similarly gets that the Fermi energy is~\cite{Mukherjee:2017} $E_{\rm F} \propto N^{1/\alpha}$.
In current experiments $p>10$ is readily achieved, and there are indications that values of $p$ up to $\sim 100$ are feasible~\cite{Hueck:2018}. 
\end{tcolorbox}
\label{fig:box}
\end{figure}}


\section{Making box traps}

In this section we briefly introduce the key tools (Fig.\,\ref{fig:box_traps}a) and ideas involved in the making of optical box traps. These methods have been used to create boxes for Bose and Fermi atomic and molecular gases in various dimensionalities~\cite{Gaunt:2013,Chomaz:2015,Mukherjee:2017,Hueck:2018,Tajik:2019,Bause:2021}, as illustrated in Fig.\,\ref{fig:box_traps}b; see also Refs.~\cite{Meyrath:2005,van2010box} for early prototypes of 1D box traps, and Ref.~\cite{Gauthier:2021} for a comprehensive review of recent advances in light shaping.

Optical box traps can be made using either red-detuned (attractive) or blue-detuned (repulsive) far-off-resonant optical fields, but the latter approach is generally simpler. In `red traps', to create a uniform trap floor one needs to sculpt high intensity light such that  the variations in the optical potential are smaller than all the relevant energy scales in the gas. On the other hand, in `blue traps', the floor is dark and the sharp walls confining the cloud just need to be high enough. To make a 3D box potential, one must also levitate the particles against gravity. This can be done using a light field of linearly varying intensity~\cite{Shibata:2020}, but usually atoms are levitated
using a static magnetic field gradient~\cite{Gaunt:2013}, while 
polar molecules can be levitated using a static electric field gradient~\cite{Bause:2021}. 

Regarding light-shaping, many key developments have been enabled
by two complementary types of programmable spatial light modulators (SLMs) - the liquid-crystal SLMs that modulate the phase of laser beams and the Digital Micromirror Devices (DMDs) that modulate their amplitude.

A liquid-crystal SLM is a rectangular array of $\sim 10^6$ pixel elements (each $\sim 10~\mu$m in size) with individually controllable indices of refraction. Using it to imprint a spatially-modulated phase delay on a laser beam, one controls the intensity pattern in the conjugate (Fourier) plane. To create an arbitrary pattern in the conjugate plane, one should in general modulate both the phase and the amplitude of the beam, but various numerical algorithms have been developed for creating good approximations of arbitrary trapping potentials using only phase control~\cite{Gauthier:2021}. Moreover, in some cases we have analytical solutions for the required phase modulations. An example of this is the often-used cylindrical 3D box~\cite{Gaunt:2013} `assembled' using one hollow-tube beam and two end-caps sheet beams (see Fig.\,\ref{fig:box_traps}a,b) - in this case all three beams can be created starting with a single Gaussian beam and using a single SLM to imprint an analytically calculated phase pattern~\cite{Gaunt:2014-th}. 

Similarly, a DMD is a rectangular array of $\sim 10^6$ mirrors (each $\sim 10~\mu$m in size) that can individually be 
turned `on' or `off' (by changing their tilt angle) to spatially modulate the amplitude of a laser beam. An arbitrary intensity pattern can then be imaged onto the cloud~\cite{Gauthier:2016}, which is convenient for making rectangular traps with sharp corners in 2D (Fig.~\ref{fig:box_traps}b) or cuboid traps in 3D. 

The two approaches have different advantages. The liquid-crystal one is more power-efficient, because the light is steered rather than partially dumped (by the `off' DMD mirrors). On the other hand, DMDs are much better for creating dynamical potentials. Due to their (currently) sub-kHz refresh rate, the liquid-crystal SLMs cannot be reprogrammed and updated with a different phase pattern during an experimental run, since this would temporarily turn off the trap and release the cloud. Meanwhile, the $\sim 10$~kHz refresh rate of DMDs is sufficiently high for them to be reprogrammed and the trapping pattern dynamically changed without the ultracold particles moving significantly during the updates.

The versatility of SLMs has been essential for experimental exploration \cite{Gaunt:2014-th}, but box traps can also be made with non-tuneable tools~\cite{Gauthier:2021} such as axicons~\cite{manek1998generation, Mukherjee:2017, Hueck:2018} and custom-manufactured masks~\cite{Chomaz:2015}. Another option is to use `painted', time-averaged potentials created by fast spatial scanning of a laser beam~\cite{Henderson:2009,Gauthier:2021}.

As a final point in this section, optical box traps are not perfect. The sharpness of their walls is limited by the optical wavelength of $\sim 1~\mu$m, which is not negligible compared to the typical box dimensions $10 -100~\mu$m. 
Moreover, additional fields used for levitation, tight out-of-plane confinement of 2D samples~\cite{Chomaz:2015,Ville:2017}, or tuning of interactions via Feshbach resonances~\cite{Chin:2010}, can lead to further imperfections. In Box 1, we discuss two complementary methods used to quantify the uniformity of box-trapped gases. The currently achieved uniformity has been good enough for the success stories we discuss in the next section, but how close to perfect a box needs to be ultimately depends on the specific scientific problem.

\section{Success stories}

Here we outline the scientific advances afforded by homogeneous atomic gases, which also illustrate the general types of problems for which the box traps are advantageous.


\setcounter{figure}{1}
\begin{figure}
\centering
\includegraphics[width=\columnwidth]{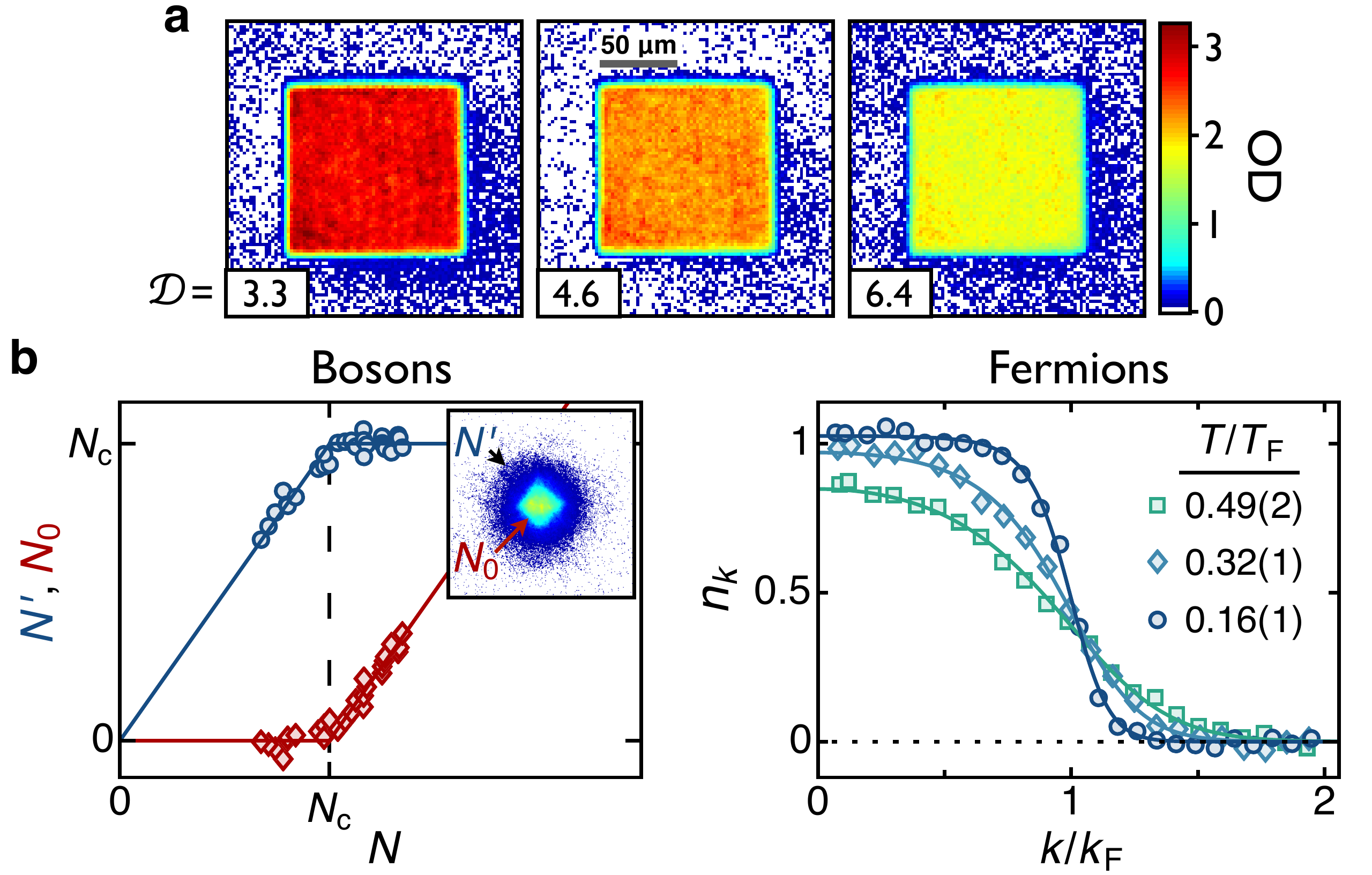}
\caption{\textbf{Quantum statistics in homogeneous gases.} (a) \emph{In situ} absorption images of homogeneous quantum-degenerate 3D gases with different phase space densities $\mathcal{D}>1$. The gas with lowest density (lowest optical density OD) is the coldest and actually has the highest $\mathcal{D}$, but one cannot tell. One cannot even tell whether the gases are fermionic or bosonic (here they are fermionic $^{6}$Li gases).
(b) While no effects of quantum statistics and degeneracy are seen in real space, they are striking in momentum space. 
For a Bose gas, the statistical nature of the BEC phase transition, driven by the saturation of the excited states, is revealed more clearly than in the corresponding harmonic-trap experiments~\cite{Tammuz:2011}. Here the plot shows the total number of atoms in the thermal cloud ($N'$) and the condensate ($N_0$), as the number of atoms in the cloud ($N$) is varied at constant temperature~\cite{Schmidutz:2014}; the inset shows an image of a partially condensed gas after release from the box and free expansion.
For a Fermi gas below the Fermi temperature $T_{\rm F}$, the effects of the Pauli exclusion principle are clearly observed - the occupation of individual momentum states are limited to $n_k \leq 1$, and the Fermi surface forms at the Fermi wavevector $k_{\rm F}$  (adapted from \cite{Mukherjee:2017}).
\label{fig:QS}}
\end{figure}

We start with experiments on purely quantum-statistical phenomena (Fig.\,\ref{fig:QS}). In harmonic traps the real and momentum space are coupled, so one can for example observe real-space effects of Bose-Einstein condensation (BEC)~\cite{Anderson:1999} and Fermi pressure~\cite{Truscott:2001}. In a box trap, the two spaces are disentangled and there are no clear signatures of quantum statistics in real space (see Fig.~\ref{fig:QS}a), but in momentum-space they are revealed more cleanly (Fig.~\ref{fig:QS}b).

For bosons, one clearly observes the statistical nature of the BEC transition, which is driven by the saturation of the total occupation of all excited (momentum) states~\cite{Schmidutz:2014}. As the total atom number in the gas is increased at a fixed temperature, the number of atoms in the thermal cloud saturates at the critical value for condensation, and all the extra atoms accumulate in the condensate. 
For fermions, one instead observes the occupation-number saturation at the level of individual single-particle states, as prescribed by the Pauli exclusion principle. As the temperature (normalised to the Fermi temperature $T_{\rm F}$) is reduced, the occupation of all momentum states remains bound to $n_k \leq 1$ and the Fermi surface forms~\cite{Mukherjee:2017}.

Experiments on the thermodynamics of nearly-ideal box-trapped gases also lead to the unexpected observation of the Joule-Thomson effect that arises solely from quantum correlations~\cite{Schmidutz:2014}. While the ideal classical gas does not change temperature under isoenthalpic rarefaction, the ideal Bose gas cools and the ideal Fermi gas is expected to heat~\cite{Kothari:1937}.


\begin{figure*}
\centering
\includegraphics[width=\textwidth]{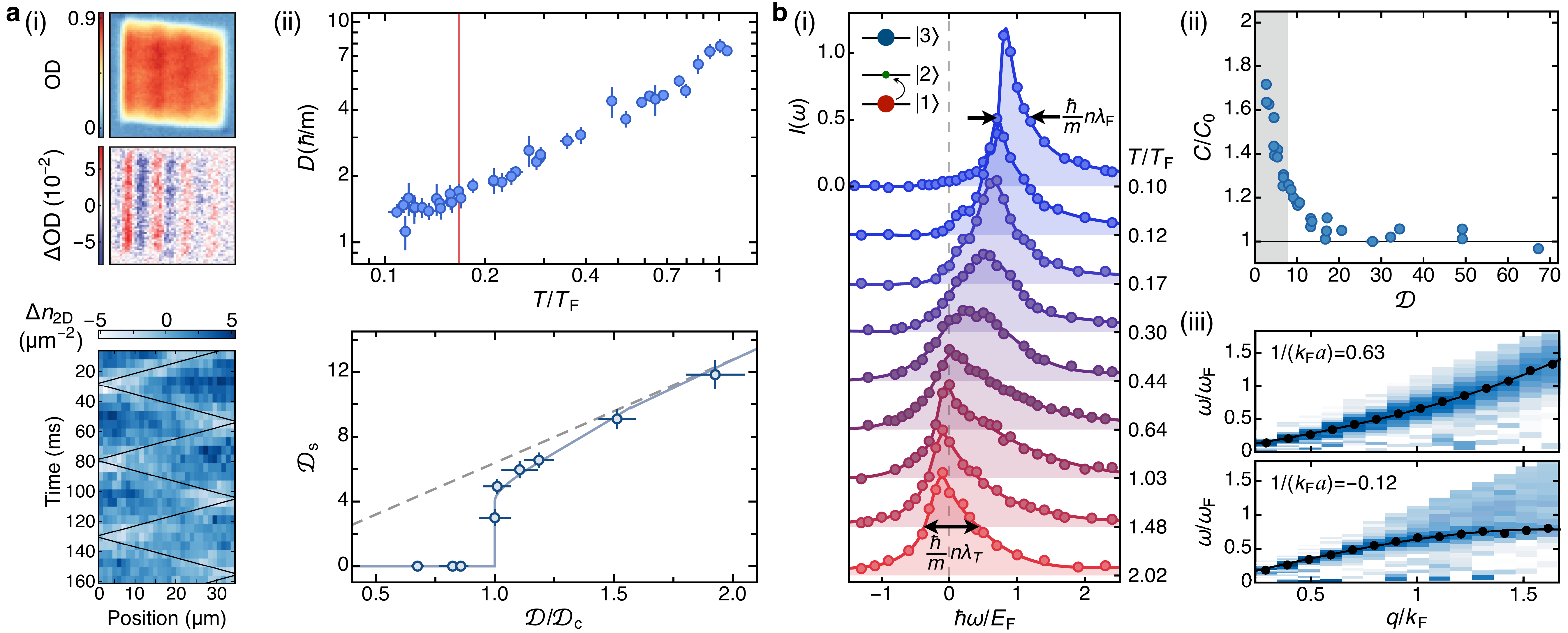}
\caption{\textbf{Sound and spectroscopy measurements on box-trapped gases} (a) Sound: (i) Low-energy sound modes can be probed by perturbing the gas with an external potential and observing the evolution of the resulting density modulations in time and space.
Images show examples of measurements in a 3D Fermi gas~\cite{Patel:2019} (top) and a 2D Bose gas~\cite{Ville:2018} (bottom). 
(ii) Top: the sound diffusivity $D$ (seen in the attenuation of the wave) in the low-temperature unitary Fermi gas approaches the universal quantum limit, $D \approx \hbar/m$; the red line indicates the critical temperature for superfluidity (adapted from~\cite{Patel:2019}).
Bottom: the superfluid phase-space density $\mathcal{D}_{\rm s}$ in a 2D Bose gas, which was deduced from the measured speeds of first and second sound, undergoes a universal jump from $0$ to $4$ at the BKT phase transition; here $\mathcal{D}$ is the total phase-space density and  $\mathcal{D}_{\rm c}$ its critical value~\cite{Christodoulou:2020}. 
(b) Spectroscopy: 
(i) Particle-ejection spectroscopy on the unitary 3D Fermi gas~\cite{Mukherjee:2019}.  
Rabi rf spectroscopy was performed on the whole cloud and the differences induced by small changes in  $T/T_{\rm F}$ are observable only due to the lack of inhomogeneous broadening.
Such measurements probe short-range correlations and have also revealed non-Fermi-liquid behaviour in the normal unitary gas, at $T> T_{\rm c}$.
(ii) Ramsey rf spectroscopy on a 2D Bose gas was used to determine the two-body contact ($C$) as a function of $\mathcal{D}$; the shaded region indicates the non-superfluid phase (adapted from~\cite{Zou:2021tan}). 
(iii) Bragg spectroscopy was used to measure the excitation spectrum in a strongly interacting 3D Fermi gas, here shown on the BEC (top) and the BCS (bottom) side of the BEC-BCS crossover~\cite{Biss:2021}.
\label{fig:sound}}
\end{figure*}

We now move on to the experiments on the many-body physics of interacting gases, starting with the broad class of spectroscopic and transport measurements in which one weakly perturbs a system in order to extract information on its equilibrium properties (Fig.~\ref{fig:sound}). 

A lot of attention has been given to long-wavelength sound waves~\cite{Navon:2016, Ville:2018, Patel:2019, Baird:2019, Garratt:2019, Christodoulou:2020,  Bohlen:2020, Zhang:2021}. In Fig.~\ref{fig:sound}a we show examples of sound propagation in 3D Fermi~\cite{Patel:2019} and 2D Bose~\cite{Ville:2018} gases, and in Fig.~\ref{fig:sound}b we illustrate two scientific highlights of such experiments. 
In both 3D and 2D unitary Fermi gases the quantum limit of sound diffusivity, set by $\hbar/m$ (where $m$ is the atom mass), was demonstrated~\cite{Patel:2019,Bohlen:2020}; this universal limit should also be relevant for other strongly interacting Fermi systems such as neutron stars. In 2D, superfluidity emerges via the infinite-order Berezinskii–Kosterlitz–Thouless (BKT) phase transition~\cite{Berezinskii:1971,Kosterlitz:1973} instead of the BEC one; the 2D box trap allowed the first observation of first and second sound in a BKT superfluid~\cite{ Christodoulou:2020}, and the measurements of the two sound speeds revealed the universal superfluid-density jump at the BKT transition~\cite{Nelson:1977}. 
Related experiments on superfluidity have investigated the Josephson effect~\cite{Luick:2020} and the critical velocity~\cite{Sobirey:2020} in a 2D Fermi gas, while a more complex trapping geometry allowed studies of a Bose-gas superfluid flow through a constriction between two reservoirs~\cite{Gauthier:2019b}.

Fig.~\ref{fig:sound}b illustrates benefits of gas homogeneity for spectroscopic measurements that globally probe the system~\cite{Gotlibovych:2014, Lopes:2017, Lopes:2017b, Mukherjee:2019,Yan:2019, Zou:2021mag, Zou:2021tan, Biss:2021}; for experiments on extraction of homogeneous-gas properties by local probing of harmonically trapped gases see~\cite{Sagi:2012,Sagi:2015,Ota:2017,Carcy:2019}. The Rabi radio-frequency (rf) spectra shown in Fig.~\ref{fig:sound}b(i) measure the energy cost of removing a particle from a spin$-1/2$ 3D Fermi gas at different reduced temperatures $T/T_{\rm F}$~\cite{Mukherjee:2019}.
Measurements were performed on the whole sample and the spectra taken at closely-spaced $T/T_{\rm F}$ values are clearly distinct only thanks to the lack of inhomogeneous broadening; in a harmonic trap global measurements would mix signals for a wide range of $T/T_{\rm F}({\bf r}) \propto n({\bf r})^{-2/3}$, where $n({\bf r})$ is the local density.
Rabi rf spectroscopy of 3D Fermi gases~\cite{Mukherjee:2019, Yan:2019} has, for example, provided an observation of non-Fermi-liquid behaviour in a normal strongly interacting gas~\cite{Mukherjee:2019}, while Ramsey rf spectroscopy of 2D Bose gases has provided measurements of short-range correlations across the BKT transition (Fig.~\ref{fig:sound}b(ii)) and an observation of magnetic-dipole interactions~\cite{Zou:2021mag,Zou:2021tan}.

Gas homogeneity is similarly beneficial for Bragg spectroscopy~\cite{Kozuma:1999a, Stenger:1999b}. Fig.~\ref{fig:sound}b(iii) shows measurements of the excitation spectrum of a strongly interacting 3D Fermi gas, which were used to extract the concavity of the dispersion relation and the pairing gap in the BEC-BCS crossover~\cite{Biss:2021}.
Bragg-spectroscopy experiments on condensed homogeneous 3D Bose gases have provided an observation of Heisenberg-limited long-range coherence~\cite{Gotlibovych:2014}, the confirmation of Bogoliubov's theory of quantum depletion~\cite{Lopes:2017b}, and an observation of the breakdown of Bogoliubov's theory of the excitation spectrum for sufficiently strong interactions~\cite{Lopes:2017}.


\begin{figure*}[t!]
\centering
\includegraphics[width=\textwidth]{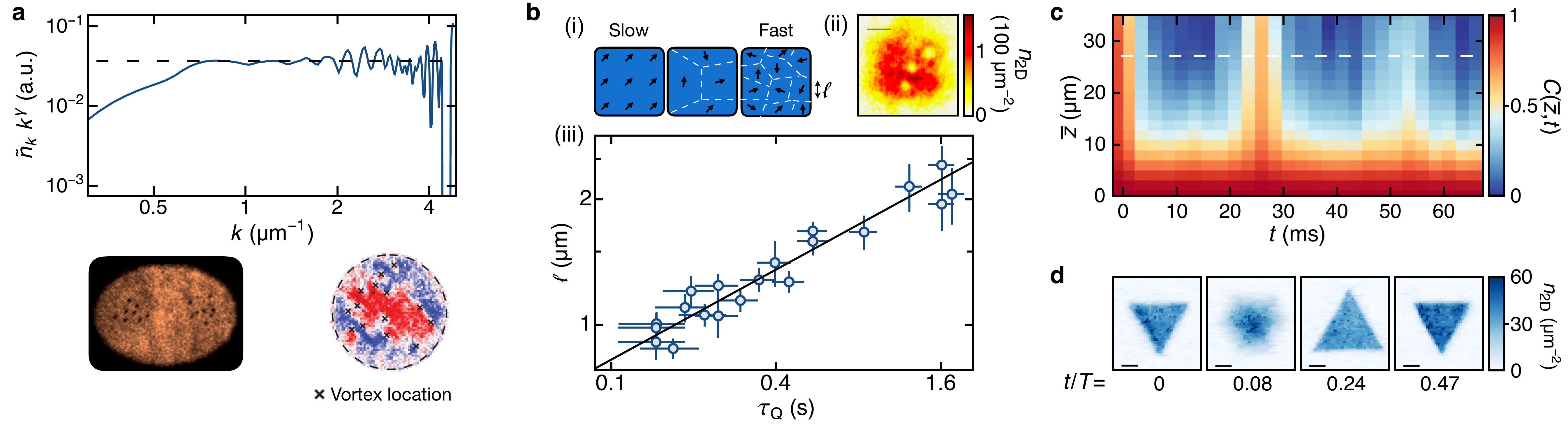}
\caption{\textbf{Non-equilibrium phenomena} (a) Wave and vortex turbulence. Top: power-law momentum distribution, $\tilde{n}_k \propto k^{-\gamma}$, in a 3D Bose gas driven on a large lengthscale reveals a wave-turbulence cascade towards smaller lengthscales~\cite{Navon:2016,Navon:2019}. Bottom: experiments on vortex turbulence in quasi-2D Bose gases showed large-scale vortex clustering, which was observed using direct absorption imaging~\cite{Gauthier:2019} (left) and circulation-sensitive imaging~\cite{Johnstone:2019} (right). 
(b) Critical dynamics. (i) Critical slowing down results in nonadiabatic crossing of the BEC transition and formation of domains with different spontaneously chosen condensate phases (indicated by the arrows).
(ii) Topological defects form at the domain boundaries; here the image shows vortices spontaneously generated in a quench-cooled gas~\cite{Chomaz:2015}.  (iii) The power-law dependence of the average domain size, $\ell$, on the quench time, $\tau_Q$, is in agreement with the Kibble-Zurek theory~\cite{Navon:2015}. 
(c) Recurrences of phase correlations were observed in a 1D Bose gas~\cite{Rauer:2018}. 
(d) A novel breather was seen in a 2D Bose gas; for particular initial density distributions, such as a uniform triangle prepared in a box trap, the cloud evolving in a harmonic potential periodically returns to its initial state~\cite{SaintJalm:2019}. 
\label{fig:noneq}} 
\end{figure*}

In another large class of experiments, interacting gases have been driven or quenched far from equilibrium (Fig.~\ref{fig:noneq}). 

One paradigmatic topic in non-equilibrium physics is turbulence in strongly driven systems. Depending on the system and the excitation protocol, turbulent dynamics can be dominated by either waves or vortices, and the advent of box traps has led to new insights in both cases (Fig.~\ref{fig:noneq}a). In `shaken' 3D Bose gases the power-law momentum distribution characteristic of a direct turbulent cascade was observed~\cite{Navon:2016}, and the elusive particle and energy fluxes through the cascade have been measured~\cite{Navon:2019}. In (quasi-)2D Bose gases vortex clustering corresponding to negative temperatures was observed~\cite{Johnstone:2019, Gauthier:2019}  (see also \cite{Stockdale:2020, Reeves:2020}), as predicted by Onsager in 1949~\cite{Onsager:1949}. In another recent experiment the interplay of vortices and waves has also been observed~\cite{Kwon:2021}.

Another major topic for which homogeneous gases have distinct advantages is the study of the diverging-range correlations associated with the critical behavior near second-order phase transitions (Fig.\,\ref{fig:noneq}b and Box 2). In a non-equilibrium context, dynamic crossing of such a transition results in causally disconnected domains that display different choices of the symmetry-breaking order parameter. The Kibble-Zurek (KZ) theory~\cite{Kibble:1976,Zurek:1985} that describes these dynamics was originally developed specifically for homogeneous systems and its key assumption is that when it comes to the choice of the order parameter all parts of the system have `equal voting rights'~\cite{delCampo:2014,Beugnon:2017}. Box-trap experiments have provided quantitative tests of the KZ predictions for how the domain size~\cite{Navon:2015} and the resulting density of defects in the ordered state~\cite{Chomaz:2015} depend on the quench rate (see also~\cite{Corman:2014,Aidelsburger:2017,Keesling:2019} for ring-trap and optical-tweezers experiments).

\begin{figure*}
\begin{tcolorbox}[colback=blue!5,colframe=blue!75!black,title=\textbf{Box 2: Critical phenomena in harmonically and box trapped gases}]
\justify
Critical behaviour near a second-order phase transition is a paradigmatic problem for which using a homogeneous system is advantageous, because the local density approximation (LDA), used to interpret harmonic-trap experiments, fundamentally breaks down. Here we illustrate this with a simple ideal-gas calculation for the BEC phase transition.

In a homogeneous system, near the critical point the correlation length $\xi$ diverges as illustrated in panel (a). 
At a fixed temperature, $\xi/\lambda= A (|n-n_{\rm c}|/n_{\rm c})^{-\nu}$, where $\lambda$ is the thermal wavelength, $n_{\rm c}$ is the critical density, $\nu$ the critical exponent, and $A$ is a non-universal prefactor; for the ideal-gas BEC transition $n_{\rm c} = 2.612/\lambda^3$, $\nu=1$, and $A = 1/2.612$.  

For a harmonically trapped gas with spatially uniform $T$, 
as $n(r=0) \rightarrow n_{\rm c}$, the spatially varying $\xi(r)$ evaluated within the LDA exceeds the lengthscale over which it changes. 
Relaxing the usual LDA assumptions, one can still, at the cost of reducing the experimental signal, focus on the central part of the cloud and assume that $n$ is constant within some non-infinitesimal volume (similar to Refs.~\cite{Drake:2012, Sagi:2012,Sagi:2015,Carcy:2019}). In reality $n$ and $\xi$ vary within this region, but one can still directly probe correlations on a lengthscale $\ell$ if $\xi(r) > \ell$ for all $r < \ell/2$; this approach was successfully used in Ref.~\cite{Donner:2007}.

\begin{center}
 \includegraphics[width=0.8\columnwidth]{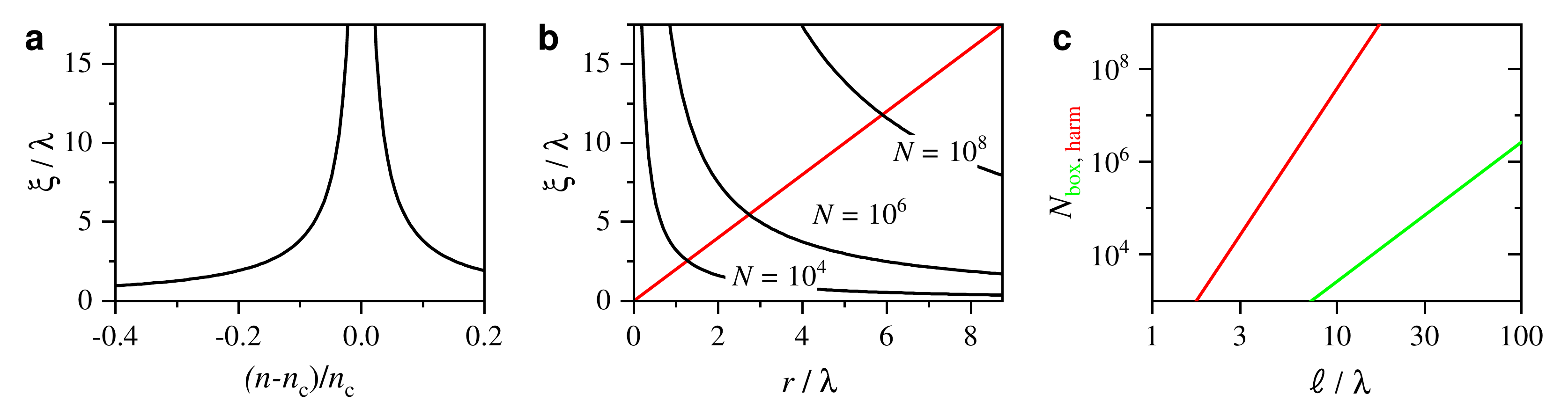} 
\end{center}

Setting $n(r=0)=n_{\rm c}$, assuming an isotropic trap of frequency $\omega$, and expanding the ideal-gas distribution~\cite{Dalfovo:1999} near $r=0$ gives $\xi(r)/\lambda \approx  k_{\rm B} T / (\hbar \omega) \times [ 2 \pi  r/\lambda]^{-1}$. 
Noting that $k_{\rm B} T/(\hbar \omega) = (N/1.202)^{1/3}$, where $N$ is the total number of atoms in the trap, in panel (b) we plot $\xi/\lambda$ versus $r/\lambda$ for different $N$ (black lines). The intersects of these curves with the line $\xi = 2r$ (red) then give the achievable $\ell = 2r$ for a given $N$, irrespective of the choices of $\omega$ and corresponding $T$. 
Conversely, the atom number needed to directly observe correlations on a lengthscale $\ell$ in a harmonic trap is
\begin{equation}
     N_{\rm harm} =1.202 \, \pi^3 (\ell/\lambda)^6 \, ,
\end{equation}
shown by the red line in panel (c). With a typical $N\sim 10^6$, one can reach only $\ell \sim 5 \, \lambda$.

On the other hand, working with a box trap, one just needs a box of size $\ell$ and the corresponding atom number 
\begin{equation}
   N_{\rm box}=n_{\rm c} \ell^3\approx 2.612 \times (\ell/\lambda)^3 \, ,
\end{equation}
shown by the green line in panel (c). In this case the same $N\sim 10^6$ is sufficient for measurements up to $\ell \sim 70\, \lambda$.

For an interacting gas, for which $\nu \approx 0.67$~\cite{Campostrini:2006,Burovski:2006,Donner:2007}, one gets similar conclusions. In this case $N_{\rm box} \propto (\ell/\lambda)^3$ is essentially the same, with just the prefactor ($\propto n_{\rm c}/\lambda^3$) changing slightly, and one can estimate $N_{\rm harm}$ in various ways: still assuming ideal-gas $n(r)$ gives $N_{\rm harm} \propto (\ell/\lambda)^{3 + 3/\nu} = (\ell/\lambda)^{7.5}$, while approximating $n(r)$ as a Gaussian gives $N_{\rm harm} \propto (\ell/\lambda)^{3 + 3/(2\nu)} = (\ell/\lambda)^{5.25}$. In either case one concludes that with the same atom-number resources one can directly observe much-longer-range correlations in a box trap.

\end{tcolorbox}
\end{figure*}

In other non-equilibrium experiments, the excitation spectrum of a homogeneous 1D Bose gas allowed the observation of recurrences in a closed quantum system~\cite{Rauer:2018,Schweigler:2020decay} (Fig.~\ref{fig:noneq}c), while homogeneous 3D Bose gases allowed observations of the universal loss and prethermalization dynamics following a quench to unitarity~\cite{Eigen:2017, Eigen:2018} and the universal dynamic scaling in a far-from-equilibrium weakly interacting gas~\cite{Glidden:2020}.

Finally, box-trapping and related technologies have also facilitated many experiments that are less directly related to the physics of homogeneous gases. One example of this is the discovery of a novel breather in a 2D Bose gas~\cite{SaintJalm:2019}; the breather shown in  Fig.~\ref{fig:noneq}d is observed in a harmonic potential, but the initial state had to be prepared in a box trap~\cite{SaintJalm:2019}. Another similar example is the deterministic preparation of a Townes soliton in Ref.~\cite{Bakkali-Hassani:2021} (see also Refs.~\cite{Chen:2020, Chen:2021b} for other observations of Townes solitons in box traps); this 2D soliton is an inhomogeneous ground state of the system, but its deterministic preparation was based on starting with a homogeneous gas and imprinting arbitrary density profiles using a DMD~\cite{Zou:2021}. A different example of a practical advantage of box traps is the observation of the transition from an atomic to a molecular condensate~\cite{Zhang:2021b}; in this case the creation of a (quasi-)equilibrium condensed gas of unstable molecules was facilitated by the use of a 2D box trap to minimise losses and heating. Further examples of box-enabled experiments include observation of the weak collapse of a condensate with attractive interactions~\cite{Eigen:2016} and the studies of the modulation instabilities that lead to emission of matter-wave jets, pattern formation and quasi-particle pair-production~\cite{Clark:2017,Fu:2018, Zhang_2020,Chen:2021}.

\section{Outlook} 

The scientific exploitation of box-trapped quantum gases is still in its infancy. Here we highlight some exciting possibilities for the future, and also some open technical challenges.

The successful studies of phase-transition dynamics could be extended to the infinite-order BKT transition~\cite{Mathey:2010,Jelic:2011,Mathey:2017,Comaron:2019,Brown:2021} and the bubble-nucleation dynamics associated with first-order transitions, including some believed to be relevant for the physics of the early universe~\cite{Fialko:2015,Braden:2018,Braden:2019,Billam:2019}.
Another general area where box traps could offer great advantages is topological physics~\cite{Goldman:2016, Ozawa:2019}; sharp boundaries could allow real-space studies of edge states~\cite{fletcher2019geometric}, which have so far been observed in cold-atom systems exploiting synthetic dimensions associated with internal (spin) degrees of freedom~\cite{Mancini:2015, Stuhl:2015, Chalopin:2020}.
It has also been predicted~\cite{Roccuzzo:2021} that the supersolid phases of gases with strong dipolar interactions~\cite{Bottcher:2019, Tanzi:2019, Chomaz:2019,Norcia:2021, Hertkorn:2021b,Bottcher:2021} should be qualitatively different in a box trap.

Further possibilities are offered by combining box traps and other trapping methods.
For example, combining box traps and optical lattices has already facilitated the observation of a low-entropy homogeneous state with long-range antiferromagnetic correlations~\cite{Mazurenko:2017} and studies of competing magnetic orders in the bilayer Hubbard model~\cite{Gall:2021}.
Further possibilities are suggested by the hybrid trap of Ref.~\cite{Mukherjee:2017}, which is box-like along two directions and harmonic along the third. In this case the harmonic direction provides tuning of the local chemical potential, while probing the system along a perpendicular direction retains many of the advantages of box traps, at least as long as the local density approximation (LDA) is 
valid. Such a setup could be used to study interfaces between different phases of matter, and could also facilitate searches for exotic states that are expected to occur only in narrow regions of bulk phase diagrams; an important example of such a still-sought-for phase is the Fulde-Ferrell-Larkin-Ovchinikov (FFLO) superfluid~\cite{Fulde:1964,Larkin:1964,Kinnunen:2018}.

While the range of scientific possibilities is broad and exciting, we can already anticipate that some of them will also require further technological developments. 

The first issue is that many interesting experiments are likely to require increasingly larger and closer to perfect box traps. 
This is particularly true for studies of critical phenomena (see Box 2), and more generally long-range correlations. 
As an illustration, a paradigmatic problem for which the current technology is still marginal is that of the critical temperature for Bose--Einstein condensation in an interacting homogeneous gas~\cite{Lee:1957a, Reppy:2000, Arnold:2001, Kashurnikov:2001, Andersen:2004,Holzmann:2004}. 
Critical fluctuations in a repulsively-interacting gas are predicted to raise $T_{\rm c}$ above the ideal-gas value. 
However, in a harmonic trap one observes the opposite~\cite{Ensher:1996, Gerbier:2004, Meppelink:2010, Smith:2011, Smith:2011b} - the beyond-mean-field correlation shift of $T_{\rm c}$ is diminished because only a small fraction of the cloud is critical at $T_{\rm c}$, and it is overpowered by a geometric mean-field (MF) effect~\cite{Giorgini:1996} that reduces $T_{\rm c}$.
For a general power-law trap, $V(r) \propto r^p$ (see Box 1), with increasing $p$ the beyond-MF shift should be more pronounced and the MF one should diminish. 
Based on an LDA estimate, we find that for the currently typical $p\sim 10$ the two effects are still comparable, and that one needs $p \gtrsim 100$ to cleanly observe the beyond-MF correlation shift of $T_{\rm c}$. 
We expect other fundamental correlation-physics problems to similarly create a moving target for the tolerable box imperfections.

The second issue is that many exciting possibilities rely on specific features of different atomic species and their mixtures, 
but the methods for the levitation of gases in 3D box traps are generally species specific.
In current experiments one can study mixtures of species that have very similar ratios of mass and magnetic moment~\cite{Mukherjee:2017, shkedrov2021absence},
but creating arbitrary homogeneous mixtures of different chemical elements, different isotopes, or even just different spin states of the same isotope, is an open challenge.

These two issues are in fact related, since even in single-species experiments the limitations for making the box traps larger and closer to uniform (with larger $p$ values) are often related to the need to levitate particles with additional fields. An exciting possibility for the future would therefore be to send box-trap setups to space and perform many-body experiments in a microgravity environment~\cite{Becker:2018,Aveline:2020,Frye:2021bose}.

We  thank C. Eigen for his help in the preparation of the figures, and M. Zwierlein, P. Patel, B. Mukherjee, J. Beugnon, J. Dalibard, R. Saint-Jalm, H. Biss, T. Lompe, and H. Moritz for sharing their data. We thank \cs for comments on the manuscript. This work was supported by the EPSRC (grant nos EP/N011759/1, EP/P009565/1, and EP/T019913/1), ERC (QBox), QuantERA (NAQUAS, EPSRC grant no. EP/R043396/1),  NSF CAREER (grant no. 1945324), and DARPA (grant no. 00010372). N. N.  acknowledges  support  from  the  David  and  Lucile  Packard  Foundation, and  the  Alfred  P.  Sloan  Foundation. R.P.S. acknowledges support from the Royal Society. Z.H. acknowledges support from the Royal Society Wolfson Fellowship.

%


\begin{thebibliography}{142}%
\makeatletter
\providecommand \@ifxundefined [1]{%
 \@ifx{#1\undefined}
}%
\providecommand \@ifnum [1]{%
 \ifnum #1\expandafter \@firstoftwo
 \else \expandafter \@secondoftwo
 \fi
}%
\providecommand \@ifx [1]{%
 \ifx #1\expandafter \@firstoftwo
 \else \expandafter \@secondoftwo
 \fi
}%
\providecommand \natexlab [1]{#1}%
\providecommand \enquote  [1]{``#1''}%
\providecommand \bibnamefont  [1]{#1}%
\providecommand \bibfnamefont [1]{#1}%
\providecommand \citenamefont [1]{#1}%
\providecommand \href@noop [0]{\@secondoftwo}%
\providecommand \href [0]{\begingroup \@sanitize@url \@href}%
\providecommand \@href[1]{\@@startlink{#1}\@@href}%
\providecommand \@@href[1]{\endgroup#1\@@endlink}%
\providecommand \@sanitize@url [0]{\catcode `\\12\catcode `\$12\catcode
  `\&12\catcode `\#12\catcode `\^12\catcode `\_12\catcode `\%12\relax}%
\providecommand \@@startlink[1]{}%
\providecommand \@@endlink[0]{}%
\providecommand \url  [0]{\begingroup\@sanitize@url \@url }%
\providecommand \@url [1]{\endgroup\@href {#1}{\urlprefix }}%
\providecommand \urlprefix  [0]{URL }%
\providecommand \Eprint [0]{\href }%
\providecommand \doibase [0]{http://dx.doi.org/}%
\providecommand \selectlanguage [0]{\@gobble}%
\providecommand \bibinfo  [0]{\@secondoftwo}%
\providecommand \bibfield  [0]{\@secondoftwo}%
\providecommand \translation [1]{[#1]}%
\providecommand \BibitemOpen [0]{}%
\providecommand \bibitemStop [0]{}%
\providecommand \bibitemNoStop [0]{.\EOS\space}%
\providecommand \EOS [0]{\spacefactor3000\relax}%
\providecommand \BibitemShut  [1]{\csname bibitem#1\endcsname}%
\let\auto@bib@innerbib\@empty
\bibitem [{\citenamefont {Dalfovo}\ \emph {et~al.}(1999)\citenamefont
  {Dalfovo}, \citenamefont {Giorgini}, \citenamefont {Pitaevskii},\ and\
  \citenamefont {Stringari}}]{Dalfovo:1999}%
  \BibitemOpen
  \bibfield  {author} {\bibinfo {author} {\bibfnamefont {F.}~\bibnamefont
  {Dalfovo}}, \bibinfo {author} {\bibfnamefont {S.}~\bibnamefont {Giorgini}},
  \bibinfo {author} {\bibfnamefont {L.~P.}\ \bibnamefont {Pitaevskii}}, \ and\
  \bibinfo {author} {\bibfnamefont {S.}~\bibnamefont {Stringari}},\ }\bibfield
  {title} {\enquote {\bibinfo {title} {Theory of {B}ose-{E}instein condensation
  in trapped gases},}\ }\href {\doibase 10.1103/RevModPhys.71.463} {\bibfield
  {journal} {\bibinfo  {journal} {Rev. Mod. Phys.}\ }\textbf {\bibinfo {volume}
  {71}},\ \bibinfo {pages} {463--512} (\bibinfo {year} {1999})}\BibitemShut
  {NoStop}%
\bibitem [{\citenamefont {Giorgini}\ \emph {et~al.}(2008)\citenamefont
  {Giorgini}, \citenamefont {Pitaevskii},\ and\ \citenamefont
  {Stringari}}]{Giorgini:2008}%
  \BibitemOpen
  \bibfield  {author} {\bibinfo {author} {\bibfnamefont {S.}~\bibnamefont
  {Giorgini}}, \bibinfo {author} {\bibfnamefont {L.~P.}\ \bibnamefont
  {Pitaevskii}}, \ and\ \bibinfo {author} {\bibfnamefont {S.}~\bibnamefont
  {Stringari}},\ }\bibfield  {title} {\enquote {\bibinfo {title} {Theory of
  ultracold atomic {F}ermi gases},}\ }\href {\doibase
  10.1103/RevModPhys.80.1215} {\bibfield  {journal} {\bibinfo  {journal} {Rev.
  Mod. Phys.}\ }\textbf {\bibinfo {volume} {80}},\ \bibinfo {pages}
  {1215--1274} (\bibinfo {year} {2008})}\BibitemShut {NoStop}%
\bibitem [{\citenamefont {Bloch}\ \emph {et~al.}(2008)\citenamefont {Bloch},
  \citenamefont {Dalibard},\ and\ \citenamefont {Zwerger}}]{Bloch:2008}%
  \BibitemOpen
  \bibfield  {author} {\bibinfo {author} {\bibfnamefont {I.}~\bibnamefont
  {Bloch}}, \bibinfo {author} {\bibfnamefont {J.}~\bibnamefont {Dalibard}}, \
  and\ \bibinfo {author} {\bibfnamefont {W.}~\bibnamefont {Zwerger}},\
  }\bibfield  {title} {\enquote {\bibinfo {title} {Many-body physics with
  ultracold gases},}\ }\href {\doibase 10.1103/RevModPhys.80.885} {\bibfield
  {journal} {\bibinfo  {journal} {Rev. Mod. Phys.}\ }\textbf {\bibinfo {volume}
  {80}},\ \bibinfo {pages} {885--964} (\bibinfo {year} {2008})}\BibitemShut
  {NoStop}%
\bibitem [{\citenamefont {G\"orlitz}\ \emph {et~al.}(2001)\citenamefont
  {G\"orlitz}, \citenamefont {Vogels}, \citenamefont {Leanhardt}, \citenamefont
  {Raman}, \citenamefont {Gustavson}, \citenamefont {Abo-Shaeer}, \citenamefont
  {Chikkatur}, \citenamefont {Gupta}, \citenamefont {Inouye}, \citenamefont
  {Rosenband},\ and\ \citenamefont {Ketterle}}]{Gorlitz:2001b}%
  \BibitemOpen
  \bibfield  {author} {\bibinfo {author} {\bibfnamefont {A.}~\bibnamefont
  {G\"orlitz}}, \bibinfo {author} {\bibfnamefont {J.~M.}\ \bibnamefont
  {Vogels}}, \bibinfo {author} {\bibfnamefont {A.~E.}\ \bibnamefont
  {Leanhardt}}, \bibinfo {author} {\bibfnamefont {C.}~\bibnamefont {Raman}},
  \bibinfo {author} {\bibfnamefont {T.~L.}\ \bibnamefont {Gustavson}}, \bibinfo
  {author} {\bibfnamefont {J.~R.}\ \bibnamefont {Abo-Shaeer}}, \bibinfo
  {author} {\bibfnamefont {A.~P.}\ \bibnamefont {Chikkatur}}, \bibinfo {author}
  {\bibfnamefont {S.}~\bibnamefont {Gupta}}, \bibinfo {author} {\bibfnamefont
  {S.}~\bibnamefont {Inouye}}, \bibinfo {author} {\bibfnamefont
  {T.}~\bibnamefont {Rosenband}}, \ and\ \bibinfo {author} {\bibfnamefont
  {W.}~\bibnamefont {Ketterle}},\ }\bibfield  {title} {\enquote {\bibinfo
  {title} {Realization of {B}ose-{E}instein condensates in lower dimensions},}\
  }\href {\doibase 10.1103/PhysRevLett.87.130402} {\bibfield  {journal}
  {\bibinfo  {journal} {Phys. Rev. Lett.}\ }\textbf {\bibinfo {volume} {87}},\
  \bibinfo {pages} {130402} (\bibinfo {year} {2001})}\BibitemShut {NoStop}%
\bibitem [{\citenamefont {Andrews}\ \emph
  {et~al.}(1997{\natexlab{a}})\citenamefont {Andrews}, \citenamefont
  {Townsend}, \citenamefont {Miesner}, \citenamefont {Durfee}, \citenamefont
  {Kurn},\ and\ \citenamefont {Ketterle}}]{Andrews:1997a}%
  \BibitemOpen
  \bibfield  {author} {\bibinfo {author} {\bibfnamefont {M.~R.}\ \bibnamefont
  {Andrews}}, \bibinfo {author} {\bibfnamefont {C.~G.}\ \bibnamefont
  {Townsend}}, \bibinfo {author} {\bibfnamefont {H.~J.}\ \bibnamefont
  {Miesner}}, \bibinfo {author} {\bibfnamefont {D.~S.}\ \bibnamefont {Durfee}},
  \bibinfo {author} {\bibfnamefont {D.~M.}\ \bibnamefont {Kurn}}, \ and\
  \bibinfo {author} {\bibfnamefont {W.}~\bibnamefont {Ketterle}},\ }\bibfield
  {title} {\enquote {\bibinfo {title} {Observation of interference between two
  {B}ose condensates},}\ }\href {\doibase 10.1126/science.275.5300.637}
  {\bibfield  {journal} {\bibinfo  {journal} {Science}\ }\textbf {\bibinfo
  {volume} {275}},\ \bibinfo {pages} {637} (\bibinfo {year}
  {1997}{\natexlab{a}})}\BibitemShut {NoStop}%
\bibitem [{\citenamefont {Anderson}\ and\ \citenamefont
  {Kasevich}(1998)}]{Anderson:1998}%
  \BibitemOpen
  \bibfield  {author} {\bibinfo {author} {\bibfnamefont {B.~P.}\ \bibnamefont
  {Anderson}}\ and\ \bibinfo {author} {\bibfnamefont {M.~A.}\ \bibnamefont
  {Kasevich}},\ }\bibfield  {title} {\enquote {\bibinfo {title} {Macroscopic
  quantum interference from atomic tunnel arrays},}\ }\href {\doibase
  10.1126/science.282.5394.1686} {\bibfield  {journal} {\bibinfo  {journal}
  {Science}\ }\textbf {\bibinfo {volume} {282}},\ \bibinfo {pages} {1686}
  (\bibinfo {year} {1998})}\BibitemShut {NoStop}%
\bibitem [{\citenamefont {Greiner}\ \emph {et~al.}(2001)\citenamefont
  {Greiner}, \citenamefont {Bloch}, \citenamefont {Mandel}, \citenamefont
  {H\"ansch},\ and\ \citenamefont {Esslinger}}]{Greiner:2001b}%
  \BibitemOpen
  \bibfield  {author} {\bibinfo {author} {\bibfnamefont {M.}~\bibnamefont
  {Greiner}}, \bibinfo {author} {\bibfnamefont {I.}~\bibnamefont {Bloch}},
  \bibinfo {author} {\bibfnamefont {O.}~\bibnamefont {Mandel}}, \bibinfo
  {author} {\bibfnamefont {T.~W.}\ \bibnamefont {H\"ansch}}, \ and\ \bibinfo
  {author} {\bibfnamefont {T.}~\bibnamefont {Esslinger}},\ }\bibfield  {title}
  {\enquote {\bibinfo {title} {Exploring phase coherence in a {2D} lattice of
  {B}ose-{E}instein condensates},}\ }\href {\doibase
  10.1103/PhysRevLett.87.160405} {\bibfield  {journal} {\bibinfo  {journal}
  {Phys. Rev. Lett.}\ }\textbf {\bibinfo {volume} {87}},\ \bibinfo {pages}
  {160405} (\bibinfo {year} {2001})}\BibitemShut {NoStop}%
\bibitem [{\citenamefont {Bloch}(2005)}]{Bloch:2005}%
  \BibitemOpen
  \bibfield  {author} {\bibinfo {author} {\bibfnamefont {I.}~\bibnamefont
  {Bloch}},\ }\bibfield  {title} {\enquote {\bibinfo {title} {Ultracold quantum
  gases in optical lattices},}\ }\href {\doibase
  https://doi.org/10.1038/nphys138} {\bibfield  {journal} {\bibinfo  {journal}
  {Nat. Phys.}\ }\textbf {\bibinfo {volume} {1}},\ \bibinfo {pages} {23--30}
  (\bibinfo {year} {2005})}\BibitemShut {NoStop}%
\bibitem [{\citenamefont {Jin}\ \emph {et~al.}(1996)\citenamefont {Jin},
  \citenamefont {Ensher}, \citenamefont {Matthews}, \citenamefont {Wieman},\
  and\ \citenamefont {Cornell}}]{Jin:1996a}%
  \BibitemOpen
  \bibfield  {author} {\bibinfo {author} {\bibfnamefont {D.~S.}\ \bibnamefont
  {Jin}}, \bibinfo {author} {\bibfnamefont {J.~R.}\ \bibnamefont {Ensher}},
  \bibinfo {author} {\bibfnamefont {M.~R.}\ \bibnamefont {Matthews}}, \bibinfo
  {author} {\bibfnamefont {C.~E.}\ \bibnamefont {Wieman}}, \ and\ \bibinfo
  {author} {\bibfnamefont {E.~A.}\ \bibnamefont {Cornell}},\ }\bibfield
  {title} {\enquote {\bibinfo {title} {Collective excitations of a
  {B}ose-{E}instein condensate in a dilute gas},}\ }\href {\doibase
  10.1103/PhysRevLett.77.420} {\bibfield  {journal} {\bibinfo  {journal} {Phys.
  Rev. Lett.}\ }\textbf {\bibinfo {volume} {77}},\ \bibinfo {pages} {420--423}
  (\bibinfo {year} {1996})}\BibitemShut {NoStop}%
\bibitem [{\citenamefont {Mewes}\ \emph {et~al.}(1996)\citenamefont {Mewes},
  \citenamefont {Andrews}, \citenamefont {van Druten}, \citenamefont {Kurn},
  \citenamefont {Durfee}, \citenamefont {Townsend},\ and\ \citenamefont
  {Ketterle}}]{Mewes:1996b}%
  \BibitemOpen
  \bibfield  {author} {\bibinfo {author} {\bibfnamefont {M.-O.}\ \bibnamefont
  {Mewes}}, \bibinfo {author} {\bibfnamefont {M.~R.}\ \bibnamefont {Andrews}},
  \bibinfo {author} {\bibfnamefont {N.~J.}\ \bibnamefont {van Druten}},
  \bibinfo {author} {\bibfnamefont {D.~M.}\ \bibnamefont {Kurn}}, \bibinfo
  {author} {\bibfnamefont {D.~S.}\ \bibnamefont {Durfee}}, \bibinfo {author}
  {\bibfnamefont {C.~G.}\ \bibnamefont {Townsend}}, \ and\ \bibinfo {author}
  {\bibfnamefont {W.}~\bibnamefont {Ketterle}},\ }\bibfield  {title} {\enquote
  {\bibinfo {title} {Collective excitations of a {B}ose-{E}instein condensate
  in a magnetic trap},}\ }\href {\doibase 10.1103/PhysRevLett.77.988}
  {\bibfield  {journal} {\bibinfo  {journal} {Phys. Rev. Lett.}\ }\textbf
  {\bibinfo {volume} {77}},\ \bibinfo {pages} {988--991} (\bibinfo {year}
  {1996})}\BibitemShut {NoStop}%
\bibitem [{\citenamefont {Andrews}\ \emph
  {et~al.}(1997{\natexlab{b}})\citenamefont {Andrews}, \citenamefont {Kurn},
  \citenamefont {Miesner}, \citenamefont {Durfee}, \citenamefont {Townsend},
  \citenamefont {Inouye},\ and\ \citenamefont {Ketterle}}]{Andrews:1997}%
  \BibitemOpen
  \bibfield  {author} {\bibinfo {author} {\bibfnamefont {M.~R.}\ \bibnamefont
  {Andrews}}, \bibinfo {author} {\bibfnamefont {D.~M.}\ \bibnamefont {Kurn}},
  \bibinfo {author} {\bibfnamefont {H.-J.}\ \bibnamefont {Miesner}}, \bibinfo
  {author} {\bibfnamefont {D.~S.}\ \bibnamefont {Durfee}}, \bibinfo {author}
  {\bibfnamefont {C.~G.}\ \bibnamefont {Townsend}}, \bibinfo {author}
  {\bibfnamefont {S.}~\bibnamefont {Inouye}}, \ and\ \bibinfo {author}
  {\bibfnamefont {W.}~\bibnamefont {Ketterle}},\ }\bibfield  {title} {\enquote
  {\bibinfo {title} {Propagation of sound in a {B}ose-{E}instein condensate},}\
  }\href {\doibase 10.1103/PhysRevLett.79.553} {\bibfield  {journal} {\bibinfo
  {journal} {Phys. Rev. Lett.}\ }\textbf {\bibinfo {volume} {79}},\ \bibinfo
  {pages} {553--556} (\bibinfo {year} {1997}{\natexlab{b}})}\BibitemShut
  {NoStop}%
\bibitem [{\citenamefont {Stamper-Kurn}\ \emph {et~al.}(1998)\citenamefont
  {Stamper-Kurn}, \citenamefont {Miesner}, \citenamefont {Chikkatur},
  \citenamefont {Inouye}, \citenamefont {Stenger},\ and\ \citenamefont
  {Ketterle}}]{Stamper-Kurn:1998c}%
  \BibitemOpen
  \bibfield  {author} {\bibinfo {author} {\bibfnamefont {D.~M.}\ \bibnamefont
  {Stamper-Kurn}}, \bibinfo {author} {\bibfnamefont {H.-J.}\ \bibnamefont
  {Miesner}}, \bibinfo {author} {\bibfnamefont {A.~P.}\ \bibnamefont
  {Chikkatur}}, \bibinfo {author} {\bibfnamefont {S.}~\bibnamefont {Inouye}},
  \bibinfo {author} {\bibfnamefont {J.}~\bibnamefont {Stenger}}, \ and\
  \bibinfo {author} {\bibfnamefont {W.}~\bibnamefont {Ketterle}},\ }\bibfield
  {title} {\enquote {\bibinfo {title} {Reversible formation of a
  {B}ose-{E}instein condensate},}\ }\href {\doibase
  10.1103/PhysRevLett.81.2194} {\bibfield  {journal} {\bibinfo  {journal}
  {Phys. Rev. Lett.}\ }\textbf {\bibinfo {volume} {81}},\ \bibinfo {pages}
  {2194--2197} (\bibinfo {year} {1998})}\BibitemShut {NoStop}%
\bibitem [{\citenamefont {Greiner}\ \emph {et~al.}(2002)\citenamefont
  {Greiner}, \citenamefont {Mandel}, \citenamefont {Esslinger}, \citenamefont
  {H{\"a}nsch},\ and\ \citenamefont {Bloch}}]{Greiner:2002a}%
  \BibitemOpen
  \bibfield  {author} {\bibinfo {author} {\bibfnamefont {M.}~\bibnamefont
  {Greiner}}, \bibinfo {author} {\bibfnamefont {M.~O.}\ \bibnamefont {Mandel}},
  \bibinfo {author} {\bibfnamefont {T.}~\bibnamefont {Esslinger}}, \bibinfo
  {author} {\bibfnamefont {T.}~\bibnamefont {H{\"a}nsch}}, \ and\ \bibinfo
  {author} {\bibfnamefont {I.}~\bibnamefont {Bloch}},\ }\bibfield  {title}
  {\enquote {\bibinfo {title} {Quantum phase transition from a superfluid to a
  {M}ott insulator in a gas of ultracold atoms},}\ }\href {\doibase
  https://doi.org/10.1038/415039a} {\bibfield  {journal} {\bibinfo  {journal}
  {Nature}\ }\textbf {\bibinfo {volume} {415}},\ \bibinfo {pages} {39}
  (\bibinfo {year} {2002})}\BibitemShut {NoStop}%
\bibitem [{\citenamefont {Gaunt}\ \emph {et~al.}(2013)\citenamefont {Gaunt},
  \citenamefont {Schmidutz}, \citenamefont {Gotlibovych}, \citenamefont
  {Smith},\ and\ \citenamefont {Hadzibabic}}]{Gaunt:2013}%
  \BibitemOpen
  \bibfield  {author} {\bibinfo {author} {\bibfnamefont {A.~L.}\ \bibnamefont
  {Gaunt}}, \bibinfo {author} {\bibfnamefont {T.~F.}\ \bibnamefont
  {Schmidutz}}, \bibinfo {author} {\bibfnamefont {I.}~\bibnamefont
  {Gotlibovych}}, \bibinfo {author} {\bibfnamefont {R.~P.}\ \bibnamefont
  {Smith}}, \ and\ \bibinfo {author} {\bibfnamefont {Z.}~\bibnamefont
  {Hadzibabic}},\ }\bibfield  {title} {\enquote {\bibinfo {title}
  {{Bose--Einstein Condensation of Atoms in a Uniform Potential}},}\ }\href
  {\doibase 10.1103/PhysRevLett.110.200406} {\bibfield  {journal} {\bibinfo
  {journal} {Phys. Rev. Lett.}\ }\textbf {\bibinfo {volume} {110}},\ \bibinfo
  {pages} {200406} (\bibinfo {year} {2013})}\BibitemShut {NoStop}%
\bibitem [{\citenamefont {Chomaz}\ \emph {et~al.}(2015)\citenamefont {Chomaz},
  \citenamefont {Corman}, \citenamefont {Bienaim{\'e}}, \citenamefont
  {Desbuquois}, \citenamefont {Weitenberg}, \citenamefont {Nascimb{\`e}ne},
  \citenamefont {Beugnon},\ and\ \citenamefont {Dalibard}}]{Chomaz:2015}%
  \BibitemOpen
  \bibfield  {author} {\bibinfo {author} {\bibfnamefont {L.}~\bibnamefont
  {Chomaz}}, \bibinfo {author} {\bibfnamefont {L.}~\bibnamefont {Corman}},
  \bibinfo {author} {\bibfnamefont {T.}~\bibnamefont {Bienaim{\'e}}}, \bibinfo
  {author} {\bibfnamefont {R.}~\bibnamefont {Desbuquois}}, \bibinfo {author}
  {\bibfnamefont {C.}~\bibnamefont {Weitenberg}}, \bibinfo {author}
  {\bibfnamefont {S.}~\bibnamefont {Nascimb{\`e}ne}}, \bibinfo {author}
  {\bibfnamefont {J.}~\bibnamefont {Beugnon}}, \ and\ \bibinfo {author}
  {\bibfnamefont {J.}~\bibnamefont {Dalibard}},\ }\bibfield  {title} {\enquote
  {\bibinfo {title} {{Emergence of coherence via transverse condensation in a
  uniform quasi-two-dimensional Bose gas}},}\ }\href
  {https://doi.org/10.1038/ncomms7162} {\bibfield  {journal} {\bibinfo
  {journal} {Nat. Commun.}\ }\textbf {\bibinfo {volume} {6}},\ \bibinfo {pages}
  {6162} (\bibinfo {year} {2015})}\BibitemShut {NoStop}%
\bibitem [{\citenamefont {Mukherjee}\ \emph {et~al.}(2017)\citenamefont
  {Mukherjee}, \citenamefont {Yan}, \citenamefont {Patel}, \citenamefont
  {Hadzibabic}, \citenamefont {Yefsah}, \citenamefont {Struck},\ and\
  \citenamefont {Zwierlein}}]{Mukherjee:2017}%
  \BibitemOpen
  \bibfield  {author} {\bibinfo {author} {\bibfnamefont {B.}~\bibnamefont
  {Mukherjee}}, \bibinfo {author} {\bibfnamefont {Z.}~\bibnamefont {Yan}},
  \bibinfo {author} {\bibfnamefont {P.~B.}\ \bibnamefont {Patel}}, \bibinfo
  {author} {\bibfnamefont {Z.}~\bibnamefont {Hadzibabic}}, \bibinfo {author}
  {\bibfnamefont {T.}~\bibnamefont {Yefsah}}, \bibinfo {author} {\bibfnamefont
  {J.}~\bibnamefont {Struck}}, \ and\ \bibinfo {author} {\bibfnamefont {M.~W.}\
  \bibnamefont {Zwierlein}},\ }\bibfield  {title} {\enquote {\bibinfo {title}
  {Homogeneous atomic {F}ermi gases},}\ }\href {\doibase
  10.1103/PhysRevLett.118.123401} {\bibfield  {journal} {\bibinfo  {journal}
  {Phys. Rev. Lett.}\ }\textbf {\bibinfo {volume} {118}},\ \bibinfo {pages}
  {123401} (\bibinfo {year} {2017})}\BibitemShut {NoStop}%
\bibitem [{\citenamefont {Hueck}\ \emph {et~al.}(2018)\citenamefont {Hueck},
  \citenamefont {Luick}, \citenamefont {Sobirey}, \citenamefont {Siegl},
  \citenamefont {Lompe},\ and\ \citenamefont {Moritz}}]{Hueck:2018}%
  \BibitemOpen
  \bibfield  {author} {\bibinfo {author} {\bibfnamefont {K.}~\bibnamefont
  {Hueck}}, \bibinfo {author} {\bibfnamefont {N.}~\bibnamefont {Luick}},
  \bibinfo {author} {\bibfnamefont {L.}~\bibnamefont {Sobirey}}, \bibinfo
  {author} {\bibfnamefont {J.}~\bibnamefont {Siegl}}, \bibinfo {author}
  {\bibfnamefont {T.}~\bibnamefont {Lompe}}, \ and\ \bibinfo {author}
  {\bibfnamefont {H.}~\bibnamefont {Moritz}},\ }\bibfield  {title} {\enquote
  {\bibinfo {title} {Two-dimensional homogeneous {F}ermi gases},}\ }\href
  {\doibase 10.1103/PhysRevLett.120.060402} {\bibfield  {journal} {\bibinfo
  {journal} {Phys. Rev. Lett.}\ }\textbf {\bibinfo {volume} {120}},\ \bibinfo
  {pages} {060402} (\bibinfo {year} {2018})}\BibitemShut {NoStop}%
\bibitem [{\citenamefont {{Tajik}}\ \emph {et~al.}(2019)\citenamefont
  {{Tajik}}, \citenamefont {{Rauer}}, \citenamefont {{Schweigler}},
  \citenamefont {{Cataldini}}, \citenamefont {{Sabino}}, \citenamefont
  {{M{\o}ller}}, \citenamefont {{Ji}}, \citenamefont {{Mazets}},\ and\
  \citenamefont {{Schmiedmayer}}}]{Tajik:2019}%
  \BibitemOpen
  \bibfield  {author} {\bibinfo {author} {\bibfnamefont {M.}~\bibnamefont
  {{Tajik}}}, \bibinfo {author} {\bibfnamefont {B.}~\bibnamefont {{Rauer}}},
  \bibinfo {author} {\bibfnamefont {T.}~\bibnamefont {{Schweigler}}}, \bibinfo
  {author} {\bibfnamefont {F.}~\bibnamefont {{Cataldini}}}, \bibinfo {author}
  {\bibfnamefont {J.}~\bibnamefont {{Sabino}}}, \bibinfo {author}
  {\bibfnamefont {F.~S.}\ \bibnamefont {{M{\o}ller}}}, \bibinfo {author}
  {\bibfnamefont {S.-C.}\ \bibnamefont {{Ji}}}, \bibinfo {author}
  {\bibfnamefont {I.~E.}\ \bibnamefont {{Mazets}}}, \ and\ \bibinfo {author}
  {\bibfnamefont {J.}~\bibnamefont {{Schmiedmayer}}},\ }\bibfield  {title}
  {\enquote {\bibinfo {title} {{Designing arbitrary one-dimensional potentials
  on an atom chip}},}\ }\href {\doibase 10.1364/OE.27.033474} {\bibfield
  {journal} {\bibinfo  {journal} {Optics Express}\ }\textbf {\bibinfo {volume}
  {27}},\ \bibinfo {pages} {33474} (\bibinfo {year} {2019})}\BibitemShut
  {NoStop}%
\bibitem [{\citenamefont {{Bause}}\ \emph {et~al.}(2021)\citenamefont
  {{Bause}}, \citenamefont {{Schindewolf}}, \citenamefont {{Tao}},
  \citenamefont {{Duda}}, \citenamefont {{Chen}}, \citenamefont
  {{Qu{\'e}m{\'e}ner}}, \citenamefont {{Karman}}, \citenamefont
  {{Christianen}}, \citenamefont {{Bloch}},\ and\ \citenamefont
  {{Luo}}}]{Bause:2021}%
  \BibitemOpen
  \bibfield  {author} {\bibinfo {author} {\bibfnamefont {R.}~\bibnamefont
  {{Bause}}}, \bibinfo {author} {\bibfnamefont {A.}~\bibnamefont
  {{Schindewolf}}}, \bibinfo {author} {\bibfnamefont {R.}~\bibnamefont
  {{Tao}}}, \bibinfo {author} {\bibfnamefont {M.}~\bibnamefont {{Duda}}},
  \bibinfo {author} {\bibfnamefont {X.-Y.}\ \bibnamefont {{Chen}}}, \bibinfo
  {author} {\bibfnamefont {G.}~\bibnamefont {{Qu{\'e}m{\'e}ner}}}, \bibinfo
  {author} {\bibfnamefont {T.}~\bibnamefont {{Karman}}}, \bibinfo {author}
  {\bibfnamefont {A.}~\bibnamefont {{Christianen}}}, \bibinfo {author}
  {\bibfnamefont {I.}~\bibnamefont {{Bloch}}}, \ and\ \bibinfo {author}
  {\bibfnamefont {X.-Y.}\ \bibnamefont {{Luo}}},\ }\href@noop {} {\enquote
  {\bibinfo {title} {{Collisions of ultracold molecules in bright and dark
  optical dipole traps}},}\ } (\bibinfo {year} {2021}),\ \Eprint
  {http://arxiv.org/abs/2103.00889} {arXiv:2103.00889} \BibitemShut {NoStop}%
\bibitem [{\citenamefont {{Amico}}\ \emph {et~al.}(2020)\citenamefont
  {{Amico}}, \citenamefont {{Boshier}}, \citenamefont {{Birkl}}, \citenamefont
  {{Minguzzi}}, \citenamefont {{Miniatura}}, \citenamefont {{Kwek}},
  \citenamefont {{Aghamalyan}}, \citenamefont {{Ahufinger}}, \citenamefont
  {{Anderson}}, \citenamefont {{Andrei}}, \citenamefont {{Arnold}},
  \citenamefont {{Baker}}, \citenamefont {{Bell}}, \citenamefont {{Bland}},
  \citenamefont {{Brantut}}, \citenamefont {{Cassettari}}, \citenamefont
  {{Chetcuti}}, \citenamefont {{Chevy}}, \citenamefont {{Citro}}, \citenamefont
  {{De Palo}}, \citenamefont {{Dumke}}, \citenamefont {{Edwards}},
  \citenamefont {{Folman}}, \citenamefont {{Fortagh}}, \citenamefont
  {{Gardiner}}, \citenamefont {{Garraway}}, \citenamefont {{Gauthier}},
  \citenamefont {{G{\"u}nther}}, \citenamefont {{Haug}}, \citenamefont
  {{Hufnagel}}, \citenamefont {{Keil}}, \citenamefont {{von Klitzing}},
  \citenamefont {{Ireland}}, \citenamefont {{Lebrat}}, \citenamefont {{Li}},
  \citenamefont {{Longchambon}}, \citenamefont {{Mompart}}, \citenamefont
  {{Morsch}}, \citenamefont {{Naldesi}}, \citenamefont {{Neely}}, \citenamefont
  {{Olshanii}}, \citenamefont {{Orignac}}, \citenamefont {{Pandey}},
  \citenamefont {{P{\'e}rez-Obiol}}, \citenamefont {{Perrin}}, \citenamefont
  {{Piroli}}, \citenamefont {{Polo}}, \citenamefont {{Pritchard}},
  \citenamefont {{Proukakis}}, \citenamefont {{Rylands}}, \citenamefont
  {{Rubinsztein-Dunlop}}, \citenamefont {{Scazza}}, \citenamefont
  {{Stringari}}, \citenamefont {{Tosto}}, \citenamefont {{Trombettoni}},
  \citenamefont {{Victorin}}, \citenamefont {{Wilkowski}}, \citenamefont
  {{Xhani}},\ and\ \citenamefont {{Yakimenko}}}]{Amico:2020}%
  \BibitemOpen
  \bibfield  {author} {\bibinfo {author} {\bibfnamefont {L.}~\bibnamefont
  {{Amico}}}, \bibinfo {author} {\bibfnamefont {M.}~\bibnamefont {{Boshier}}},
  \bibinfo {author} {\bibfnamefont {G.}~\bibnamefont {{Birkl}}}, \bibinfo
  {author} {\bibfnamefont {A.}~\bibnamefont {{Minguzzi}}}, \bibinfo {author}
  {\bibfnamefont {C.}~\bibnamefont {{Miniatura}}}, \bibinfo {author}
  {\bibfnamefont {L.~C.}\ \bibnamefont {{Kwek}}}, \bibinfo {author}
  {\bibfnamefont {D.}~\bibnamefont {{Aghamalyan}}}, \bibinfo {author}
  {\bibfnamefont {V.}~\bibnamefont {{Ahufinger}}}, \bibinfo {author}
  {\bibfnamefont {D.}~\bibnamefont {{Anderson}}}, \bibinfo {author}
  {\bibfnamefont {N.}~\bibnamefont {{Andrei}}}, \bibinfo {author}
  {\bibfnamefont {A.~S.}\ \bibnamefont {{Arnold}}}, \bibinfo {author}
  {\bibfnamefont {M.}~\bibnamefont {{Baker}}}, \bibinfo {author} {\bibfnamefont
  {T.~A.}\ \bibnamefont {{Bell}}}, \bibinfo {author} {\bibfnamefont
  {T.}~\bibnamefont {{Bland}}}, \bibinfo {author} {\bibfnamefont {J.~P.}\
  \bibnamefont {{Brantut}}}, \bibinfo {author} {\bibfnamefont {D.}~\bibnamefont
  {{Cassettari}}}, \bibinfo {author} {\bibfnamefont {W.~J.}\ \bibnamefont
  {{Chetcuti}}}, \bibinfo {author} {\bibfnamefont {F.}~\bibnamefont {{Chevy}}},
  \bibinfo {author} {\bibfnamefont {R.}~\bibnamefont {{Citro}}}, \bibinfo
  {author} {\bibfnamefont {S.}~\bibnamefont {{De Palo}}}, \bibinfo {author}
  {\bibfnamefont {R.}~\bibnamefont {{Dumke}}}, \bibinfo {author} {\bibfnamefont
  {M.}~\bibnamefont {{Edwards}}}, \bibinfo {author} {\bibfnamefont
  {R.}~\bibnamefont {{Folman}}}, \bibinfo {author} {\bibfnamefont
  {J.}~\bibnamefont {{Fortagh}}}, \bibinfo {author} {\bibfnamefont {S.~A.}\
  \bibnamefont {{Gardiner}}}, \bibinfo {author} {\bibfnamefont {B.~M.}\
  \bibnamefont {{Garraway}}}, \bibinfo {author} {\bibfnamefont
  {G.}~\bibnamefont {{Gauthier}}}, \bibinfo {author} {\bibfnamefont
  {A.}~\bibnamefont {{G{\"u}nther}}}, \bibinfo {author} {\bibfnamefont
  {T.}~\bibnamefont {{Haug}}}, \bibinfo {author} {\bibfnamefont
  {C.}~\bibnamefont {{Hufnagel}}}, \bibinfo {author} {\bibfnamefont
  {M.}~\bibnamefont {{Keil}}}, \bibinfo {author} {\bibfnamefont
  {W.}~\bibnamefont {{von Klitzing}}}, \bibinfo {author} {\bibfnamefont
  {P.}~\bibnamefont {{Ireland}}}, \bibinfo {author} {\bibfnamefont
  {M.}~\bibnamefont {{Lebrat}}}, \bibinfo {author} {\bibfnamefont
  {W.}~\bibnamefont {{Li}}}, \bibinfo {author} {\bibfnamefont {L.}~\bibnamefont
  {{Longchambon}}}, \bibinfo {author} {\bibfnamefont {J.}~\bibnamefont
  {{Mompart}}}, \bibinfo {author} {\bibfnamefont {O.}~\bibnamefont {{Morsch}}},
  \bibinfo {author} {\bibfnamefont {P.}~\bibnamefont {{Naldesi}}}, \bibinfo
  {author} {\bibfnamefont {T.~W.}\ \bibnamefont {{Neely}}}, \bibinfo {author}
  {\bibfnamefont {M.}~\bibnamefont {{Olshanii}}}, \bibinfo {author}
  {\bibfnamefont {E.}~\bibnamefont {{Orignac}}}, \bibinfo {author}
  {\bibfnamefont {S.}~\bibnamefont {{Pandey}}}, \bibinfo {author}
  {\bibfnamefont {A.}~\bibnamefont {{P{\'e}rez-Obiol}}}, \bibinfo {author}
  {\bibfnamefont {H.}~\bibnamefont {{Perrin}}}, \bibinfo {author}
  {\bibfnamefont {L.}~\bibnamefont {{Piroli}}}, \bibinfo {author}
  {\bibfnamefont {J.}~\bibnamefont {{Polo}}}, \bibinfo {author} {\bibfnamefont
  {A.~L.}\ \bibnamefont {{Pritchard}}}, \bibinfo {author} {\bibfnamefont
  {N.~P.}\ \bibnamefont {{Proukakis}}}, \bibinfo {author} {\bibfnamefont
  {C.}~\bibnamefont {{Rylands}}}, \bibinfo {author} {\bibfnamefont
  {H.}~\bibnamefont {{Rubinsztein-Dunlop}}}, \bibinfo {author} {\bibfnamefont
  {F.}~\bibnamefont {{Scazza}}}, \bibinfo {author} {\bibfnamefont
  {S.}~\bibnamefont {{Stringari}}}, \bibinfo {author} {\bibfnamefont
  {F.}~\bibnamefont {{Tosto}}}, \bibinfo {author} {\bibfnamefont
  {A.}~\bibnamefont {{Trombettoni}}}, \bibinfo {author} {\bibfnamefont
  {N.}~\bibnamefont {{Victorin}}}, \bibinfo {author} {\bibfnamefont
  {D.}~\bibnamefont {{Wilkowski}}}, \bibinfo {author} {\bibfnamefont
  {K.}~\bibnamefont {{Xhani}}}, \ and\ \bibinfo {author} {\bibfnamefont
  {A.}~\bibnamefont {{Yakimenko}}},\ }\href@noop {} {\enquote {\bibinfo {title}
  {{State of the art and perspective on Atomtronics}},}\ } (\bibinfo {year}
  {2020}),\ \Eprint {http://arxiv.org/abs/2008.04439} {arXiv:2008.04439}
  \BibitemShut {NoStop}%
\bibitem [{\citenamefont {{Ni}}\ and\ \citenamefont
  {{Kaufman}}(2021)}]{Ni:2021}%
  \BibitemOpen
  \bibfield  {author} {\bibinfo {author} {\bibfnamefont {K.-K.}\ \bibnamefont
  {{Ni}}}\ and\ \bibinfo {author} {\bibfnamefont {A.}~\bibnamefont
  {{Kaufman}}},\ }\bibfield  {title} {\enquote {\bibinfo {title} {{Optical
  tweezer arrays of ultracold atoms and molecules}},}\ }\href@noop {}
  {\bibfield  {journal} {\bibinfo  {journal} {in preparation.}\ }\ }%
\bibitem [{\citenamefont {Corman}\ \emph {et~al.}(2014)\citenamefont {Corman},
  \citenamefont {Chomaz}, \citenamefont {Bienaim\'e}, \citenamefont
  {Desbuquois}, \citenamefont {Weitenberg}, \citenamefont {Nascimb\`ene},
  \citenamefont {Dalibard},\ and\ \citenamefont {Beugnon}}]{Corman:2014}%
  \BibitemOpen
  \bibfield  {author} {\bibinfo {author} {\bibfnamefont {L.}~\bibnamefont
  {Corman}}, \bibinfo {author} {\bibfnamefont {L.}~\bibnamefont {Chomaz}},
  \bibinfo {author} {\bibfnamefont {T.}~\bibnamefont {Bienaim\'e}}, \bibinfo
  {author} {\bibfnamefont {R.}~\bibnamefont {Desbuquois}}, \bibinfo {author}
  {\bibfnamefont {C.}~\bibnamefont {Weitenberg}}, \bibinfo {author}
  {\bibfnamefont {S.}~\bibnamefont {Nascimb\`ene}}, \bibinfo {author}
  {\bibfnamefont {J.}~\bibnamefont {Dalibard}}, \ and\ \bibinfo {author}
  {\bibfnamefont {J.}~\bibnamefont {Beugnon}},\ }\bibfield  {title} {\enquote
  {\bibinfo {title} {{Quench-Induced Supercurrents in an Annular Bose Gas}},}\
  }\href {\doibase 10.1103/PhysRevLett.113.135302} {\bibfield  {journal}
  {\bibinfo  {journal} {Phys. Rev. Lett.}\ }\textbf {\bibinfo {volume} {113}},\
  \bibinfo {pages} {135302} (\bibinfo {year} {2014})}\BibitemShut {NoStop}%
\bibitem [{\citenamefont {Navon}\ \emph {et~al.}(2015)\citenamefont {Navon},
  \citenamefont {Gaunt}, \citenamefont {Smith},\ and\ \citenamefont
  {Hadzibabic}}]{Navon:2015}%
  \BibitemOpen
  \bibfield  {author} {\bibinfo {author} {\bibfnamefont {N.}~\bibnamefont
  {Navon}}, \bibinfo {author} {\bibfnamefont {A.~L.}\ \bibnamefont {Gaunt}},
  \bibinfo {author} {\bibfnamefont {R.~P.}\ \bibnamefont {Smith}}, \ and\
  \bibinfo {author} {\bibfnamefont {Z.}~\bibnamefont {Hadzibabic}},\ }\bibfield
   {title} {\enquote {\bibinfo {title} {{Critical dynamics of spontaneous
  symmetry breaking in a homogeneous Bose gas}},}\ }\href {\doibase
  10.1126/science.1258676} {\bibfield  {journal} {\bibinfo  {journal}
  {Science}\ }\textbf {\bibinfo {volume} {347}},\ \bibinfo {pages} {167}
  (\bibinfo {year} {2015})}\BibitemShut {NoStop}%
\bibitem [{\citenamefont {Keesling}\ \emph {et~al.}(2019)\citenamefont
  {Keesling}, \citenamefont {Omran}, \citenamefont {Levine}, \citenamefont
  {Bernien}, \citenamefont {Pichler}, \citenamefont {Choi}, \citenamefont
  {Samajdar}, \citenamefont {Schwartz}, \citenamefont {Silvi}, \citenamefont
  {Sachdev}, \citenamefont {Zoller}, \citenamefont {Endres}, \citenamefont
  {Greiner}, \citenamefont {Vuleti{\'c}},\ and\ \citenamefont
  {Lukin}}]{Keesling:2019}%
  \BibitemOpen
  \bibfield  {author} {\bibinfo {author} {\bibfnamefont {A.}~\bibnamefont
  {Keesling}}, \bibinfo {author} {\bibfnamefont {A.}~\bibnamefont {Omran}},
  \bibinfo {author} {\bibfnamefont {H.}~\bibnamefont {Levine}}, \bibinfo
  {author} {\bibfnamefont {H.}~\bibnamefont {Bernien}}, \bibinfo {author}
  {\bibfnamefont {H.}~\bibnamefont {Pichler}}, \bibinfo {author} {\bibfnamefont
  {S.}~\bibnamefont {Choi}}, \bibinfo {author} {\bibfnamefont {R.}~\bibnamefont
  {Samajdar}}, \bibinfo {author} {\bibfnamefont {S.}~\bibnamefont {Schwartz}},
  \bibinfo {author} {\bibfnamefont {P.}~\bibnamefont {Silvi}}, \bibinfo
  {author} {\bibfnamefont {S.}~\bibnamefont {Sachdev}}, \bibinfo {author}
  {\bibfnamefont {P.}~\bibnamefont {Zoller}}, \bibinfo {author} {\bibfnamefont
  {M.}~\bibnamefont {Endres}}, \bibinfo {author} {\bibfnamefont
  {M.}~\bibnamefont {Greiner}}, \bibinfo {author} {\bibfnamefont
  {V.}~\bibnamefont {Vuleti{\'c}}}, \ and\ \bibinfo {author} {\bibfnamefont
  {M.~D.}\ \bibnamefont {Lukin}},\ }\bibfield  {title} {\enquote {\bibinfo
  {title} {{Quantum Kibble--Zurek mechanism and critical dynamics on a
  programmable Rydberg simulator}},}\ }\href {\doibase
  10.1038/s41586-019-1070-1} {\bibfield  {journal} {\bibinfo  {journal}
  {Nature}\ }\textbf {\bibinfo {volume} {568}},\ \bibinfo {pages} {207}
  (\bibinfo {year} {2019})}\BibitemShut {NoStop}%
\bibitem [{\citenamefont {Schmidutz}\ \emph {et~al.}(2014)\citenamefont
  {Schmidutz}, \citenamefont {Gotlibovych}, \citenamefont {Gaunt},
  \citenamefont {Smith}, \citenamefont {Navon},\ and\ \citenamefont
  {Hadzibabic}}]{Schmidutz:2014}%
  \BibitemOpen
  \bibfield  {author} {\bibinfo {author} {\bibfnamefont {T.~F.}\ \bibnamefont
  {Schmidutz}}, \bibinfo {author} {\bibfnamefont {I.}~\bibnamefont
  {Gotlibovych}}, \bibinfo {author} {\bibfnamefont {A.~L.}\ \bibnamefont
  {Gaunt}}, \bibinfo {author} {\bibfnamefont {R.~P.}\ \bibnamefont {Smith}},
  \bibinfo {author} {\bibfnamefont {N.}~\bibnamefont {Navon}}, \ and\ \bibinfo
  {author} {\bibfnamefont {Z.}~\bibnamefont {Hadzibabic}},\ }\bibfield  {title}
  {\enquote {\bibinfo {title} {Quantum {J}oule--{T}homson effect in a saturated
  homogeneous {B}ose gas},}\ }\href {\doibase 10.1103/PhysRevLett.112.040403}
  {\bibfield  {journal} {\bibinfo  {journal} {Phys. Rev. Lett.}\ }\textbf
  {\bibinfo {volume} {112}},\ \bibinfo {pages} {040403} (\bibinfo {year}
  {2014})}\BibitemShut {NoStop}%
\bibitem [{\citenamefont {Meyrath}\ \emph {et~al.}(2005)\citenamefont
  {Meyrath}, \citenamefont {Schreck}, \citenamefont {Hanssen}, \citenamefont
  {Chuu},\ and\ \citenamefont {Raizen}}]{Meyrath:2005}%
  \BibitemOpen
  \bibfield  {author} {\bibinfo {author} {\bibfnamefont {T.~P.}\ \bibnamefont
  {Meyrath}}, \bibinfo {author} {\bibfnamefont {F.}~\bibnamefont {Schreck}},
  \bibinfo {author} {\bibfnamefont {J.~L.}\ \bibnamefont {Hanssen}}, \bibinfo
  {author} {\bibfnamefont {C.-S.}\ \bibnamefont {Chuu}}, \ and\ \bibinfo
  {author} {\bibfnamefont {M.~G.}\ \bibnamefont {Raizen}},\ }\bibfield  {title}
  {\enquote {\bibinfo {title} {Bose--{E}instein condensate in a box},}\ }\href
  {\doibase 10.1103/PhysRevA.71.041604} {\bibfield  {journal} {\bibinfo
  {journal} {Phys. Rev. A}\ }\textbf {\bibinfo {volume} {71}},\ \bibinfo
  {pages} {041604} (\bibinfo {year} {2005})}\BibitemShut {NoStop}%
\bibitem [{\citenamefont {Van~Es}\ \emph {et~al.}(2010)\citenamefont {Van~Es},
  \citenamefont {Wicke}, \citenamefont {Van~Amerongen}, \citenamefont
  {R{\'e}tif}, \citenamefont {Whitlock},\ and\ \citenamefont
  {Van~Druten}}]{van2010box}%
  \BibitemOpen
  \bibfield  {author} {\bibinfo {author} {\bibfnamefont {J.}~\bibnamefont
  {Van~Es}}, \bibinfo {author} {\bibfnamefont {P.}~\bibnamefont {Wicke}},
  \bibinfo {author} {\bibfnamefont {A.}~\bibnamefont {Van~Amerongen}}, \bibinfo
  {author} {\bibfnamefont {C.}~\bibnamefont {R{\'e}tif}}, \bibinfo {author}
  {\bibfnamefont {S.}~\bibnamefont {Whitlock}}, \ and\ \bibinfo {author}
  {\bibfnamefont {N.}~\bibnamefont {Van~Druten}},\ }\bibfield  {title}
  {\enquote {\bibinfo {title} {Box traps on an atom chip for one-dimensional
  quantum gases},}\ }\href {\doibase
  https://doi.org/10.1088/0953-4075/43/15/155002} {\bibfield  {journal}
  {\bibinfo  {journal} {Journal of Physics B: Atomic, Molecular and Optical
  Physics}\ }\textbf {\bibinfo {volume} {43}},\ \bibinfo {pages} {155002}
  (\bibinfo {year} {2010})}\BibitemShut {NoStop}%
\bibitem [{\citenamefont {{Gauthier}}\ \emph {et~al.}(2021)\citenamefont
  {{Gauthier}}, \citenamefont {{Bell}}, \citenamefont {{Stilgoe}},
  \citenamefont {{Baker}}, \citenamefont {{Rubinsztein-Dunlop}},\ and\
  \citenamefont {{Neely}}}]{Gauthier:2021}%
  \BibitemOpen
  \bibfield  {author} {\bibinfo {author} {\bibfnamefont {G.}~\bibnamefont
  {{Gauthier}}}, \bibinfo {author} {\bibfnamefont {T.~A.}\ \bibnamefont
  {{Bell}}}, \bibinfo {author} {\bibfnamefont {A.~B.}\ \bibnamefont
  {{Stilgoe}}}, \bibinfo {author} {\bibfnamefont {M.}~\bibnamefont {{Baker}}},
  \bibinfo {author} {\bibfnamefont {H.}~\bibnamefont {{Rubinsztein-Dunlop}}}, \
  and\ \bibinfo {author} {\bibfnamefont {T.~W.}\ \bibnamefont {{Neely}}},\
  }\href@noop {} {\enquote {\bibinfo {title} {{Dynamic high-resolution optical
  trapping of ultracold atoms}},}\ } (\bibinfo {year} {2021}),\ \Eprint
  {http://arxiv.org/abs/2103.10020} {arXiv:2103.10020} \BibitemShut {NoStop}%
\bibitem [{\citenamefont {{Shibata}}\ \emph {et~al.}(2020)\citenamefont
  {{Shibata}}, \citenamefont {{Ikeda}}, \citenamefont {{Suzuki}},\ and\
  \citenamefont {{Hirano}}}]{Shibata:2020}%
  \BibitemOpen
  \bibfield  {author} {\bibinfo {author} {\bibfnamefont {K.}~\bibnamefont
  {{Shibata}}}, \bibinfo {author} {\bibfnamefont {H.}~\bibnamefont {{Ikeda}}},
  \bibinfo {author} {\bibfnamefont {R.}~\bibnamefont {{Suzuki}}}, \ and\
  \bibinfo {author} {\bibfnamefont {T.}~\bibnamefont {{Hirano}}},\ }\bibfield
  {title} {\enquote {\bibinfo {title} {{Compensation of gravity on cold atoms
  by a linear optical potential}},}\ }\href {\doibase
  10.1103/PhysRevResearch.2.013068} {\bibfield  {journal} {\bibinfo  {journal}
  {Physical Review Research}\ }\textbf {\bibinfo {volume} {2}},\ \bibinfo {eid}
  {013068} (\bibinfo {year} {2020})}\BibitemShut {NoStop}%
\bibitem [{\citenamefont {Gaunt}(2014)}]{Gaunt:2014-th}%
  \BibitemOpen
  \bibfield  {author} {\bibinfo {author} {\bibfnamefont {A.~L.}\ \bibnamefont
  {Gaunt}},\ }\emph {\bibinfo {title} {{Degenerate Bose Gases: Tuning
  Interactions $\&$ Geometry}}},\ \href@noop {} {Ph.D. thesis},\ \bibinfo
  {school} {University of Cambridge} (\bibinfo {year} {2014})\BibitemShut
  {NoStop}%
\bibitem [{\citenamefont {Gauthier}\ \emph {et~al.}(2016)\citenamefont
  {Gauthier}, \citenamefont {Lenton}, \citenamefont {Parry}, \citenamefont
  {Baker}, \citenamefont {Davis}, \citenamefont {Rubinsztein-Dunlop},\ and\
  \citenamefont {Neely}}]{Gauthier:2016}%
  \BibitemOpen
  \bibfield  {author} {\bibinfo {author} {\bibfnamefont {G.}~\bibnamefont
  {Gauthier}}, \bibinfo {author} {\bibfnamefont {I.}~\bibnamefont {Lenton}},
  \bibinfo {author} {\bibfnamefont {N.~M.}\ \bibnamefont {Parry}}, \bibinfo
  {author} {\bibfnamefont {M.}~\bibnamefont {Baker}}, \bibinfo {author}
  {\bibfnamefont {M.~J.}\ \bibnamefont {Davis}}, \bibinfo {author}
  {\bibfnamefont {H.}~\bibnamefont {Rubinsztein-Dunlop}}, \ and\ \bibinfo
  {author} {\bibfnamefont {T.~W.}\ \bibnamefont {Neely}},\ }\bibfield  {title}
  {\enquote {\bibinfo {title} {Direct imaging of a digital-micromirror device
  for configurable microscopic optical potentials},}\ }\href {\doibase
  10.1364/OPTICA.3.001136} {\bibfield  {journal} {\bibinfo  {journal} {Optica}\
  }\textbf {\bibinfo {volume} {3}},\ \bibinfo {pages} {1136--1143} (\bibinfo
  {year} {2016})}\BibitemShut {NoStop}%
\bibitem [{\citenamefont {Manek}\ \emph {et~al.}(1998)\citenamefont {Manek},
  \citenamefont {Ovchinnikov},\ and\ \citenamefont
  {Grimm}}]{manek1998generation}%
  \BibitemOpen
  \bibfield  {author} {\bibinfo {author} {\bibfnamefont {I.}~\bibnamefont
  {Manek}}, \bibinfo {author} {\bibfnamefont {Y.~B.}\ \bibnamefont
  {Ovchinnikov}}, \ and\ \bibinfo {author} {\bibfnamefont {R.}~\bibnamefont
  {Grimm}},\ }\bibfield  {title} {\enquote {\bibinfo {title} {Generation of a
  hollow laser beam for atom trapping using an axicon},}\ }\href {\doibase
  https://doi.org/10.1016/S0030-4018(97)00645-7} {\bibfield  {journal}
  {\bibinfo  {journal} {Opt. Comm.}\ }\textbf {\bibinfo {volume} {147}},\
  \bibinfo {pages} {67--70} (\bibinfo {year} {1998})}\BibitemShut {NoStop}%
\bibitem [{\citenamefont {{Henderson}}\ \emph {et~al.}(2009)\citenamefont
  {{Henderson}}, \citenamefont {{Ryu}}, \citenamefont {{MacCormick}},\ and\
  \citenamefont {{Boshier}}}]{Henderson:2009}%
  \BibitemOpen
  \bibfield  {author} {\bibinfo {author} {\bibfnamefont {K.}~\bibnamefont
  {{Henderson}}}, \bibinfo {author} {\bibfnamefont {C.}~\bibnamefont {{Ryu}}},
  \bibinfo {author} {\bibfnamefont {C.}~\bibnamefont {{MacCormick}}}, \ and\
  \bibinfo {author} {\bibfnamefont {M.~G.}\ \bibnamefont {{Boshier}}},\
  }\bibfield  {title} {\enquote {\bibinfo {title} {{Experimental demonstration
  of painting arbitrary and dynamic potentials for Bose--Einstein
  condensates}},}\ }\href {\doibase 10.1088/1367-2630/11/4/043030} {\bibfield
  {journal} {\bibinfo  {journal} {New Journal of Physics}\ }\textbf {\bibinfo
  {volume} {11}},\ \bibinfo {eid} {043030} (\bibinfo {year}
  {2009})}\BibitemShut {NoStop}%
\bibitem [{\citenamefont {Ville}\ \emph {et~al.}(2017)\citenamefont {Ville},
  \citenamefont {Bienaim\'e}, \citenamefont {Saint-Jalm}, \citenamefont
  {Corman}, \citenamefont {Aidelsburger}, \citenamefont {Chomaz}, \citenamefont
  {Kleinlein}, \citenamefont {Perconte}, \citenamefont {Nascimb\`ene},
  \citenamefont {Dalibard},\ and\ \citenamefont {Beugnon}}]{Ville:2017}%
  \BibitemOpen
  \bibfield  {author} {\bibinfo {author} {\bibfnamefont {J.~L.}\ \bibnamefont
  {Ville}}, \bibinfo {author} {\bibfnamefont {T.}~\bibnamefont {Bienaim\'e}},
  \bibinfo {author} {\bibfnamefont {R.}~\bibnamefont {Saint-Jalm}}, \bibinfo
  {author} {\bibfnamefont {L.}~\bibnamefont {Corman}}, \bibinfo {author}
  {\bibfnamefont {M.}~\bibnamefont {Aidelsburger}}, \bibinfo {author}
  {\bibfnamefont {L.}~\bibnamefont {Chomaz}}, \bibinfo {author} {\bibfnamefont
  {K.}~\bibnamefont {Kleinlein}}, \bibinfo {author} {\bibfnamefont
  {D.}~\bibnamefont {Perconte}}, \bibinfo {author} {\bibfnamefont
  {S.}~\bibnamefont {Nascimb\`ene}}, \bibinfo {author} {\bibfnamefont
  {J.}~\bibnamefont {Dalibard}}, \ and\ \bibinfo {author} {\bibfnamefont
  {J.}~\bibnamefont {Beugnon}},\ }\bibfield  {title} {\enquote {\bibinfo
  {title} {Loading and compression of a single two-dimensional {B}ose gas in an
  optical accordion},}\ }\href {\doibase 10.1103/PhysRevA.95.013632} {\bibfield
   {journal} {\bibinfo  {journal} {Phys. Rev. A}\ }\textbf {\bibinfo {volume}
  {95}},\ \bibinfo {pages} {013632} (\bibinfo {year} {2017})}\BibitemShut
  {NoStop}%
\bibitem [{\citenamefont {Chin}\ \emph {et~al.}(2010)\citenamefont {Chin},
  \citenamefont {Grimm}, \citenamefont {Julienne},\ and\ \citenamefont
  {Tiesinga}}]{Chin:2010}%
  \BibitemOpen
  \bibfield  {author} {\bibinfo {author} {\bibfnamefont {C.}~\bibnamefont
  {Chin}}, \bibinfo {author} {\bibfnamefont {R.}~\bibnamefont {Grimm}},
  \bibinfo {author} {\bibfnamefont {P.}~\bibnamefont {Julienne}}, \ and\
  \bibinfo {author} {\bibfnamefont {E.}~\bibnamefont {Tiesinga}},\ }\bibfield
  {title} {\enquote {\bibinfo {title} {Feshbach resonances in ultracold
  gases},}\ }\href {\doibase 10.1103/RevModPhys.82.1225} {\bibfield  {journal}
  {\bibinfo  {journal} {Rev. Mod. Phys.}\ }\textbf {\bibinfo {volume} {82}},\
  \bibinfo {pages} {1225--1286} (\bibinfo {year} {2010})}\BibitemShut {NoStop}%
\bibitem [{\citenamefont {Tammuz}\ \emph {et~al.}(2011)\citenamefont {Tammuz},
  \citenamefont {Smith}, \citenamefont {Campbell}, \citenamefont {Beattie},
  \citenamefont {Moulder}, \citenamefont {Dalibard},\ and\ \citenamefont
  {Hadzibabic}}]{Tammuz:2011}%
  \BibitemOpen
  \bibfield  {author} {\bibinfo {author} {\bibfnamefont {N.}~\bibnamefont
  {Tammuz}}, \bibinfo {author} {\bibfnamefont {R.~P.}\ \bibnamefont {Smith}},
  \bibinfo {author} {\bibfnamefont {R.~L.~D.}\ \bibnamefont {Campbell}},
  \bibinfo {author} {\bibfnamefont {S.}~\bibnamefont {Beattie}}, \bibinfo
  {author} {\bibfnamefont {S.}~\bibnamefont {Moulder}}, \bibinfo {author}
  {\bibfnamefont {J.}~\bibnamefont {Dalibard}}, \ and\ \bibinfo {author}
  {\bibfnamefont {Z.}~\bibnamefont {Hadzibabic}},\ }\bibfield  {title}
  {\enquote {\bibinfo {title} {Can a {B}ose gas be saturated?}}\ }\href
  {\doibase 10.1103/PhysRevLett.106.230401} {\bibfield  {journal} {\bibinfo
  {journal} {Phys. Rev. Lett.}\ }\textbf {\bibinfo {volume} {106}},\ \bibinfo
  {pages} {230401} (\bibinfo {year} {2011})}\BibitemShut {NoStop}%
\bibitem [{\citenamefont {Anderson}\ and\ \citenamefont
  {Kasevich}(1999)}]{Anderson:1999}%
  \BibitemOpen
  \bibfield  {author} {\bibinfo {author} {\bibfnamefont {B.~P.}\ \bibnamefont
  {Anderson}}\ and\ \bibinfo {author} {\bibfnamefont {M.~A.}\ \bibnamefont
  {Kasevich}},\ }\bibfield  {title} {\enquote {\bibinfo {title} {Spatial
  observation of {B}ose-{E}instein condensation of ${}^{87}\mathrm{Rb}$ in a
  confining potential},}\ }\href {\doibase 10.1103/PhysRevA.59.R938} {\bibfield
   {journal} {\bibinfo  {journal} {Phys. Rev. A}\ }\textbf {\bibinfo {volume}
  {59}},\ \bibinfo {pages} {R938--R941} (\bibinfo {year} {1999})}\BibitemShut
  {NoStop}%
\bibitem [{\citenamefont {Truscott}\ \emph {et~al.}(2001)\citenamefont
  {Truscott}, \citenamefont {Strecker}, \citenamefont {McAlexander},
  \citenamefont {Partridge},\ and\ \citenamefont {Hulet}}]{Truscott:2001}%
  \BibitemOpen
  \bibfield  {author} {\bibinfo {author} {\bibfnamefont {A.}~\bibnamefont
  {Truscott}}, \bibinfo {author} {\bibfnamefont {K.}~\bibnamefont {Strecker}},
  \bibinfo {author} {\bibfnamefont {W.}~\bibnamefont {McAlexander}}, \bibinfo
  {author} {\bibfnamefont {G.}~\bibnamefont {Partridge}}, \ and\ \bibinfo
  {author} {\bibfnamefont {R.~G.}\ \bibnamefont {Hulet}},\ }\bibfield  {title}
  {\enquote {\bibinfo {title} {Observation of {F}ermi pressure in a gas of
  trapped atoms},}\ }\href {\doibase 10.1126/science.1059318} {\bibfield
  {journal} {\bibinfo  {journal} {Science}\ }\textbf {\bibinfo {volume}
  {291}},\ \bibinfo {pages} {2570} (\bibinfo {year} {2001})}\BibitemShut
  {NoStop}%
\bibitem [{\citenamefont {{Kothari}}\ and\ \citenamefont
  {{Srivasava}}(1937)}]{Kothari:1937}%
  \BibitemOpen
  \bibfield  {author} {\bibinfo {author} {\bibfnamefont {D.~S.}\ \bibnamefont
  {{Kothari}}}\ and\ \bibinfo {author} {\bibfnamefont {B.~N.}\ \bibnamefont
  {{Srivasava}}},\ }\bibfield  {title} {\enquote {\bibinfo {title}
  {{Joule-Thomson Effect and Quantum Statistics}},}\ }\href {\doibase
  10.1038/140970b0} {\bibfield  {journal} {\bibinfo  {journal} {Nature}\
  }\textbf {\bibinfo {volume} {140}},\ \bibinfo {pages} {970--971} (\bibinfo
  {year} {1937})}\BibitemShut {NoStop}%
\bibitem [{\citenamefont {{Patel}}\ \emph {et~al.}(2020)\citenamefont
  {{Patel}}, \citenamefont {{Yan}}, \citenamefont {{Mukherjee}}, \citenamefont
  {{Fletcher}}, \citenamefont {{Struck}},\ and\ \citenamefont
  {{Zwierlein}}}]{Patel:2019}%
  \BibitemOpen
  \bibfield  {author} {\bibinfo {author} {\bibfnamefont {P.~B.}\ \bibnamefont
  {{Patel}}}, \bibinfo {author} {\bibfnamefont {Z.}~\bibnamefont {{Yan}}},
  \bibinfo {author} {\bibfnamefont {B.}~\bibnamefont {{Mukherjee}}}, \bibinfo
  {author} {\bibfnamefont {R.~J.}\ \bibnamefont {{Fletcher}}}, \bibinfo
  {author} {\bibfnamefont {J.}~\bibnamefont {{Struck}}}, \ and\ \bibinfo
  {author} {\bibfnamefont {M.~W.}\ \bibnamefont {{Zwierlein}}},\ }\bibfield
  {title} {\enquote {\bibinfo {title} {{Universal sound diffusion in a strongly
  interacting Fermi gas}},}\ }\href {\doibase 10.1126/science.aaz5756}
  {\bibfield  {journal} {\bibinfo  {journal} {Science}\ }\textbf {\bibinfo
  {volume} {370}},\ \bibinfo {pages} {1222--1226} (\bibinfo {year}
  {2020})}\BibitemShut {NoStop}%
\bibitem [{\citenamefont {Ville}\ \emph {et~al.}(2018)\citenamefont {Ville},
  \citenamefont {Saint-Jalm}, \citenamefont {Le~Cerf}, \citenamefont
  {Aidelsburger}, \citenamefont {Nascimb\`ene}, \citenamefont {Dalibard},\ and\
  \citenamefont {Beugnon}}]{Ville:2018}%
  \BibitemOpen
  \bibfield  {author} {\bibinfo {author} {\bibfnamefont {J.~L.}\ \bibnamefont
  {Ville}}, \bibinfo {author} {\bibfnamefont {R.}~\bibnamefont {Saint-Jalm}},
  \bibinfo {author} {\bibfnamefont {E.}~\bibnamefont {Le~Cerf}}, \bibinfo
  {author} {\bibfnamefont {M.}~\bibnamefont {Aidelsburger}}, \bibinfo {author}
  {\bibfnamefont {S.}~\bibnamefont {Nascimb\`ene}}, \bibinfo {author}
  {\bibfnamefont {J.}~\bibnamefont {Dalibard}}, \ and\ \bibinfo {author}
  {\bibfnamefont {J.}~\bibnamefont {Beugnon}},\ }\bibfield  {title} {\enquote
  {\bibinfo {title} {Sound propagation in a uniform superfluid two-dimensional
  {B}ose gas},}\ }\href {\doibase 10.1103/PhysRevLett.121.145301} {\bibfield
  {journal} {\bibinfo  {journal} {Phys. Rev. Lett.}\ }\textbf {\bibinfo
  {volume} {121}},\ \bibinfo {pages} {145301} (\bibinfo {year}
  {2018})}\BibitemShut {NoStop}%
\bibitem [{\citenamefont {Christodoulou}\ \emph {et~al.}(2021)\citenamefont
  {Christodoulou}, \citenamefont {Gałka}, \citenamefont {Dogra}, \citenamefont
  {Lopes}, \citenamefont {Schmitt},\ and\ \citenamefont
  {Hadzibabic}}]{Christodoulou:2020}%
  \BibitemOpen
  \bibfield  {author} {\bibinfo {author} {\bibfnamefont {P.}~\bibnamefont
  {Christodoulou}}, \bibinfo {author} {\bibfnamefont {M.}~\bibnamefont
  {Gałka}}, \bibinfo {author} {\bibfnamefont {N.}~\bibnamefont {Dogra}},
  \bibinfo {author} {\bibfnamefont {R.}~\bibnamefont {Lopes}}, \bibinfo
  {author} {\bibfnamefont {J.}~\bibnamefont {Schmitt}}, \ and\ \bibinfo
  {author} {\bibfnamefont {Z.}~\bibnamefont {Hadzibabic}},\ }\bibfield  {title}
  {\enquote {\bibinfo {title} {Observation of first and second sound in a {BKT}
  superfluid},}\ }\href {\doibase https://doi.org/10.1038/s41586-021-03537-9}
  {\bibfield  {journal} {\bibinfo  {journal} {Nature}\ }\textbf {\bibinfo
  {volume} {594}},\ \bibinfo {pages} {191--194} (\bibinfo {year}
  {2021})}\BibitemShut {NoStop}%
\bibitem [{\citenamefont {Mukherjee}\ \emph {et~al.}(2019)\citenamefont
  {Mukherjee}, \citenamefont {Patel}, \citenamefont {Yan}, \citenamefont
  {Fletcher}, \citenamefont {Struck},\ and\ \citenamefont
  {Zwierlein}}]{Mukherjee:2019}%
  \BibitemOpen
  \bibfield  {author} {\bibinfo {author} {\bibfnamefont {B.}~\bibnamefont
  {Mukherjee}}, \bibinfo {author} {\bibfnamefont {P.~B.}\ \bibnamefont
  {Patel}}, \bibinfo {author} {\bibfnamefont {Z.}~\bibnamefont {Yan}}, \bibinfo
  {author} {\bibfnamefont {R.~J.}\ \bibnamefont {Fletcher}}, \bibinfo {author}
  {\bibfnamefont {J.}~\bibnamefont {Struck}}, \ and\ \bibinfo {author}
  {\bibfnamefont {M.~W.}\ \bibnamefont {Zwierlein}},\ }\bibfield  {title}
  {\enquote {\bibinfo {title} {Spectral response and contact of the unitary
  {F}ermi gas},}\ }\href {\doibase 10.1103/PhysRevLett.122.203402} {\bibfield
  {journal} {\bibinfo  {journal} {Phys. Rev. Lett.}\ }\textbf {\bibinfo
  {volume} {122}},\ \bibinfo {pages} {203402} (\bibinfo {year}
  {2019})}\BibitemShut {NoStop}%
\bibitem [{\citenamefont {Zou}\ \emph {et~al.}(2021)\citenamefont {Zou},
  \citenamefont {Bakkali-Hassani}, \citenamefont {Maury}, \citenamefont {Cerf},
  \citenamefont {Nascimbene}, \citenamefont {Dalibard},\ and\ \citenamefont
  {Beugnon}}]{Zou:2021tan}%
  \BibitemOpen
  \bibfield  {author} {\bibinfo {author} {\bibfnamefont {Y.}~\bibnamefont
  {Zou}}, \bibinfo {author} {\bibfnamefont {B.}~\bibnamefont
  {Bakkali-Hassani}}, \bibinfo {author} {\bibfnamefont {C.}~\bibnamefont
  {Maury}}, \bibinfo {author} {\bibfnamefont {{\'E}.~L.}\ \bibnamefont {Cerf}},
  \bibinfo {author} {\bibfnamefont {S.}~\bibnamefont {Nascimbene}}, \bibinfo
  {author} {\bibfnamefont {J.}~\bibnamefont {Dalibard}}, \ and\ \bibinfo
  {author} {\bibfnamefont {J.}~\bibnamefont {Beugnon}},\ }\bibfield  {title}
  {\enquote {\bibinfo {title} {Tan's two-body contact across the superfluid
  transition of a planar {B}ose gas.}}\ }\href {\doibase
  https://doi.org/10.1038/s41467-020-20647-6} {\bibfield  {journal} {\bibinfo
  {journal} {Nature communications}\ }\textbf {\bibinfo {volume} {12 1}},\
  \bibinfo {pages} {760} (\bibinfo {year} {2021})}\BibitemShut {NoStop}%
\bibitem [{\citenamefont {{Biss}}\ \emph {et~al.}(2021)\citenamefont {{Biss}},
  \citenamefont {{Sobirey}}, \citenamefont {{Luick}}, \citenamefont {{Bohlen}},
  \citenamefont {{Kinnunen}}, \citenamefont {{Bruun}}, \citenamefont
  {{Lompe}},\ and\ \citenamefont {{Moritz}}}]{Biss:2021}%
  \BibitemOpen
  \bibfield  {author} {\bibinfo {author} {\bibfnamefont {H.}~\bibnamefont
  {{Biss}}}, \bibinfo {author} {\bibfnamefont {L.}~\bibnamefont {{Sobirey}}},
  \bibinfo {author} {\bibfnamefont {N.}~\bibnamefont {{Luick}}}, \bibinfo
  {author} {\bibfnamefont {M.}~\bibnamefont {{Bohlen}}}, \bibinfo {author}
  {\bibfnamefont {J.~J.}\ \bibnamefont {{Kinnunen}}}, \bibinfo {author}
  {\bibfnamefont {G.~M.}\ \bibnamefont {{Bruun}}}, \bibinfo {author}
  {\bibfnamefont {T.}~\bibnamefont {{Lompe}}}, \ and\ \bibinfo {author}
  {\bibfnamefont {H.}~\bibnamefont {{Moritz}}},\ }\href@noop {} {\enquote
  {\bibinfo {title} {{Excitation Spectrum and Superfluid Gap of an Ultracold
  Fermi Gas}},}\ } (\bibinfo {year} {2021}),\ \Eprint
  {http://arxiv.org/abs/2105.09820} {arXiv:2105.09820} \BibitemShut {NoStop}%
\bibitem [{\citenamefont {Navon}\ \emph {et~al.}(2016)\citenamefont {Navon},
  \citenamefont {Gaunt}, \citenamefont {Smith},\ and\ \citenamefont
  {Hadzibabic}}]{Navon:2016}%
  \BibitemOpen
  \bibfield  {author} {\bibinfo {author} {\bibfnamefont {N.}~\bibnamefont
  {Navon}}, \bibinfo {author} {\bibfnamefont {A.~L.}\ \bibnamefont {Gaunt}},
  \bibinfo {author} {\bibfnamefont {R.~P.}\ \bibnamefont {Smith}}, \ and\
  \bibinfo {author} {\bibfnamefont {Z.}~\bibnamefont {Hadzibabic}},\ }\bibfield
   {title} {\enquote {\bibinfo {title} {Emergence of a turbulent cascade in a
  quantum gas},}\ }\href {https://doi.org/10.1038/nature20114} {\bibfield
  {journal} {\bibinfo  {journal} {Nature}\ }\textbf {\bibinfo {volume} {539}},\
  \bibinfo {pages} {72} (\bibinfo {year} {2016})}\BibitemShut {NoStop}%
\bibitem [{\citenamefont {Baird}\ \emph {et~al.}(2019)\citenamefont {Baird},
  \citenamefont {Wang}, \citenamefont {Roof},\ and\ \citenamefont
  {Thomas}}]{Baird:2019}%
  \BibitemOpen
  \bibfield  {author} {\bibinfo {author} {\bibfnamefont {L.}~\bibnamefont
  {Baird}}, \bibinfo {author} {\bibfnamefont {X.}~\bibnamefont {Wang}},
  \bibinfo {author} {\bibfnamefont {S.}~\bibnamefont {Roof}}, \ and\ \bibinfo
  {author} {\bibfnamefont {J.~E.}\ \bibnamefont {Thomas}},\ }\bibfield  {title}
  {\enquote {\bibinfo {title} {Measuring the hydrodynamic linear response of a
  unitary {F}ermi gas},}\ }\href {\doibase 10.1103/PhysRevLett.123.160402}
  {\bibfield  {journal} {\bibinfo  {journal} {Phys. Rev. Lett.}\ }\textbf
  {\bibinfo {volume} {123}},\ \bibinfo {pages} {160402} (\bibinfo {year}
  {2019})}\BibitemShut {NoStop}%
\bibitem [{\citenamefont {Garratt}\ \emph {et~al.}(2019)\citenamefont
  {Garratt}, \citenamefont {Eigen}, \citenamefont {Zhang}, \citenamefont
  {Turz\'ak}, \citenamefont {Lopes}, \citenamefont {Smith}, \citenamefont
  {Hadzibabic},\ and\ \citenamefont {Navon}}]{Garratt:2019}%
  \BibitemOpen
  \bibfield  {author} {\bibinfo {author} {\bibfnamefont {S.~J.}\ \bibnamefont
  {Garratt}}, \bibinfo {author} {\bibfnamefont {C.}~\bibnamefont {Eigen}},
  \bibinfo {author} {\bibfnamefont {J.}~\bibnamefont {Zhang}}, \bibinfo
  {author} {\bibfnamefont {P.}~\bibnamefont {Turz\'ak}}, \bibinfo {author}
  {\bibfnamefont {R.}~\bibnamefont {Lopes}}, \bibinfo {author} {\bibfnamefont
  {R.~P.}\ \bibnamefont {Smith}}, \bibinfo {author} {\bibfnamefont
  {Z.}~\bibnamefont {Hadzibabic}}, \ and\ \bibinfo {author} {\bibfnamefont
  {N.}~\bibnamefont {Navon}},\ }\bibfield  {title} {\enquote {\bibinfo {title}
  {From single-particle excitations to sound waves in a box-trapped atomic
  {B}ose--{E}instein condensate},}\ }\href {\doibase
  10.1103/PhysRevA.99.021601} {\bibfield  {journal} {\bibinfo  {journal} {Phys.
  Rev. A}\ }\textbf {\bibinfo {volume} {99}},\ \bibinfo {pages} {021601}
  (\bibinfo {year} {2019})}\BibitemShut {NoStop}%
\bibitem [{\citenamefont {Bohlen}\ \emph {et~al.}(2020)\citenamefont {Bohlen},
  \citenamefont {Sobirey}, \citenamefont {Luick}, \citenamefont {Biss},
  \citenamefont {Enss}, \citenamefont {Lompe},\ and\ \citenamefont
  {Moritz}}]{Bohlen:2020}%
  \BibitemOpen
  \bibfield  {author} {\bibinfo {author} {\bibfnamefont {M.}~\bibnamefont
  {Bohlen}}, \bibinfo {author} {\bibfnamefont {L.}~\bibnamefont {Sobirey}},
  \bibinfo {author} {\bibfnamefont {N.}~\bibnamefont {Luick}}, \bibinfo
  {author} {\bibfnamefont {H.}~\bibnamefont {Biss}}, \bibinfo {author}
  {\bibfnamefont {T.}~\bibnamefont {Enss}}, \bibinfo {author} {\bibfnamefont
  {T.}~\bibnamefont {Lompe}}, \ and\ \bibinfo {author} {\bibfnamefont
  {H.}~\bibnamefont {Moritz}},\ }\bibfield  {title} {\enquote {\bibinfo {title}
  {Sound propagation and quantum-limited damping in a two-dimensional {F}ermi
  gas},}\ }\href {\doibase 10.1103/PhysRevLett.124.240403} {\bibfield
  {journal} {\bibinfo  {journal} {Phys. Rev. Lett.}\ }\textbf {\bibinfo
  {volume} {124}},\ \bibinfo {pages} {240403} (\bibinfo {year}
  {2020})}\BibitemShut {NoStop}%
\bibitem [{\citenamefont {{Zhang}}\ \emph {et~al.}(2021)\citenamefont
  {{Zhang}}, \citenamefont {{Eigen}}, \citenamefont {{Zheng}}, \citenamefont
  {{Glidden}}, \citenamefont {{Hilker}}, \citenamefont {{Garratt}},
  \citenamefont {{Lopes}}, \citenamefont {{Cooper}}, \citenamefont
  {{Hadzibabic}},\ and\ \citenamefont {{Navon}}}]{Zhang:2021}%
  \BibitemOpen
  \bibfield  {author} {\bibinfo {author} {\bibfnamefont {J.}~\bibnamefont
  {{Zhang}}}, \bibinfo {author} {\bibfnamefont {C.}~\bibnamefont {{Eigen}}},
  \bibinfo {author} {\bibfnamefont {W.}~\bibnamefont {{Zheng}}}, \bibinfo
  {author} {\bibfnamefont {J.~A.~P.}\ \bibnamefont {{Glidden}}}, \bibinfo
  {author} {\bibfnamefont {T.~A.}\ \bibnamefont {{Hilker}}}, \bibinfo {author}
  {\bibfnamefont {S.~J.}\ \bibnamefont {{Garratt}}}, \bibinfo {author}
  {\bibfnamefont {R.}~\bibnamefont {{Lopes}}}, \bibinfo {author} {\bibfnamefont
  {N.~R.}\ \bibnamefont {{Cooper}}}, \bibinfo {author} {\bibfnamefont
  {Z.}~\bibnamefont {{Hadzibabic}}}, \ and\ \bibinfo {author} {\bibfnamefont
  {N.}~\bibnamefont {{Navon}}},\ }\bibfield  {title} {\enquote {\bibinfo
  {title} {{Many-Body Decay of the Gapped Lowest Excitation of a Bose-Einstein
  Condensate}},}\ }\href {\doibase 10.1103/PhysRevLett.126.060402} {\bibfield
  {journal} {\bibinfo  {journal} {\prl}\ }\textbf {\bibinfo {volume} {126}},\
  \bibinfo {eid} {060402} (\bibinfo {year} {2021})}\BibitemShut {NoStop}%
\bibitem [{\citenamefont {Berezinskii}(1971)}]{Berezinskii:1971}%
  \BibitemOpen
  \bibfield  {author} {\bibinfo {author} {\bibfnamefont {V.~L.}\ \bibnamefont
  {Berezinskii}},\ }\bibfield  {title} {\enquote {\bibinfo {title} {Destruction
  of long-range order in one-dimensional and two-dimensional system possessing
  a continous symmetry group - ii. quantum systems},}\ }\href@noop {}
  {\bibfield  {journal} {\bibinfo  {journal} {Soviet Physics JETP}\ }\textbf
  {\bibinfo {volume} {34}},\ \bibinfo {pages} {610} (\bibinfo {year}
  {1971})}\BibitemShut {NoStop}%
\bibitem [{\citenamefont {{K}osterlitz}\ and\ \citenamefont
  {{T}houless}(1973)}]{Kosterlitz:1973}%
  \BibitemOpen
  \bibfield  {author} {\bibinfo {author} {\bibfnamefont {J.~M.}\ \bibnamefont
  {{K}osterlitz}}\ and\ \bibinfo {author} {\bibfnamefont {D.~J.}\ \bibnamefont
  {{T}houless}},\ }\bibfield  {title} {\enquote {\bibinfo {title} {Ordering,
  metastability and phase transitions in two dimensional systems},}\
  }\href@noop {} {\bibfield  {journal} {\bibinfo  {journal} {J. Phys. C: Solid
  State Physics}\ }\textbf {\bibinfo {volume} {6}},\ \bibinfo {pages} {1181}
  (\bibinfo {year} {1973})}\BibitemShut {NoStop}%
\bibitem [{\citenamefont {Nelson}\ and\ \citenamefont
  {Kosterlitz}(1977)}]{Nelson:1977}%
  \BibitemOpen
  \bibfield  {author} {\bibinfo {author} {\bibfnamefont {D.~R.}\ \bibnamefont
  {Nelson}}\ and\ \bibinfo {author} {\bibfnamefont {J.~M.}\ \bibnamefont
  {Kosterlitz}},\ }\bibfield  {title} {\enquote {\bibinfo {title} {Universal
  jump in the superfluid density of two-dimensional superfluids},}\ }\href
  {\doibase 10.1103/PhysRevLett.39.1201} {\bibfield  {journal} {\bibinfo
  {journal} {Phys. Rev. Lett.}\ }\textbf {\bibinfo {volume} {39}},\ \bibinfo
  {pages} {1201--1205} (\bibinfo {year} {1977})}\BibitemShut {NoStop}%
\bibitem [{\citenamefont {Luick}\ \emph {et~al.}(2020)\citenamefont {Luick},
  \citenamefont {Sobirey}, \citenamefont {Bohlen}, \citenamefont {Singh},
  \citenamefont {Mathey}, \citenamefont {Lompe},\ and\ \citenamefont
  {Moritz}}]{Luick:2020}%
  \BibitemOpen
  \bibfield  {author} {\bibinfo {author} {\bibfnamefont {N.}~\bibnamefont
  {Luick}}, \bibinfo {author} {\bibfnamefont {L.}~\bibnamefont {Sobirey}},
  \bibinfo {author} {\bibfnamefont {M.}~\bibnamefont {Bohlen}}, \bibinfo
  {author} {\bibfnamefont {V.~P.}\ \bibnamefont {Singh}}, \bibinfo {author}
  {\bibfnamefont {L.}~\bibnamefont {Mathey}}, \bibinfo {author} {\bibfnamefont
  {T.}~\bibnamefont {Lompe}}, \ and\ \bibinfo {author} {\bibfnamefont
  {H.}~\bibnamefont {Moritz}},\ }\bibfield  {title} {\enquote {\bibinfo {title}
  {An ideal {J}osephson junction in an ultracold two-dimensional {F}ermi
  gas},}\ }\href {\doibase 10.1126/science.aaz2342} {\bibfield  {journal}
  {\bibinfo  {journal} {Science}\ }\textbf {\bibinfo {volume} {369}},\ \bibinfo
  {pages} {89--91} (\bibinfo {year} {2020})}\BibitemShut {NoStop}%
\bibitem [{\citenamefont {{Sobirey}}\ \emph {et~al.}(2021)\citenamefont
  {{Sobirey}}, \citenamefont {{Luick}}, \citenamefont {{Bohlen}}, \citenamefont
  {{Biss}}, \citenamefont {{Moritz}},\ and\ \citenamefont
  {{Lompe}}}]{Sobirey:2020}%
  \BibitemOpen
  \bibfield  {author} {\bibinfo {author} {\bibfnamefont {L.}~\bibnamefont
  {{Sobirey}}}, \bibinfo {author} {\bibfnamefont {N.}~\bibnamefont {{Luick}}},
  \bibinfo {author} {\bibfnamefont {M.}~\bibnamefont {{Bohlen}}}, \bibinfo
  {author} {\bibfnamefont {H.}~\bibnamefont {{Biss}}}, \bibinfo {author}
  {\bibfnamefont {H.}~\bibnamefont {{Moritz}}}, \ and\ \bibinfo {author}
  {\bibfnamefont {T.}~\bibnamefont {{Lompe}}},\ }\bibfield  {title} {\enquote
  {\bibinfo {title} {{Observation of superfluidity in a strongly correlated
  two-dimensional Fermi gas}},}\ }\href {\doibase 10.1126/science.abc8793}
  {\bibfield  {journal} {\bibinfo  {journal} {Science}\ }\textbf {\bibinfo
  {volume} {372}},\ \bibinfo {pages} {844--846} (\bibinfo {year}
  {2021})}\BibitemShut {NoStop}%
\bibitem [{\citenamefont {Gauthier}\ \emph
  {et~al.}(2019{\natexlab{a}})\citenamefont {Gauthier}, \citenamefont
  {Szigeti}, \citenamefont {Reeves}, \citenamefont {Baker}, \citenamefont
  {Bell}, \citenamefont {Rubinsztein-Dunlop}, \citenamefont {Davis},\ and\
  \citenamefont {Neely}}]{Gauthier:2019b}%
  \BibitemOpen
  \bibfield  {author} {\bibinfo {author} {\bibfnamefont {G.}~\bibnamefont
  {Gauthier}}, \bibinfo {author} {\bibfnamefont {S.~S.}\ \bibnamefont
  {Szigeti}}, \bibinfo {author} {\bibfnamefont {M.~T.}\ \bibnamefont {Reeves}},
  \bibinfo {author} {\bibfnamefont {M.}~\bibnamefont {Baker}}, \bibinfo
  {author} {\bibfnamefont {T.~A.}\ \bibnamefont {Bell}}, \bibinfo {author}
  {\bibfnamefont {H.}~\bibnamefont {Rubinsztein-Dunlop}}, \bibinfo {author}
  {\bibfnamefont {M.~J.}\ \bibnamefont {Davis}}, \ and\ \bibinfo {author}
  {\bibfnamefont {T.~W.}\ \bibnamefont {Neely}},\ }\bibfield  {title} {\enquote
  {\bibinfo {title} {Quantitative acoustic models for superfluid circuits},}\
  }\href {\doibase 10.1103/PhysRevLett.123.260402} {\bibfield  {journal}
  {\bibinfo  {journal} {Phys. Rev. Lett.}\ }\textbf {\bibinfo {volume} {123}},\
  \bibinfo {pages} {260402} (\bibinfo {year} {2019}{\natexlab{a}})}\BibitemShut
  {NoStop}%
\bibitem [{\citenamefont {Gotlibovych}\ \emph {et~al.}(2014)\citenamefont
  {Gotlibovych}, \citenamefont {Schmidutz}, \citenamefont {Gaunt},
  \citenamefont {Navon}, \citenamefont {Smith},\ and\ \citenamefont
  {Hadzibabic}}]{Gotlibovych:2014}%
  \BibitemOpen
  \bibfield  {author} {\bibinfo {author} {\bibfnamefont {I.}~\bibnamefont
  {Gotlibovych}}, \bibinfo {author} {\bibfnamefont {T.~F.}\ \bibnamefont
  {Schmidutz}}, \bibinfo {author} {\bibfnamefont {A.~L.}\ \bibnamefont
  {Gaunt}}, \bibinfo {author} {\bibfnamefont {N.}~\bibnamefont {Navon}},
  \bibinfo {author} {\bibfnamefont {R.~P.}\ \bibnamefont {Smith}}, \ and\
  \bibinfo {author} {\bibfnamefont {Z.}~\bibnamefont {Hadzibabic}},\ }\bibfield
   {title} {\enquote {\bibinfo {title} {Observing properties of an interacting
  homogeneous {B}ose--{E}instein condensate: {H}eisenberg-limited momentum
  spread, interaction energy, and free-expansion dynamics},}\ }\href {\doibase
  10.1103/PhysRevA.89.061604} {\bibfield  {journal} {\bibinfo  {journal} {Phys.
  Rev. A}\ }\textbf {\bibinfo {volume} {89}},\ \bibinfo {pages} {061604}
  (\bibinfo {year} {2014})}\BibitemShut {NoStop}%
\bibitem [{\citenamefont {Lopes}\ \emph
  {et~al.}(2017{\natexlab{a}})\citenamefont {Lopes}, \citenamefont {Eigen},
  \citenamefont {Barker}, \citenamefont {Viebahn}, \citenamefont {Robert-de
  Saint-Vincent}, \citenamefont {Navon}, \citenamefont {Hadzibabic},\ and\
  \citenamefont {Smith}}]{Lopes:2017}%
  \BibitemOpen
  \bibfield  {author} {\bibinfo {author} {\bibfnamefont {R.}~\bibnamefont
  {Lopes}}, \bibinfo {author} {\bibfnamefont {C.}~\bibnamefont {Eigen}},
  \bibinfo {author} {\bibfnamefont {A.}~\bibnamefont {Barker}}, \bibinfo
  {author} {\bibfnamefont {K.~G.~H.}\ \bibnamefont {Viebahn}}, \bibinfo
  {author} {\bibfnamefont {M.}~\bibnamefont {Robert-de Saint-Vincent}},
  \bibinfo {author} {\bibfnamefont {N.}~\bibnamefont {Navon}}, \bibinfo
  {author} {\bibfnamefont {Z.}~\bibnamefont {Hadzibabic}}, \ and\ \bibinfo
  {author} {\bibfnamefont {R.~P.}\ \bibnamefont {Smith}},\ }\bibfield  {title}
  {\enquote {\bibinfo {title} {Quasiparticle energy in a strongly interacting
  homogeneous {Bose--Einstein} condensate},}\ }\href {\doibase
  10.1103/PhysRevLett.118.210401} {\bibfield  {journal} {\bibinfo  {journal}
  {Phys. Rev. Lett.}\ }\textbf {\bibinfo {volume} {118}},\ \bibinfo {pages}
  {210401} (\bibinfo {year} {2017}{\natexlab{a}})}\BibitemShut {NoStop}%
\bibitem [{\citenamefont {Lopes}\ \emph
  {et~al.}(2017{\natexlab{b}})\citenamefont {Lopes}, \citenamefont {Eigen},
  \citenamefont {Navon}, \citenamefont {Cl\'ement}, \citenamefont {Smith},\
  and\ \citenamefont {Hadzibabic}}]{Lopes:2017b}%
  \BibitemOpen
  \bibfield  {author} {\bibinfo {author} {\bibfnamefont {R.}~\bibnamefont
  {Lopes}}, \bibinfo {author} {\bibfnamefont {C.}~\bibnamefont {Eigen}},
  \bibinfo {author} {\bibfnamefont {N.}~\bibnamefont {Navon}}, \bibinfo
  {author} {\bibfnamefont {D.}~\bibnamefont {Cl\'ement}}, \bibinfo {author}
  {\bibfnamefont {R.~P.}\ \bibnamefont {Smith}}, \ and\ \bibinfo {author}
  {\bibfnamefont {Z.}~\bibnamefont {Hadzibabic}},\ }\bibfield  {title}
  {\enquote {\bibinfo {title} {Quantum depletion of a homogeneous
  {B}ose--{E}instein condensate},}\ }\href {\doibase
  10.1103/PhysRevLett.119.190404} {\bibfield  {journal} {\bibinfo  {journal}
  {Phys. Rev. Lett.}\ }\textbf {\bibinfo {volume} {119}},\ \bibinfo {pages}
  {190404} (\bibinfo {year} {2017}{\natexlab{b}})}\BibitemShut {NoStop}%
\bibitem [{\citenamefont {Yan}\ \emph {et~al.}(2019)\citenamefont {Yan},
  \citenamefont {Patel}, \citenamefont {Mukherjee}, \citenamefont {Fletcher},
  \citenamefont {Struck},\ and\ \citenamefont {Zwierlein}}]{Yan:2019}%
  \BibitemOpen
  \bibfield  {author} {\bibinfo {author} {\bibfnamefont {Z.}~\bibnamefont
  {Yan}}, \bibinfo {author} {\bibfnamefont {P.~B.}\ \bibnamefont {Patel}},
  \bibinfo {author} {\bibfnamefont {B.}~\bibnamefont {Mukherjee}}, \bibinfo
  {author} {\bibfnamefont {R.~J.}\ \bibnamefont {Fletcher}}, \bibinfo {author}
  {\bibfnamefont {J.}~\bibnamefont {Struck}}, \ and\ \bibinfo {author}
  {\bibfnamefont {M.~W.}\ \bibnamefont {Zwierlein}},\ }\bibfield  {title}
  {\enquote {\bibinfo {title} {Boiling a unitary {F}ermi liquid},}\ }\href
  {\doibase 10.1103/PhysRevLett.122.093401} {\bibfield  {journal} {\bibinfo
  {journal} {Phys. Rev. Lett.}\ }\textbf {\bibinfo {volume} {122}},\ \bibinfo
  {pages} {093401} (\bibinfo {year} {2019})}\BibitemShut {NoStop}%
\bibitem [{\citenamefont {Zou}\ \emph {et~al.}(2020)\citenamefont {Zou},
  \citenamefont {Bakkali-Hassani}, \citenamefont {Maury}, \citenamefont
  {Le~Cerf}, \citenamefont {Nascimbene}, \citenamefont {Dalibard},\ and\
  \citenamefont {Beugnon}}]{Zou:2021mag}%
  \BibitemOpen
  \bibfield  {author} {\bibinfo {author} {\bibfnamefont {Y.-Q.}\ \bibnamefont
  {Zou}}, \bibinfo {author} {\bibfnamefont {B.}~\bibnamefont
  {Bakkali-Hassani}}, \bibinfo {author} {\bibfnamefont {C.}~\bibnamefont
  {Maury}}, \bibinfo {author} {\bibfnamefont {E.}~\bibnamefont {Le~Cerf}},
  \bibinfo {author} {\bibfnamefont {S.}~\bibnamefont {Nascimbene}}, \bibinfo
  {author} {\bibfnamefont {J.}~\bibnamefont {Dalibard}}, \ and\ \bibinfo
  {author} {\bibfnamefont {J.}~\bibnamefont {Beugnon}},\ }\bibfield  {title}
  {\enquote {\bibinfo {title} {Magnetic dipolar interaction between hyperfine
  clock states in a planar alkali {B}ose gas},}\ }\href {\doibase
  10.1103/PhysRevLett.125.233604} {\bibfield  {journal} {\bibinfo  {journal}
  {Phys. Rev. Lett.}\ }\textbf {\bibinfo {volume} {125}},\ \bibinfo {pages}
  {233604} (\bibinfo {year} {2020})}\BibitemShut {NoStop}%
\bibitem [{\citenamefont {Sagi}\ \emph {et~al.}(2012)\citenamefont {Sagi},
  \citenamefont {Drake}, \citenamefont {Paudel},\ and\ \citenamefont
  {Jin}}]{Sagi:2012}%
  \BibitemOpen
  \bibfield  {author} {\bibinfo {author} {\bibfnamefont {Y.}~\bibnamefont
  {Sagi}}, \bibinfo {author} {\bibfnamefont {T.~E.}\ \bibnamefont {Drake}},
  \bibinfo {author} {\bibfnamefont {R.}~\bibnamefont {Paudel}}, \ and\ \bibinfo
  {author} {\bibfnamefont {D.~S.}\ \bibnamefont {Jin}},\ }\bibfield  {title}
  {\enquote {\bibinfo {title} {Measurement of the homogeneous contact of a
  unitary {F}ermi gas},}\ }\href {\doibase 10.1103/PhysRevLett.109.220402}
  {\bibfield  {journal} {\bibinfo  {journal} {Phys. Rev. Lett.}\ }\textbf
  {\bibinfo {volume} {109}},\ \bibinfo {pages} {220402} (\bibinfo {year}
  {2012})}\BibitemShut {NoStop}%
\bibitem [{\citenamefont {Sagi}\ \emph {et~al.}(2015)\citenamefont {Sagi},
  \citenamefont {Drake}, \citenamefont {Paudel}, \citenamefont {Chapurin},\
  and\ \citenamefont {Jin}}]{Sagi:2015}%
  \BibitemOpen
  \bibfield  {author} {\bibinfo {author} {\bibfnamefont {Y.}~\bibnamefont
  {Sagi}}, \bibinfo {author} {\bibfnamefont {T.~E.}\ \bibnamefont {Drake}},
  \bibinfo {author} {\bibfnamefont {R.}~\bibnamefont {Paudel}}, \bibinfo
  {author} {\bibfnamefont {R.}~\bibnamefont {Chapurin}}, \ and\ \bibinfo
  {author} {\bibfnamefont {D.~S.}\ \bibnamefont {Jin}},\ }\bibfield  {title}
  {\enquote {\bibinfo {title} {Breakdown of the {F}ermi--liquid description for
  strongly interacting fermions},}\ }\href {\doibase
  10.1103/PhysRevLett.114.075301} {\bibfield  {journal} {\bibinfo  {journal}
  {Phys. Rev. Lett.}\ }\textbf {\bibinfo {volume} {114}},\ \bibinfo {pages}
  {075301} (\bibinfo {year} {2015})}\BibitemShut {NoStop}%
\bibitem [{\citenamefont {Ota}\ \emph {et~al.}(2017)\citenamefont {Ota},
  \citenamefont {Tajima}, \citenamefont {Hanai}, \citenamefont {Inotani},\ and\
  \citenamefont {Ohashi}}]{Ota:2017}%
  \BibitemOpen
  \bibfield  {author} {\bibinfo {author} {\bibfnamefont {M.}~\bibnamefont
  {Ota}}, \bibinfo {author} {\bibfnamefont {H.}~\bibnamefont {Tajima}},
  \bibinfo {author} {\bibfnamefont {R.}~\bibnamefont {Hanai}}, \bibinfo
  {author} {\bibfnamefont {D.}~\bibnamefont {Inotani}}, \ and\ \bibinfo
  {author} {\bibfnamefont {Y.}~\bibnamefont {Ohashi}},\ }\bibfield  {title}
  {\enquote {\bibinfo {title} {Local photoemission spectra and effects of
  spatial inhomogeneity in the {BCS}-{BEC}-crossover regime of a trapped
  ultracold {F}ermi gas},}\ }\href {\doibase 10.1103/PhysRevA.95.053623}
  {\bibfield  {journal} {\bibinfo  {journal} {Phys. Rev. A}\ }\textbf {\bibinfo
  {volume} {95}},\ \bibinfo {pages} {053623} (\bibinfo {year}
  {2017})}\BibitemShut {NoStop}%
\bibitem [{\citenamefont {Carcy}\ \emph {et~al.}(2019)\citenamefont {Carcy},
  \citenamefont {Hoinka}, \citenamefont {Lingham}, \citenamefont {Dyke},
  \citenamefont {Kuhn}, \citenamefont {Hu},\ and\ \citenamefont
  {Vale}}]{Carcy:2019}%
  \BibitemOpen
  \bibfield  {author} {\bibinfo {author} {\bibfnamefont {C.}~\bibnamefont
  {Carcy}}, \bibinfo {author} {\bibfnamefont {S.}~\bibnamefont {Hoinka}},
  \bibinfo {author} {\bibfnamefont {M.~G.}\ \bibnamefont {Lingham}}, \bibinfo
  {author} {\bibfnamefont {P.}~\bibnamefont {Dyke}}, \bibinfo {author}
  {\bibfnamefont {C.~C.~N.}\ \bibnamefont {Kuhn}}, \bibinfo {author}
  {\bibfnamefont {H.}~\bibnamefont {Hu}}, \ and\ \bibinfo {author}
  {\bibfnamefont {C.~J.}\ \bibnamefont {Vale}},\ }\bibfield  {title} {\enquote
  {\bibinfo {title} {Contact and sum rules in a near-uniform {F}ermi gas at
  unitarity},}\ }\href {\doibase 10.1103/PhysRevLett.122.203401} {\bibfield
  {journal} {\bibinfo  {journal} {Phys. Rev. Lett.}\ }\textbf {\bibinfo
  {volume} {122}},\ \bibinfo {pages} {203401} (\bibinfo {year}
  {2019})}\BibitemShut {NoStop}%
\bibitem [{\citenamefont {Kozuma}\ \emph {et~al.}(1999)\citenamefont {Kozuma},
  \citenamefont {Deng}, \citenamefont {Hagley}, \citenamefont {Wen},
  \citenamefont {Lutwak}, \citenamefont {Helmerson}, \citenamefont {Rolston},\
  and\ \citenamefont {Phillips}}]{Kozuma:1999a}%
  \BibitemOpen
  \bibfield  {author} {\bibinfo {author} {\bibfnamefont {M.}~\bibnamefont
  {Kozuma}}, \bibinfo {author} {\bibfnamefont {L.}~\bibnamefont {Deng}},
  \bibinfo {author} {\bibfnamefont {E.~W.}\ \bibnamefont {Hagley}}, \bibinfo
  {author} {\bibfnamefont {J.}~\bibnamefont {Wen}}, \bibinfo {author}
  {\bibfnamefont {R.}~\bibnamefont {Lutwak}}, \bibinfo {author} {\bibfnamefont
  {K.}~\bibnamefont {Helmerson}}, \bibinfo {author} {\bibfnamefont {S.~L.}\
  \bibnamefont {Rolston}}, \ and\ \bibinfo {author} {\bibfnamefont {W.~D.}\
  \bibnamefont {Phillips}},\ }\bibfield  {title} {\enquote {\bibinfo {title}
  {Coherent splitting of {B}ose-{E}instein condensed atoms with optically
  induced {B}ragg diffraction},}\ }\href {\doibase 10.1103/PhysRevLett.82.871}
  {\bibfield  {journal} {\bibinfo  {journal} {Phys. Rev. Lett.}\ }\textbf
  {\bibinfo {volume} {82}},\ \bibinfo {pages} {871--875} (\bibinfo {year}
  {1999})}\BibitemShut {NoStop}%
\bibitem [{\citenamefont {Stenger}\ \emph {et~al.}(1999)\citenamefont
  {Stenger}, \citenamefont {Inouye}, \citenamefont {Chikkatur}, \citenamefont
  {Stamper-Kurn}, \citenamefont {Pritchard},\ and\ \citenamefont
  {Ketterle}}]{Stenger:1999b}%
  \BibitemOpen
  \bibfield  {author} {\bibinfo {author} {\bibfnamefont {J.}~\bibnamefont
  {Stenger}}, \bibinfo {author} {\bibfnamefont {S.}~\bibnamefont {Inouye}},
  \bibinfo {author} {\bibfnamefont {A.~P.}\ \bibnamefont {Chikkatur}}, \bibinfo
  {author} {\bibfnamefont {D.~M.}\ \bibnamefont {Stamper-Kurn}}, \bibinfo
  {author} {\bibfnamefont {D.~E.}\ \bibnamefont {Pritchard}}, \ and\ \bibinfo
  {author} {\bibfnamefont {W.}~\bibnamefont {Ketterle}},\ }\bibfield  {title}
  {\enquote {\bibinfo {title} {Bragg spectroscopy of a {B}ose-{E}instein
  condensate},}\ }\href {\doibase 10.1103/PhysRevLett.82.4569} {\bibfield
  {journal} {\bibinfo  {journal} {Phys. Rev. Lett.}\ }\textbf {\bibinfo
  {volume} {82}},\ \bibinfo {pages} {4569--4573} (\bibinfo {year}
  {1999})}\BibitemShut {NoStop}%
\bibitem [{\citenamefont {Navon}\ \emph {et~al.}(2019)\citenamefont {Navon},
  \citenamefont {Eigen}, \citenamefont {Zhang}, \citenamefont {Lopes},
  \citenamefont {Gaunt}, \citenamefont {Fujimoto}, \citenamefont {Tsubota},
  \citenamefont {Smith},\ and\ \citenamefont {Hadzibabic}}]{Navon:2019}%
  \BibitemOpen
  \bibfield  {author} {\bibinfo {author} {\bibfnamefont {N.}~\bibnamefont
  {Navon}}, \bibinfo {author} {\bibfnamefont {C.}~\bibnamefont {Eigen}},
  \bibinfo {author} {\bibfnamefont {J.}~\bibnamefont {Zhang}}, \bibinfo
  {author} {\bibfnamefont {R.}~\bibnamefont {Lopes}}, \bibinfo {author}
  {\bibfnamefont {A.~L.}\ \bibnamefont {Gaunt}}, \bibinfo {author}
  {\bibfnamefont {K.}~\bibnamefont {Fujimoto}}, \bibinfo {author}
  {\bibfnamefont {M.}~\bibnamefont {Tsubota}}, \bibinfo {author} {\bibfnamefont
  {R.~P.}\ \bibnamefont {Smith}}, \ and\ \bibinfo {author} {\bibfnamefont
  {Z.}~\bibnamefont {Hadzibabic}},\ }\bibfield  {title} {\enquote {\bibinfo
  {title} {Synthetic dissipation and cascade fluxes in a turbulent quantum
  gas},}\ }\href {\doibase 10.1126/science.aau6103} {\bibfield  {journal}
  {\bibinfo  {journal} {Science}\ }\textbf {\bibinfo {volume} {366}},\ \bibinfo
  {pages} {382} (\bibinfo {year} {2019})}\BibitemShut {NoStop}%
\bibitem [{\citenamefont {Gauthier}\ \emph
  {et~al.}(2019{\natexlab{b}})\citenamefont {Gauthier}, \citenamefont {Reeves},
  \citenamefont {Yu}, \citenamefont {Bradley}, \citenamefont {Baker},
  \citenamefont {Bell}, \citenamefont {Rubinsztein-Dunlop}, \citenamefont
  {Davis},\ and\ \citenamefont {Neely}}]{Gauthier:2019}%
  \BibitemOpen
  \bibfield  {author} {\bibinfo {author} {\bibfnamefont {G.}~\bibnamefont
  {Gauthier}}, \bibinfo {author} {\bibfnamefont {M.~T.}\ \bibnamefont
  {Reeves}}, \bibinfo {author} {\bibfnamefont {X.}~\bibnamefont {Yu}}, \bibinfo
  {author} {\bibfnamefont {A.~S.}\ \bibnamefont {Bradley}}, \bibinfo {author}
  {\bibfnamefont {M.~A.}\ \bibnamefont {Baker}}, \bibinfo {author}
  {\bibfnamefont {T.~A.}\ \bibnamefont {Bell}}, \bibinfo {author}
  {\bibfnamefont {H.}~\bibnamefont {Rubinsztein-Dunlop}}, \bibinfo {author}
  {\bibfnamefont {M.~J.}\ \bibnamefont {Davis}}, \ and\ \bibinfo {author}
  {\bibfnamefont {T.~W.}\ \bibnamefont {Neely}},\ }\bibfield  {title} {\enquote
  {\bibinfo {title} {Giant vortex clusters in a two-dimensional quantum
  fluid},}\ }\href {\doibase 10.1126/science.aat5718} {\bibfield  {journal}
  {\bibinfo  {journal} {Science}\ }\textbf {\bibinfo {volume} {364}},\ \bibinfo
  {pages} {1264} (\bibinfo {year} {2019}{\natexlab{b}})}\BibitemShut {NoStop}%
\bibitem [{\citenamefont {Johnstone}\ \emph {et~al.}(2019)\citenamefont
  {Johnstone}, \citenamefont {Groszek}, \citenamefont {Starkey}, \citenamefont
  {Billington}, \citenamefont {Simula},\ and\ \citenamefont
  {Helmerson}}]{Johnstone:2019}%
  \BibitemOpen
  \bibfield  {author} {\bibinfo {author} {\bibfnamefont {S.~P.}\ \bibnamefont
  {Johnstone}}, \bibinfo {author} {\bibfnamefont {A.~J.}\ \bibnamefont
  {Groszek}}, \bibinfo {author} {\bibfnamefont {P.~T.}\ \bibnamefont
  {Starkey}}, \bibinfo {author} {\bibfnamefont {C.~J.}\ \bibnamefont
  {Billington}}, \bibinfo {author} {\bibfnamefont {T.~P.}\ \bibnamefont
  {Simula}}, \ and\ \bibinfo {author} {\bibfnamefont {K.}~\bibnamefont
  {Helmerson}},\ }\bibfield  {title} {\enquote {\bibinfo {title} {{Evolution of
  large-scale flow from turbulence in a two-dimensional superfluid}},}\ }\href
  {\doibase 10.1126/science.aat5793} {\bibfield  {journal} {\bibinfo  {journal}
  {Science}\ }\textbf {\bibinfo {volume} {364}},\ \bibinfo {pages} {1267}
  (\bibinfo {year} {2019})}\BibitemShut {NoStop}%
\bibitem [{\citenamefont {Rauer}\ \emph {et~al.}(2018)\citenamefont {Rauer},
  \citenamefont {Erne}, \citenamefont {Schweigler}, \citenamefont {Cataldini},
  \citenamefont {Tajik},\ and\ \citenamefont {Schmiedmayer}}]{Rauer:2018}%
  \BibitemOpen
  \bibfield  {author} {\bibinfo {author} {\bibfnamefont {B.}~\bibnamefont
  {Rauer}}, \bibinfo {author} {\bibfnamefont {S.}~\bibnamefont {Erne}},
  \bibinfo {author} {\bibfnamefont {T.}~\bibnamefont {Schweigler}}, \bibinfo
  {author} {\bibfnamefont {F.}~\bibnamefont {Cataldini}}, \bibinfo {author}
  {\bibfnamefont {M.}~\bibnamefont {Tajik}}, \ and\ \bibinfo {author}
  {\bibfnamefont {J.}~\bibnamefont {Schmiedmayer}},\ }\bibfield  {title}
  {\enquote {\bibinfo {title} {Recurrences in an isolated quantum many-body
  system},}\ }\href {\doibase 10.1126/science.aan7938} {\bibfield  {journal}
  {\bibinfo  {journal} {Science}\ }\textbf {\bibinfo {volume} {360}},\ \bibinfo
  {pages} {307} (\bibinfo {year} {2018})}\BibitemShut {NoStop}%
\bibitem [{\citenamefont {Saint-Jalm}\ \emph {et~al.}(2019)\citenamefont
  {Saint-Jalm}, \citenamefont {Castilho}, \citenamefont {Le~Cerf},
  \citenamefont {Bakkali-Hassani}, \citenamefont {Ville}, \citenamefont
  {Nascimbene}, \citenamefont {Beugnon},\ and\ \citenamefont
  {Dalibard}}]{SaintJalm:2019}%
  \BibitemOpen
  \bibfield  {author} {\bibinfo {author} {\bibfnamefont {R.}~\bibnamefont
  {Saint-Jalm}}, \bibinfo {author} {\bibfnamefont {P.~C.~M.}\ \bibnamefont
  {Castilho}}, \bibinfo {author} {\bibfnamefont {E.}~\bibnamefont {Le~Cerf}},
  \bibinfo {author} {\bibfnamefont {B.}~\bibnamefont {Bakkali-Hassani}},
  \bibinfo {author} {\bibfnamefont {J.-L.}\ \bibnamefont {Ville}}, \bibinfo
  {author} {\bibfnamefont {S.}~\bibnamefont {Nascimbene}}, \bibinfo {author}
  {\bibfnamefont {J.}~\bibnamefont {Beugnon}}, \ and\ \bibinfo {author}
  {\bibfnamefont {J.}~\bibnamefont {Dalibard}},\ }\bibfield  {title} {\enquote
  {\bibinfo {title} {{Dynamical Symmetry and Breathers in a Two-Dimensional
  Bose Gas}},}\ }\href {\doibase 10.1103/PhysRevX.9.021035} {\bibfield
  {journal} {\bibinfo  {journal} {Phys. Rev. X}\ }\textbf {\bibinfo {volume}
  {9}},\ \bibinfo {pages} {021035} (\bibinfo {year} {2019})}\BibitemShut
  {NoStop}%
\bibitem [{\citenamefont {Stockdale}\ \emph {et~al.}(2020)\citenamefont
  {Stockdale}, \citenamefont {Reeves}, \citenamefont {Yu}, \citenamefont
  {Gauthier}, \citenamefont {Goddard-Lee}, \citenamefont {Bowen}, \citenamefont
  {Neely},\ and\ \citenamefont {Davis}}]{Stockdale:2020}%
  \BibitemOpen
  \bibfield  {author} {\bibinfo {author} {\bibfnamefont {O.~R.}\ \bibnamefont
  {Stockdale}}, \bibinfo {author} {\bibfnamefont {M.~T.}\ \bibnamefont
  {Reeves}}, \bibinfo {author} {\bibfnamefont {X.}~\bibnamefont {Yu}}, \bibinfo
  {author} {\bibfnamefont {G.}~\bibnamefont {Gauthier}}, \bibinfo {author}
  {\bibfnamefont {K.}~\bibnamefont {Goddard-Lee}}, \bibinfo {author}
  {\bibfnamefont {W.~P.}\ \bibnamefont {Bowen}}, \bibinfo {author}
  {\bibfnamefont {T.~W.}\ \bibnamefont {Neely}}, \ and\ \bibinfo {author}
  {\bibfnamefont {M.~J.}\ \bibnamefont {Davis}},\ }\bibfield  {title} {\enquote
  {\bibinfo {title} {Universal dynamics in the expansion of vortex clusters in
  a dissipative two-dimensional superfluid},}\ }\href {\doibase
  10.1103/PhysRevResearch.2.033138} {\bibfield  {journal} {\bibinfo  {journal}
  {Phys. Rev. Research}\ }\textbf {\bibinfo {volume} {2}},\ \bibinfo {pages}
  {033138} (\bibinfo {year} {2020})}\BibitemShut {NoStop}%
\bibitem [{\citenamefont {{Reeves}}\ \emph {et~al.}(2020)\citenamefont
  {{Reeves}}, \citenamefont {{Goddard-Lee}}, \citenamefont {{Gauthier}},
  \citenamefont {{Stockdale}}, \citenamefont {{Salman}}, \citenamefont
  {{Edmonds}}, \citenamefont {{Yu}}, \citenamefont {{Bradley}}, \citenamefont
  {{Baker}}, \citenamefont {{Rubinsztein-Dunlop}}, \citenamefont {{Davis}},\
  and\ \citenamefont {{Neely}}}]{Reeves:2020}%
  \BibitemOpen
  \bibfield  {author} {\bibinfo {author} {\bibfnamefont {M.~T.}\ \bibnamefont
  {{Reeves}}}, \bibinfo {author} {\bibfnamefont {K.}~\bibnamefont
  {{Goddard-Lee}}}, \bibinfo {author} {\bibfnamefont {G.}~\bibnamefont
  {{Gauthier}}}, \bibinfo {author} {\bibfnamefont {O.~R.}\ \bibnamefont
  {{Stockdale}}}, \bibinfo {author} {\bibfnamefont {H.}~\bibnamefont
  {{Salman}}}, \bibinfo {author} {\bibfnamefont {T.}~\bibnamefont {{Edmonds}}},
  \bibinfo {author} {\bibfnamefont {X.}~\bibnamefont {{Yu}}}, \bibinfo {author}
  {\bibfnamefont {A.~S.}\ \bibnamefont {{Bradley}}}, \bibinfo {author}
  {\bibfnamefont {M.}~\bibnamefont {{Baker}}}, \bibinfo {author} {\bibfnamefont
  {H.}~\bibnamefont {{Rubinsztein-Dunlop}}}, \bibinfo {author} {\bibfnamefont
  {M.~J.}\ \bibnamefont {{Davis}}}, \ and\ \bibinfo {author} {\bibfnamefont
  {T.~W.}\ \bibnamefont {{Neely}}},\ }\href@noop {} {\enquote {\bibinfo {title}
  {{Emergence of off-axis equilibria in a quantum vortex gas}},}\ } (\bibinfo
  {year} {2020}),\ \Eprint {http://arxiv.org/abs/2010.10049} {arXiv:2010.10049}
  \BibitemShut {NoStop}%
\bibitem [{\citenamefont {{Onsager}}(1949)}]{Onsager:1949}%
  \BibitemOpen
  \bibfield  {author} {\bibinfo {author} {\bibfnamefont {L.}~\bibnamefont
  {{Onsager}}},\ }\bibfield  {title} {\enquote {\bibinfo {title} {{Statistical
  hydrodynamics}},}\ }\href@noop {} {\bibfield  {journal} {\bibinfo  {journal}
  {Il Nuovo Cimento}\ }\textbf {\bibinfo {volume} {6}},\ \bibinfo {pages}
  {279--287} (\bibinfo {year} {1949})}\BibitemShut {NoStop}%
\bibitem [{\citenamefont {{Kwon}}\ \emph {et~al.}(2021)\citenamefont {{Kwon}},
  \citenamefont {{Del Pace}}, \citenamefont {{Xhani}}, \citenamefont
  {{Galantucci}}, \citenamefont {{Muzi Falconi}}, \citenamefont {{Inguscio}},
  \citenamefont {{Scazza}},\ and\ \citenamefont {{Roati}}}]{Kwon:2021}%
  \BibitemOpen
  \bibfield  {author} {\bibinfo {author} {\bibfnamefont {W.~J.}\ \bibnamefont
  {{Kwon}}}, \bibinfo {author} {\bibfnamefont {G.}~\bibnamefont {{Del Pace}}},
  \bibinfo {author} {\bibfnamefont {K.}~\bibnamefont {{Xhani}}}, \bibinfo
  {author} {\bibfnamefont {L.}~\bibnamefont {{Galantucci}}}, \bibinfo {author}
  {\bibfnamefont {A.}~\bibnamefont {{Muzi Falconi}}}, \bibinfo {author}
  {\bibfnamefont {M.}~\bibnamefont {{Inguscio}}}, \bibinfo {author}
  {\bibfnamefont {F.}~\bibnamefont {{Scazza}}}, \ and\ \bibinfo {author}
  {\bibfnamefont {G.}~\bibnamefont {{Roati}}},\ }\href@noop {} {\enquote
  {\bibinfo {title} {{Sound emission and annihilations in a programmable
  quantum vortex collider}},}\ } (\bibinfo {year} {2021}),\ \Eprint
  {http://arxiv.org/abs/2105.15180} {arXiv:2105.15180} \BibitemShut {NoStop}%
\bibitem [{\citenamefont {Kibble}(1976)}]{Kibble:1976}%
  \BibitemOpen
  \bibfield  {author} {\bibinfo {author} {\bibfnamefont {T.~W.~B.}\
  \bibnamefont {Kibble}},\ }\bibfield  {title} {\enquote {\bibinfo {title}
  {Topology of cosmic domains and strings},}\ }\href {\doibase
  10.1088/0305-4470/9/8/029} {\bibfield  {journal} {\bibinfo  {journal} {J.
  Phys. A}\ }\textbf {\bibinfo {volume} {9}},\ \bibinfo {pages} {1387--1398}
  (\bibinfo {year} {1976})}\BibitemShut {NoStop}%
\bibitem [{\citenamefont {Zurek}(1985)}]{Zurek:1985}%
  \BibitemOpen
  \bibfield  {author} {\bibinfo {author} {\bibfnamefont {W.~H.}\ \bibnamefont
  {Zurek}},\ }\bibfield  {title} {\enquote {\bibinfo {title} {Cosmological
  experiments in superfluid helium?}}\ }\href {\doibase 10.1038/317505a0}
  {\bibfield  {journal} {\bibinfo  {journal} {Nature}\ }\textbf {\bibinfo
  {volume} {317}},\ \bibinfo {pages} {505--508} (\bibinfo {year}
  {1985})}\BibitemShut {NoStop}%
\bibitem [{\citenamefont {del Campo}\ and\ \citenamefont
  {Zurek}(2014)}]{delCampo:2014}%
  \BibitemOpen
  \bibfield  {author} {\bibinfo {author} {\bibfnamefont {A.}~\bibnamefont {del
  Campo}}\ and\ \bibinfo {author} {\bibfnamefont {W.~H.}\ \bibnamefont
  {Zurek}},\ }\bibfield  {title} {\enquote {\bibinfo {title} {{Universality of
  phase transition dynamics: Topological defects from symmetry breaking}},}\
  }\href {\doibase https://doi.org/10.1142/S0217751X1430018X} {\bibfield
  {journal} {\bibinfo  {journal} {Int. J. Mod. Phys. A}\ }\textbf {\bibinfo
  {volume} {29}},\ \bibinfo {pages} {1430018} (\bibinfo {year}
  {2014})}\BibitemShut {NoStop}%
\bibitem [{\citenamefont {Beugnon}\ and\ \citenamefont
  {Navon}(2017)}]{Beugnon:2017}%
  \BibitemOpen
  \bibfield  {author} {\bibinfo {author} {\bibfnamefont {J.}~\bibnamefont
  {Beugnon}}\ and\ \bibinfo {author} {\bibfnamefont {N.}~\bibnamefont
  {Navon}},\ }\bibfield  {title} {\enquote {\bibinfo {title} {{Exploring the
  Kibble{\textendash}Zurek mechanism with homogeneous Bose gases}},}\ }\href
  {\doibase 10.1088/1361-6455/50/2/022002} {\bibfield  {journal} {\bibinfo
  {journal} {J. Phys. B: At. Mol. Opt. Phys.}\ }\textbf {\bibinfo {volume}
  {50}},\ \bibinfo {pages} {022002} (\bibinfo {year} {2017})}\BibitemShut
  {NoStop}%
\bibitem [{\citenamefont {Aidelsburger}\ \emph {et~al.}(2017)\citenamefont
  {Aidelsburger}, \citenamefont {Ville}, \citenamefont {Saint-Jalm},
  \citenamefont {Nascimb\`ene}, \citenamefont {Dalibard},\ and\ \citenamefont
  {Beugnon}}]{Aidelsburger:2017}%
  \BibitemOpen
  \bibfield  {author} {\bibinfo {author} {\bibfnamefont {M.}~\bibnamefont
  {Aidelsburger}}, \bibinfo {author} {\bibfnamefont {J.~L.}\ \bibnamefont
  {Ville}}, \bibinfo {author} {\bibfnamefont {R.}~\bibnamefont {Saint-Jalm}},
  \bibinfo {author} {\bibfnamefont {S.}~\bibnamefont {Nascimb\`ene}}, \bibinfo
  {author} {\bibfnamefont {J.}~\bibnamefont {Dalibard}}, \ and\ \bibinfo
  {author} {\bibfnamefont {J.}~\bibnamefont {Beugnon}},\ }\bibfield  {title}
  {\enquote {\bibinfo {title} {Relaxation dynamics in the merging of $n$
  independent condensates},}\ }\href {\doibase 10.1103/PhysRevLett.119.190403}
  {\bibfield  {journal} {\bibinfo  {journal} {Phys. Rev. Lett.}\ }\textbf
  {\bibinfo {volume} {119}},\ \bibinfo {pages} {190403} (\bibinfo {year}
  {2017})}\BibitemShut {NoStop}%
\bibitem [{\citenamefont {Drake}\ \emph {et~al.}(2012)\citenamefont {Drake},
  \citenamefont {Sagi}, \citenamefont {Paudel}, \citenamefont {Stewart},
  \citenamefont {Gaebler},\ and\ \citenamefont {Jin}}]{Drake:2012}%
  \BibitemOpen
  \bibfield  {author} {\bibinfo {author} {\bibfnamefont {T.~E.}\ \bibnamefont
  {Drake}}, \bibinfo {author} {\bibfnamefont {Y.}~\bibnamefont {Sagi}},
  \bibinfo {author} {\bibfnamefont {R.}~\bibnamefont {Paudel}}, \bibinfo
  {author} {\bibfnamefont {J.~T.}\ \bibnamefont {Stewart}}, \bibinfo {author}
  {\bibfnamefont {J.~P.}\ \bibnamefont {Gaebler}}, \ and\ \bibinfo {author}
  {\bibfnamefont {D.~S.}\ \bibnamefont {Jin}},\ }\bibfield  {title} {\enquote
  {\bibinfo {title} {Direct observation of the {F}ermi surface in an ultracold
  atomic gas},}\ }\href {\doibase 10.1103/PhysRevA.86.031601} {\bibfield
  {journal} {\bibinfo  {journal} {Phys. Rev. A}\ }\textbf {\bibinfo {volume}
  {86}},\ \bibinfo {pages} {031601} (\bibinfo {year} {2012})}\BibitemShut
  {NoStop}%
\bibitem [{\citenamefont {Donner}\ \emph {et~al.}(2007)\citenamefont {Donner},
  \citenamefont {Ritter}, \citenamefont {Bourdel}, \citenamefont {Ottl},
  \citenamefont {K{\"o}hl},\ and\ \citenamefont {Esslinger}}]{Donner:2007}%
  \BibitemOpen
  \bibfield  {author} {\bibinfo {author} {\bibfnamefont {T.}~\bibnamefont
  {Donner}}, \bibinfo {author} {\bibfnamefont {S.}~\bibnamefont {Ritter}},
  \bibinfo {author} {\bibfnamefont {T.}~\bibnamefont {Bourdel}}, \bibinfo
  {author} {\bibfnamefont {A.}~\bibnamefont {Ottl}}, \bibinfo {author}
  {\bibfnamefont {M.}~\bibnamefont {K{\"o}hl}}, \ and\ \bibinfo {author}
  {\bibfnamefont {T.}~\bibnamefont {Esslinger}},\ }\bibfield  {title} {\enquote
  {\bibinfo {title} {Critical behavior of a trapped interacting {B}ose gas},}\
  }\href {\doibase 10.1126/science.1138807} {\bibfield  {journal} {\bibinfo
  {journal} {Science}\ }\textbf {\bibinfo {volume} {315}},\ \bibinfo {pages}
  {1556--1558} (\bibinfo {year} {2007})}\BibitemShut {NoStop}%
\bibitem [{\citenamefont {Campostrini}\ \emph {et~al.}(2006)\citenamefont
  {Campostrini}, \citenamefont {Hasenbusch}, \citenamefont {Pelissetto},\ and\
  \citenamefont {Vicari}}]{Campostrini:2006}%
  \BibitemOpen
  \bibfield  {author} {\bibinfo {author} {\bibfnamefont {M.}~\bibnamefont
  {Campostrini}}, \bibinfo {author} {\bibfnamefont {M.}~\bibnamefont
  {Hasenbusch}}, \bibinfo {author} {\bibfnamefont {A.}~\bibnamefont
  {Pelissetto}}, \ and\ \bibinfo {author} {\bibfnamefont {E.}~\bibnamefont
  {Vicari}},\ }\bibfield  {title} {\enquote {\bibinfo {title} {Theoretical
  estimates of the critical exponents of the superfluid transition in
  $^{4}\mathrm{He}$ by lattice methods},}\ }\href {\doibase
  10.1103/PhysRevB.74.144506} {\bibfield  {journal} {\bibinfo  {journal} {Phys.
  Rev. B}\ }\textbf {\bibinfo {volume} {74}},\ \bibinfo {pages} {144506}
  (\bibinfo {year} {2006})}\BibitemShut {NoStop}%
\bibitem [{\citenamefont {Burovski}\ \emph {et~al.}(2006)\citenamefont
  {Burovski}, \citenamefont {Prokof'ev}, \citenamefont {Svistunov},\ and\
  \citenamefont {Troyer}}]{Burovski:2006}%
  \BibitemOpen
  \bibfield  {author} {\bibinfo {author} {\bibfnamefont {E.}~\bibnamefont
  {Burovski}}, \bibinfo {author} {\bibfnamefont {N.}~\bibnamefont {Prokof'ev}},
  \bibinfo {author} {\bibfnamefont {B.}~\bibnamefont {Svistunov}}, \ and\
  \bibinfo {author} {\bibfnamefont {M.}~\bibnamefont {Troyer}},\ }\bibfield
  {title} {\enquote {\bibinfo {title} {Critical temperature and thermodynamics
  of attractive fermions at unitarity},}\ }\href {\doibase
  10.1103/PhysRevLett.96.160402} {\bibfield  {journal} {\bibinfo  {journal}
  {Phys. Rev. Lett.}\ }\textbf {\bibinfo {volume} {96}},\ \bibinfo {pages}
  {160402} (\bibinfo {year} {2006})}\BibitemShut {NoStop}%
\bibitem [{\citenamefont {{Schweigler}}\ \emph {et~al.}(2021)\citenamefont
  {{Schweigler}}, \citenamefont {{Gluza}}, \citenamefont {{Tajik}},
  \citenamefont {{Sotiriadis}}, \citenamefont {{Cataldini}}, \citenamefont
  {{Ji}}, \citenamefont {{M{\o}ller}}, \citenamefont {{Sabino}}, \citenamefont
  {{Rauer}}, \citenamefont {{Eisert}},\ and\ \citenamefont
  {{Schmiedmayer}}}]{Schweigler:2020decay}%
  \BibitemOpen
  \bibfield  {author} {\bibinfo {author} {\bibfnamefont {T.}~\bibnamefont
  {{Schweigler}}}, \bibinfo {author} {\bibfnamefont {M.}~\bibnamefont
  {{Gluza}}}, \bibinfo {author} {\bibfnamefont {M.}~\bibnamefont {{Tajik}}},
  \bibinfo {author} {\bibfnamefont {S.}~\bibnamefont {{Sotiriadis}}}, \bibinfo
  {author} {\bibfnamefont {F.}~\bibnamefont {{Cataldini}}}, \bibinfo {author}
  {\bibfnamefont {S.-C.}\ \bibnamefont {{Ji}}}, \bibinfo {author}
  {\bibfnamefont {F.~S.}\ \bibnamefont {{M{\o}ller}}}, \bibinfo {author}
  {\bibfnamefont {J.}~\bibnamefont {{Sabino}}}, \bibinfo {author}
  {\bibfnamefont {B.}~\bibnamefont {{Rauer}}}, \bibinfo {author} {\bibfnamefont
  {J.}~\bibnamefont {{Eisert}}}, \ and\ \bibinfo {author} {\bibfnamefont
  {J.}~\bibnamefont {{Schmiedmayer}}},\ }\bibfield  {title} {\enquote {\bibinfo
  {title} {{Decay and recurrence of non-Gaussian correlations in a quantum
  many-body system}},}\ }\href {\doibase
  https://doi.org/10.1038/s41567-020-01139-2} {\bibfield  {journal} {\bibinfo
  {journal} {Nature Physics}\ }\textbf {\bibinfo {volume} {17}},\ \bibinfo
  {pages} {559–563} (\bibinfo {year} {2021})}\BibitemShut {NoStop}%
\bibitem [{\citenamefont {{Eigen}}\ \emph {et~al.}(2017)\citenamefont
  {{Eigen}}, \citenamefont {{Glidden}}, \citenamefont {{Lopes}}, \citenamefont
  {{Navon}}, \citenamefont {{Hadzibabic}},\ and\ \citenamefont
  {{Smith}}}]{Eigen:2017}%
  \BibitemOpen
  \bibfield  {author} {\bibinfo {author} {\bibfnamefont {C.}~\bibnamefont
  {{Eigen}}}, \bibinfo {author} {\bibfnamefont {J.~A.~P.}\ \bibnamefont
  {{Glidden}}}, \bibinfo {author} {\bibfnamefont {R.}~\bibnamefont {{Lopes}}},
  \bibinfo {author} {\bibfnamefont {N.}~\bibnamefont {{Navon}}}, \bibinfo
  {author} {\bibfnamefont {Z.}~\bibnamefont {{Hadzibabic}}}, \ and\ \bibinfo
  {author} {\bibfnamefont {R.~P.}\ \bibnamefont {{Smith}}},\ }\bibfield
  {title} {\enquote {\bibinfo {title} {{Universal Scaling Laws in the Dynamics
  of a Homogeneous Unitary Bose Gas}},}\ }\href {\doibase
  10.1103/PhysRevLett.119.250404} {\bibfield  {journal} {\bibinfo  {journal}
  {Phys. Rev. Lett.}\ }\textbf {\bibinfo {volume} {119}},\ \bibinfo {eid}
  {250404} (\bibinfo {year} {2017})}\BibitemShut {NoStop}%
\bibitem [{\citenamefont {Eigen}\ \emph {et~al.}(2018)\citenamefont {Eigen},
  \citenamefont {Glidden}, \citenamefont {Lopes}, \citenamefont {Cornell},
  \citenamefont {Smith},\ and\ \citenamefont {Hadzibabic}}]{Eigen:2018}%
  \BibitemOpen
  \bibfield  {author} {\bibinfo {author} {\bibfnamefont {C.}~\bibnamefont
  {Eigen}}, \bibinfo {author} {\bibfnamefont {J.~A.~P.}\ \bibnamefont
  {Glidden}}, \bibinfo {author} {\bibfnamefont {R.}~\bibnamefont {Lopes}},
  \bibinfo {author} {\bibfnamefont {E.~A.}\ \bibnamefont {Cornell}}, \bibinfo
  {author} {\bibfnamefont {R.~P.}\ \bibnamefont {Smith}}, \ and\ \bibinfo
  {author} {\bibfnamefont {Z.}~\bibnamefont {Hadzibabic}},\ }\bibfield  {title}
  {\enquote {\bibinfo {title} {Universal prethermal dynamics of {B}ose gases
  quenched to unitarity},}\ }\href {\doibase 10.1038/s41586-018-0674-1}
  {\bibfield  {journal} {\bibinfo  {journal} {Nature}\ }\textbf {\bibinfo
  {volume} {563}},\ \bibinfo {pages} {221} (\bibinfo {year}
  {2018})}\BibitemShut {NoStop}%
\bibitem [{\citenamefont {{Glidden}}\ \emph {et~al.}(2021)\citenamefont
  {{Glidden}}, \citenamefont {{Eigen}}, \citenamefont {{Dogra}}, \citenamefont
  {{Hilker}}, \citenamefont {{Smith}},\ and\ \citenamefont
  {{Hadzibabic}}}]{Glidden:2020}%
  \BibitemOpen
  \bibfield  {author} {\bibinfo {author} {\bibfnamefont {J.~A.~P.}\
  \bibnamefont {{Glidden}}}, \bibinfo {author} {\bibfnamefont {C.}~\bibnamefont
  {{Eigen}}}, \bibinfo {author} {\bibfnamefont {L.~H.}\ \bibnamefont
  {{Dogra}}}, \bibinfo {author} {\bibfnamefont {T.~A.}\ \bibnamefont
  {{Hilker}}}, \bibinfo {author} {\bibfnamefont {R.~P.}\ \bibnamefont
  {{Smith}}}, \ and\ \bibinfo {author} {\bibfnamefont {Z.}~\bibnamefont
  {{Hadzibabic}}},\ }\bibfield  {title} {\enquote {\bibinfo {title}
  {{Bidirectional dynamic scaling in an isolated Bose gas far from
  equilibrium}},}\ }\href {\doibase 10.1038/s41567-020-01114-x} {\bibfield
  {journal} {\bibinfo  {journal} {Nature Physics}\ }\textbf {\bibinfo {volume}
  {17}},\ \bibinfo {pages} {457--461} (\bibinfo {year} {2021})}\BibitemShut
  {NoStop}%
\bibitem [{\citenamefont {{Bakkali-Hassani}}\ \emph {et~al.}(2021)\citenamefont
  {{Bakkali-Hassani}}, \citenamefont {{Maury}}, \citenamefont {{Zou}},
  \citenamefont {{Le Cerf}}, \citenamefont {{Saint-Jalm}}, \citenamefont
  {{Castilho}}, \citenamefont {{Nascimbene}}, \citenamefont {{Dalibard}},\ and\
  \citenamefont {{Beugnon}}}]{Bakkali-Hassani:2021}%
  \BibitemOpen
  \bibfield  {author} {\bibinfo {author} {\bibfnamefont {B.}~\bibnamefont
  {{Bakkali-Hassani}}}, \bibinfo {author} {\bibfnamefont {C.}~\bibnamefont
  {{Maury}}}, \bibinfo {author} {\bibfnamefont {Y.~Q.}\ \bibnamefont {{Zou}}},
  \bibinfo {author} {\bibfnamefont {{\'E}.}~\bibnamefont {{Le Cerf}}}, \bibinfo
  {author} {\bibfnamefont {R.}~\bibnamefont {{Saint-Jalm}}}, \bibinfo {author}
  {\bibfnamefont {P.~C.~M.}\ \bibnamefont {{Castilho}}}, \bibinfo {author}
  {\bibfnamefont {S.}~\bibnamefont {{Nascimbene}}}, \bibinfo {author}
  {\bibfnamefont {J.}~\bibnamefont {{Dalibard}}}, \ and\ \bibinfo {author}
  {\bibfnamefont {J.}~\bibnamefont {{Beugnon}}},\ }\href@noop {} {\enquote
  {\bibinfo {title} {{Realization of a Townes soliton in a two-component planar
  Bose gas}},}\ } (\bibinfo {year} {2021}),\ \Eprint
  {http://arxiv.org/abs/2103.01605} {arXiv:2103.01605} \BibitemShut {NoStop}%
\bibitem [{\citenamefont {{Chen}}\ and\ \citenamefont
  {{Hung}}(2020)}]{Chen:2020}%
  \BibitemOpen
  \bibfield  {author} {\bibinfo {author} {\bibfnamefont {C.-A.}\ \bibnamefont
  {{Chen}}}\ and\ \bibinfo {author} {\bibfnamefont {C.-L.}\ \bibnamefont
  {{Hung}}},\ }\bibfield  {title} {\enquote {\bibinfo {title} {{Observation of
  Universal Quench Dynamics and Townes Soliton Formation from Modulational
  Instability in Two-Dimensional Bose Gases}},}\ }\href {\doibase
  10.1103/PhysRevLett.125.250401} {\bibfield  {journal} {\bibinfo  {journal}
  {\prl}\ }\textbf {\bibinfo {volume} {125}},\ \bibinfo {eid} {250401}
  (\bibinfo {year} {2020})}\BibitemShut {NoStop}%
\bibitem [{\citenamefont {{Chen}}\ and\ \citenamefont
  {{Hung}}(2021)}]{Chen:2021b}%
  \BibitemOpen
  \bibfield  {author} {\bibinfo {author} {\bibfnamefont {C.-A.}\ \bibnamefont
  {{Chen}}}\ and\ \bibinfo {author} {\bibfnamefont {C.-L.}\ \bibnamefont
  {{Hung}}},\ }\bibfield  {title} {\enquote {\bibinfo {title} {{Observation of
  scale invariance in two-dimensional matter-wave Townes solitons}},}\
  }\href@noop {} {\bibfield  {journal} {\bibinfo  {journal} {arXiv e-prints}\
  ,\ \bibinfo {eid} {arXiv:2103.03156}} (\bibinfo {year} {2021})},\ \Eprint
  {http://arxiv.org/abs/2103.03156} {arXiv:2103.03156 [cond-mat.quant-gas]}
  \BibitemShut {NoStop}%
\bibitem [{\citenamefont {{Zou}}\ \emph {et~al.}(2021)\citenamefont {{Zou}},
  \citenamefont {{Le Cerf}}, \citenamefont {{Bakkali-Hassani}}, \citenamefont
  {{Maury}}, \citenamefont {{Chauveau}}, \citenamefont {{Castilho}},
  \citenamefont {{Saint-Jalm}}, \citenamefont {{Nascimbene}}, \citenamefont
  {{Dalibard}},\ and\ \citenamefont {{Beugnon}}}]{Zou:2021}%
  \BibitemOpen
  \bibfield  {author} {\bibinfo {author} {\bibfnamefont {Y.~Q.}\ \bibnamefont
  {{Zou}}}, \bibinfo {author} {\bibfnamefont {{\'E}.}~\bibnamefont {{Le
  Cerf}}}, \bibinfo {author} {\bibfnamefont {B.}~\bibnamefont
  {{Bakkali-Hassani}}}, \bibinfo {author} {\bibfnamefont {C.}~\bibnamefont
  {{Maury}}}, \bibinfo {author} {\bibfnamefont {G.}~\bibnamefont {{Chauveau}}},
  \bibinfo {author} {\bibfnamefont {P.~C.~M.}\ \bibnamefont {{Castilho}}},
  \bibinfo {author} {\bibfnamefont {R.}~\bibnamefont {{Saint-Jalm}}}, \bibinfo
  {author} {\bibfnamefont {S.}~\bibnamefont {{Nascimbene}}}, \bibinfo {author}
  {\bibfnamefont {J.}~\bibnamefont {{Dalibard}}}, \ and\ \bibinfo {author}
  {\bibfnamefont {J.}~\bibnamefont {{Beugnon}}},\ }\href@noop {} {\enquote
  {\bibinfo {title} {{Optical control of the density and spin spatial profiles
  of a planar {B}ose gas}},}\ } (\bibinfo {year} {2021}),\ \Eprint
  {http://arxiv.org/abs/2102.05492} {arXiv:2102.05492} \BibitemShut {NoStop}%
\bibitem [{\citenamefont {Zhang}\ \emph {et~al.}(2021)\citenamefont {Zhang},
  \citenamefont {Chen}, \citenamefont {Yao},\ and\ \citenamefont
  {Chin}}]{Zhang:2021b}%
  \BibitemOpen
  \bibfield  {author} {\bibinfo {author} {\bibfnamefont {Z.}~\bibnamefont
  {Zhang}}, \bibinfo {author} {\bibfnamefont {L.}~\bibnamefont {Chen}},
  \bibinfo {author} {\bibfnamefont {K.}~\bibnamefont {Yao}}, \ and\ \bibinfo
  {author} {\bibfnamefont {C.}~\bibnamefont {Chin}},\ }\bibfield  {title}
  {\enquote {\bibinfo {title} {Transition from an atomic to a molecular
  {B}ose–{E}instein condensate},}\ }\href {\doibase
  https://doi.org/10.1038/s41586-021-03443-0} {\bibfield  {journal} {\bibinfo
  {journal} {Nature}\ }\textbf {\bibinfo {volume} {592}},\ \bibinfo {pages}
  {708–711} (\bibinfo {year} {2021})}\BibitemShut {NoStop}%
\bibitem [{\citenamefont {Eigen}\ \emph {et~al.}(2016)\citenamefont {Eigen},
  \citenamefont {Gaunt}, \citenamefont {Suleymanzade}, \citenamefont {Navon},
  \citenamefont {Hadzibabic},\ and\ \citenamefont {Smith}}]{Eigen:2016}%
  \BibitemOpen
  \bibfield  {author} {\bibinfo {author} {\bibfnamefont {C.}~\bibnamefont
  {Eigen}}, \bibinfo {author} {\bibfnamefont {A.~L.}\ \bibnamefont {Gaunt}},
  \bibinfo {author} {\bibfnamefont {A.}~\bibnamefont {Suleymanzade}}, \bibinfo
  {author} {\bibfnamefont {N.}~\bibnamefont {Navon}}, \bibinfo {author}
  {\bibfnamefont {Z.}~\bibnamefont {Hadzibabic}}, \ and\ \bibinfo {author}
  {\bibfnamefont {R.~P.}\ \bibnamefont {Smith}},\ }\bibfield  {title} {\enquote
  {\bibinfo {title} {{O}bservation of {W}eak {C}ollapse in a {B}ose--{E}instein
  {C}ondensate},}\ }\href {\doibase 10.1103/PhysRevX.6.041058} {\bibfield
  {journal} {\bibinfo  {journal} {Phys. Rev. X}\ }\textbf {\bibinfo {volume}
  {6}},\ \bibinfo {pages} {041058} (\bibinfo {year} {2016})}\BibitemShut
  {NoStop}%
\bibitem [{\citenamefont {Clark}\ \emph {et~al.}(2017)\citenamefont {Clark},
  \citenamefont {Gaj}, \citenamefont {Feng},\ and\ \citenamefont
  {Chin}}]{Clark:2017}%
  \BibitemOpen
  \bibfield  {author} {\bibinfo {author} {\bibfnamefont {L.~W.}\ \bibnamefont
  {Clark}}, \bibinfo {author} {\bibfnamefont {A.}~\bibnamefont {Gaj}}, \bibinfo
  {author} {\bibfnamefont {L.}~\bibnamefont {Feng}}, \ and\ \bibinfo {author}
  {\bibfnamefont {C.}~\bibnamefont {Chin}},\ }\bibfield  {title} {\enquote
  {\bibinfo {title} {Collective emission of matter-wave jets from driven
  {B}ose--{E}instein condensates},}\ }\href
  {https://doi.org/10.1038/nature24272} {\bibfield  {journal} {\bibinfo
  {journal} {Nature}\ }\textbf {\bibinfo {volume} {551}},\ \bibinfo {pages}
  {356} (\bibinfo {year} {2017})}\BibitemShut {NoStop}%
\bibitem [{\citenamefont {Fu}\ \emph {et~al.}(2018)\citenamefont {Fu},
  \citenamefont {Feng}, \citenamefont {Anderson}, \citenamefont {Clark},
  \citenamefont {Hu}, \citenamefont {Andrade}, \citenamefont {Chin},\ and\
  \citenamefont {Levin}}]{Fu:2018}%
  \BibitemOpen
  \bibfield  {author} {\bibinfo {author} {\bibfnamefont {H.}~\bibnamefont
  {Fu}}, \bibinfo {author} {\bibfnamefont {L.}~\bibnamefont {Feng}}, \bibinfo
  {author} {\bibfnamefont {B.~M.}\ \bibnamefont {Anderson}}, \bibinfo {author}
  {\bibfnamefont {L.~W.}\ \bibnamefont {Clark}}, \bibinfo {author}
  {\bibfnamefont {J.}~\bibnamefont {Hu}}, \bibinfo {author} {\bibfnamefont
  {J.~W.}\ \bibnamefont {Andrade}}, \bibinfo {author} {\bibfnamefont
  {C.}~\bibnamefont {Chin}}, \ and\ \bibinfo {author} {\bibfnamefont
  {K.}~\bibnamefont {Levin}},\ }\bibfield  {title} {\enquote {\bibinfo {title}
  {Density waves and jet emission asymmetry in {B}ose fireworks},}\ }\href
  {\doibase 10.1103/PhysRevLett.121.243001} {\bibfield  {journal} {\bibinfo
  {journal} {Phys. Rev. Lett.}\ }\textbf {\bibinfo {volume} {121}},\ \bibinfo
  {pages} {243001} (\bibinfo {year} {2018})}\BibitemShut {NoStop}%
\bibitem [{\citenamefont {Zhang}\ \emph {et~al.}(2020)\citenamefont {Zhang},
  \citenamefont {Yao}, \citenamefont {Feng}, \citenamefont {Hu},\ and\
  \citenamefont {Chin}}]{Zhang_2020}%
  \BibitemOpen
  \bibfield  {author} {\bibinfo {author} {\bibfnamefont {Z.}~\bibnamefont
  {Zhang}}, \bibinfo {author} {\bibfnamefont {K.-X.}\ \bibnamefont {Yao}},
  \bibinfo {author} {\bibfnamefont {L.}~\bibnamefont {Feng}}, \bibinfo {author}
  {\bibfnamefont {J.}~\bibnamefont {Hu}}, \ and\ \bibinfo {author}
  {\bibfnamefont {C.}~\bibnamefont {Chin}},\ }\bibfield  {title} {\enquote
  {\bibinfo {title} {Pattern formation in a driven {B}ose–{E}instein
  condensate},}\ }\href {\doibase 10.1038/s41567-020-0839-3} {\bibfield
  {journal} {\bibinfo  {journal} {Nature Physics}\ }\textbf {\bibinfo {volume}
  {16}},\ \bibinfo {pages} {652–656} (\bibinfo {year} {2020})}\BibitemShut
  {NoStop}%
\bibitem [{\citenamefont {{Chen}}\ \emph {et~al.}(2021)\citenamefont {{Chen}},
  \citenamefont {{Khlebnikov}},\ and\ \citenamefont {{Hung}}}]{Chen:2021}%
  \BibitemOpen
  \bibfield  {author} {\bibinfo {author} {\bibfnamefont {C.-A.}\ \bibnamefont
  {{Chen}}}, \bibinfo {author} {\bibfnamefont {S.}~\bibnamefont
  {{Khlebnikov}}}, \ and\ \bibinfo {author} {\bibfnamefont {C.-L.}\
  \bibnamefont {{Hung}}},\ }\href@noop {} {\enquote {\bibinfo {title}
  {{Observation of quasiparticle pair-production and quantum entanglement in
  atomic quantum gases quenched to an attractive interaction}},}\ } (\bibinfo
  {year} {2021}),\ \Eprint {http://arxiv.org/abs/2102.11215} {arXiv:2102.11215}
  \BibitemShut {NoStop}%
\bibitem [{\citenamefont {{Mathey}}\ and\ \citenamefont
  {{Polkovnikov}}(2010)}]{Mathey:2010}%
  \BibitemOpen
  \bibfield  {author} {\bibinfo {author} {\bibfnamefont {L.}~\bibnamefont
  {{Mathey}}}\ and\ \bibinfo {author} {\bibfnamefont {A.}~\bibnamefont
  {{Polkovnikov}}},\ }\bibfield  {title} {\enquote {\bibinfo {title} {{Light
  cone dynamics and reverse Kibble-Zurek mechanism in two-dimensional
  superfluids following a quantum quench}},}\ }\href {\doibase
  10.1103/PhysRevA.81.033605} {\bibfield  {journal} {\bibinfo  {journal}
  {\pra}\ }\textbf {\bibinfo {volume} {81}},\ \bibinfo {eid} {033605} (\bibinfo
  {year} {2010})}\BibitemShut {NoStop}%
\bibitem [{\citenamefont {{Jeli{\'c}}}\ and\ \citenamefont
  {{Cugliandolo}}(2011)}]{Jelic:2011}%
  \BibitemOpen
  \bibfield  {author} {\bibinfo {author} {\bibfnamefont {A.}~\bibnamefont
  {{Jeli{\'c}}}}\ and\ \bibinfo {author} {\bibfnamefont {L.~F.}\ \bibnamefont
  {{Cugliandolo}}},\ }\bibfield  {title} {\enquote {\bibinfo {title} {{Quench
  dynamics of the 2d XY model}},}\ }\href {\doibase
  10.1088/1742-5468/2011/02/P02032} {\bibfield  {journal} {\bibinfo  {journal}
  {Journal of Statistical Mechanics: Theory and Experiment}\ }\textbf {\bibinfo
  {volume} {2011}},\ \bibinfo {pages} {02032} (\bibinfo {year}
  {2011})}\BibitemShut {NoStop}%
\bibitem [{\citenamefont {Mathey}\ \emph {et~al.}(2017)\citenamefont {Mathey},
  \citenamefont {G\"unter}, \citenamefont {Dalibard},\ and\ \citenamefont
  {Polkovnikov}}]{Mathey:2017}%
  \BibitemOpen
  \bibfield  {author} {\bibinfo {author} {\bibfnamefont {L.}~\bibnamefont
  {Mathey}}, \bibinfo {author} {\bibfnamefont {K.~J.}\ \bibnamefont
  {G\"unter}}, \bibinfo {author} {\bibfnamefont {J.}~\bibnamefont {Dalibard}},
  \ and\ \bibinfo {author} {\bibfnamefont {A.}~\bibnamefont {Polkovnikov}},\
  }\bibfield  {title} {\enquote {\bibinfo {title} {Dynamic
  {K}osterlitz--{T}houless transition in two-dimensional {B}ose mixtures of
  ultracold atoms},}\ }\href {\doibase 10.1103/PhysRevA.95.053630} {\bibfield
  {journal} {\bibinfo  {journal} {Phys. Rev. A}\ }\textbf {\bibinfo {volume}
  {95}},\ \bibinfo {pages} {053630} (\bibinfo {year} {2017})}\BibitemShut
  {NoStop}%
\bibitem [{\citenamefont {{Comaron}}\ \emph {et~al.}(2019)\citenamefont
  {{Comaron}}, \citenamefont {{Larcher}}, \citenamefont {{Dalfovo}},\ and\
  \citenamefont {{Proukakis}}}]{Comaron:2019}%
  \BibitemOpen
  \bibfield  {author} {\bibinfo {author} {\bibfnamefont {P.}~\bibnamefont
  {{Comaron}}}, \bibinfo {author} {\bibfnamefont {F.}~\bibnamefont
  {{Larcher}}}, \bibinfo {author} {\bibfnamefont {F.}~\bibnamefont
  {{Dalfovo}}}, \ and\ \bibinfo {author} {\bibfnamefont {N.~P.}\ \bibnamefont
  {{Proukakis}}},\ }\bibfield  {title} {\enquote {\bibinfo {title} {{Quench
  dynamics of an ultracold two-dimensional Bose gas}},}\ }\href {\doibase
  10.1103/PhysRevA.100.033618} {\bibfield  {journal} {\bibinfo  {journal}
  {\pra}\ }\textbf {\bibinfo {volume} {100}},\ \bibinfo {eid} {033618}
  (\bibinfo {year} {2019})}\BibitemShut {NoStop}%
\bibitem [{\citenamefont {{Brown}}\ \emph {et~al.}(2021)\citenamefont
  {{Brown}}, \citenamefont {{Bland}}, \citenamefont {{Comaron}},\ and\
  \citenamefont {{Proukakis}}}]{Brown:2021}%
  \BibitemOpen
  \bibfield  {author} {\bibinfo {author} {\bibfnamefont {K.}~\bibnamefont
  {{Brown}}}, \bibinfo {author} {\bibfnamefont {T.}~\bibnamefont {{Bland}}},
  \bibinfo {author} {\bibfnamefont {P.}~\bibnamefont {{Comaron}}}, \ and\
  \bibinfo {author} {\bibfnamefont {N.~P.}\ \bibnamefont {{Proukakis}}},\
  }\bibfield  {title} {\enquote {\bibinfo {title} {{Periodic quenches across
  the Berezinskii-Kosterlitz-Thouless phase transition}},}\ }\href {\doibase
  10.1103/PhysRevResearch.3.013097} {\bibfield  {journal} {\bibinfo  {journal}
  {Phys. Rev. Research}\ }\textbf {\bibinfo {volume} {3}},\ \bibinfo {eid}
  {013097} (\bibinfo {year} {2021})}\BibitemShut {NoStop}%
\bibitem [{\citenamefont {{Fialko}}\ \emph {et~al.}(2015)\citenamefont
  {{Fialko}}, \citenamefont {{Opanchuk}}, \citenamefont {{Sidorov}},
  \citenamefont {{Drummond}},\ and\ \citenamefont {{Brand}}}]{Fialko:2015}%
  \BibitemOpen
  \bibfield  {author} {\bibinfo {author} {\bibfnamefont {O.}~\bibnamefont
  {{Fialko}}}, \bibinfo {author} {\bibfnamefont {B.}~\bibnamefont
  {{Opanchuk}}}, \bibinfo {author} {\bibfnamefont {A.~I.}\ \bibnamefont
  {{Sidorov}}}, \bibinfo {author} {\bibfnamefont {P.~D.}\ \bibnamefont
  {{Drummond}}}, \ and\ \bibinfo {author} {\bibfnamefont {J.}~\bibnamefont
  {{Brand}}},\ }\bibfield  {title} {\enquote {\bibinfo {title} {{Fate of the
  false vacuum: Towards realization with ultra-cold atoms}},}\ }\href {\doibase
  10.1209/0295-5075/110/56001} {\bibfield  {journal} {\bibinfo  {journal}
  {EPL}\ }\textbf {\bibinfo {volume} {110}},\ \bibinfo {eid} {56001} (\bibinfo
  {year} {2015})}\BibitemShut {NoStop}%
\bibitem [{\citenamefont {{Braden}}\ \emph {et~al.}(2018)\citenamefont
  {{Braden}}, \citenamefont {{Johnson}}, \citenamefont {{Peiris}},\ and\
  \citenamefont {{Weinfurtner}}}]{Braden:2018}%
  \BibitemOpen
  \bibfield  {author} {\bibinfo {author} {\bibfnamefont {J.}~\bibnamefont
  {{Braden}}}, \bibinfo {author} {\bibfnamefont {M.~C.}\ \bibnamefont
  {{Johnson}}}, \bibinfo {author} {\bibfnamefont {H.~V.}\ \bibnamefont
  {{Peiris}}}, \ and\ \bibinfo {author} {\bibfnamefont {S.}~\bibnamefont
  {{Weinfurtner}}},\ }\bibfield  {title} {\enquote {\bibinfo {title} {{Towards
  the cold atom analog false vacuum}},}\ }\href {\doibase
  10.1007/JHEP07(2018)014} {\bibfield  {journal} {\bibinfo  {journal} {Journal
  of High Energy Physics}\ }\textbf {\bibinfo {volume} {2018}},\ \bibinfo {eid}
  {14} (\bibinfo {year} {2018})}\BibitemShut {NoStop}%
\bibitem [{\citenamefont {{Braden}}\ \emph {et~al.}(2019)\citenamefont
  {{Braden}}, \citenamefont {{Johnson}}, \citenamefont {{Peiris}},
  \citenamefont {{Pontzen}},\ and\ \citenamefont
  {{Weinfurtner}}}]{Braden:2019}%
  \BibitemOpen
  \bibfield  {author} {\bibinfo {author} {\bibfnamefont {J.}~\bibnamefont
  {{Braden}}}, \bibinfo {author} {\bibfnamefont {M.~C.}\ \bibnamefont
  {{Johnson}}}, \bibinfo {author} {\bibfnamefont {H.~V.}\ \bibnamefont
  {{Peiris}}}, \bibinfo {author} {\bibfnamefont {A.}~\bibnamefont {{Pontzen}}},
  \ and\ \bibinfo {author} {\bibfnamefont {S.}~\bibnamefont {{Weinfurtner}}},\
  }\bibfield  {title} {\enquote {\bibinfo {title} {{Nonlinear dynamics of the
  cold atom analog false vacuum}},}\ }\href {\doibase 10.1007/JHEP10(2019)174}
  {\bibfield  {journal} {\bibinfo  {journal} {Journal of High Energy Physics}\
  }\textbf {\bibinfo {volume} {2019}},\ \bibinfo {eid} {174} (\bibinfo {year}
  {2019})}\BibitemShut {NoStop}%
\bibitem [{\citenamefont {Billam}\ \emph {et~al.}(2019)\citenamefont {Billam},
  \citenamefont {Gregory}, \citenamefont {Michel},\ and\ \citenamefont
  {Moss}}]{Billam:2019}%
  \BibitemOpen
  \bibfield  {author} {\bibinfo {author} {\bibfnamefont {T.~P.}\ \bibnamefont
  {Billam}}, \bibinfo {author} {\bibfnamefont {R.}~\bibnamefont {Gregory}},
  \bibinfo {author} {\bibfnamefont {F.}~\bibnamefont {Michel}}, \ and\ \bibinfo
  {author} {\bibfnamefont {I.~G.}\ \bibnamefont {Moss}},\ }\bibfield  {title}
  {\enquote {\bibinfo {title} {Simulating seeded vacuum decay in a cold atom
  system},}\ }\href {\doibase 10.1103/PhysRevD.100.065016} {\bibfield
  {journal} {\bibinfo  {journal} {Phys. Rev. D}\ }\textbf {\bibinfo {volume}
  {100}},\ \bibinfo {pages} {065016} (\bibinfo {year} {2019})}\BibitemShut
  {NoStop}%
\bibitem [{\citenamefont {{Goldman}}\ \emph {et~al.}(2016)\citenamefont
  {{Goldman}}, \citenamefont {{Budich}},\ and\ \citenamefont
  {{Zoller}}}]{Goldman:2016}%
  \BibitemOpen
  \bibfield  {author} {\bibinfo {author} {\bibfnamefont {N.}~\bibnamefont
  {{Goldman}}}, \bibinfo {author} {\bibfnamefont {J.~C.}\ \bibnamefont
  {{Budich}}}, \ and\ \bibinfo {author} {\bibfnamefont {P.}~\bibnamefont
  {{Zoller}}},\ }\bibfield  {title} {\enquote {\bibinfo {title} {{Topological
  quantum matter with ultracold gases in optical lattices}},}\ }\href {\doibase
  10.1038/nphys3803} {\bibfield  {journal} {\bibinfo  {journal} {Nature
  Physics}\ }\textbf {\bibinfo {volume} {12}},\ \bibinfo {pages} {639--645}
  (\bibinfo {year} {2016})}\BibitemShut {NoStop}%
\bibitem [{\citenamefont {{Ozawa}}\ \emph {et~al.}(2019)\citenamefont
  {{Ozawa}}, \citenamefont {{Price}}, \citenamefont {{Amo}}, \citenamefont
  {{Goldman}}, \citenamefont {{Hafezi}}, \citenamefont {{Lu}}, \citenamefont
  {{Rechtsman}}, \citenamefont {{Schuster}}, \citenamefont {{Simon}},
  \citenamefont {{Zilberberg}},\ and\ \citenamefont
  {{Carusotto}}}]{Ozawa:2019}%
  \BibitemOpen
  \bibfield  {author} {\bibinfo {author} {\bibfnamefont {T.}~\bibnamefont
  {{Ozawa}}}, \bibinfo {author} {\bibfnamefont {H.~M.}\ \bibnamefont
  {{Price}}}, \bibinfo {author} {\bibfnamefont {A.}~\bibnamefont {{Amo}}},
  \bibinfo {author} {\bibfnamefont {N.}~\bibnamefont {{Goldman}}}, \bibinfo
  {author} {\bibfnamefont {M.}~\bibnamefont {{Hafezi}}}, \bibinfo {author}
  {\bibfnamefont {L.}~\bibnamefont {{Lu}}}, \bibinfo {author} {\bibfnamefont
  {M.~C.}\ \bibnamefont {{Rechtsman}}}, \bibinfo {author} {\bibfnamefont
  {D.}~\bibnamefont {{Schuster}}}, \bibinfo {author} {\bibfnamefont
  {J.}~\bibnamefont {{Simon}}}, \bibinfo {author} {\bibfnamefont
  {O.}~\bibnamefont {{Zilberberg}}}, \ and\ \bibinfo {author} {\bibfnamefont
  {I.}~\bibnamefont {{Carusotto}}},\ }\bibfield  {title} {\enquote {\bibinfo
  {title} {{Topological photonics}},}\ }\href {\doibase
  10.1103/RevModPhys.91.015006} {\bibfield  {journal} {\bibinfo  {journal}
  {Reviews of Modern Physics}\ }\textbf {\bibinfo {volume} {91}},\ \bibinfo
  {eid} {015006} (\bibinfo {year} {2019})}\BibitemShut {NoStop}%
\bibitem [{\citenamefont {{Fletcher}}\ \emph {et~al.}(2019)\citenamefont
  {{Fletcher}}, \citenamefont {{Shaffer}}, \citenamefont {{Wilson}},
  \citenamefont {{Patel}}, \citenamefont {{Yan}}, \citenamefont {{Cr{\'e}pel}},
  \citenamefont {{Mukherjee}},\ and\ \citenamefont
  {{Zwierlein}}}]{fletcher2019geometric}%
  \BibitemOpen
  \bibfield  {author} {\bibinfo {author} {\bibfnamefont {R.~J.}\ \bibnamefont
  {{Fletcher}}}, \bibinfo {author} {\bibfnamefont {A.}~\bibnamefont
  {{Shaffer}}}, \bibinfo {author} {\bibfnamefont {C.~C.}\ \bibnamefont
  {{Wilson}}}, \bibinfo {author} {\bibfnamefont {P.~B.}\ \bibnamefont
  {{Patel}}}, \bibinfo {author} {\bibfnamefont {Z.}~\bibnamefont {{Yan}}},
  \bibinfo {author} {\bibfnamefont {V.}~\bibnamefont {{Cr{\'e}pel}}}, \bibinfo
  {author} {\bibfnamefont {B.}~\bibnamefont {{Mukherjee}}}, \ and\ \bibinfo
  {author} {\bibfnamefont {M.~W.}\ \bibnamefont {{Zwierlein}}},\ }\href@noop {}
  {\enquote {\bibinfo {title} {{Geometric squeezing into the lowest Landau
  level}},}\ } (\bibinfo {year} {2019}),\ \Eprint
  {http://arxiv.org/abs/1911.12347} {arXiv:1911.12347} \BibitemShut {NoStop}%
\bibitem [{\citenamefont {{Mancini}}\ \emph {et~al.}(2015)\citenamefont
  {{Mancini}}, \citenamefont {{Pagano}}, \citenamefont {{Cappellini}},
  \citenamefont {{Livi}}, \citenamefont {{Rider}}, \citenamefont {{Catani}},
  \citenamefont {{Sias}}, \citenamefont {{Zoller}}, \citenamefont {{Inguscio}},
  \citenamefont {{Dalmonte}},\ and\ \citenamefont {{Fallani}}}]{Mancini:2015}%
  \BibitemOpen
  \bibfield  {author} {\bibinfo {author} {\bibfnamefont {M.}~\bibnamefont
  {{Mancini}}}, \bibinfo {author} {\bibfnamefont {G.}~\bibnamefont {{Pagano}}},
  \bibinfo {author} {\bibfnamefont {G.}~\bibnamefont {{Cappellini}}}, \bibinfo
  {author} {\bibfnamefont {L.}~\bibnamefont {{Livi}}}, \bibinfo {author}
  {\bibfnamefont {M.}~\bibnamefont {{Rider}}}, \bibinfo {author} {\bibfnamefont
  {J.}~\bibnamefont {{Catani}}}, \bibinfo {author} {\bibfnamefont
  {C.}~\bibnamefont {{Sias}}}, \bibinfo {author} {\bibfnamefont
  {P.}~\bibnamefont {{Zoller}}}, \bibinfo {author} {\bibfnamefont
  {M.}~\bibnamefont {{Inguscio}}}, \bibinfo {author} {\bibfnamefont
  {M.}~\bibnamefont {{Dalmonte}}}, \ and\ \bibinfo {author} {\bibfnamefont
  {L.}~\bibnamefont {{Fallani}}},\ }\bibfield  {title} {\enquote {\bibinfo
  {title} {{Observation of chiral edge states with neutral fermions in
  synthetic Hall ribbons}},}\ }\href {\doibase 10.1126/science.aaa8736}
  {\bibfield  {journal} {\bibinfo  {journal} {Science}\ }\textbf {\bibinfo
  {volume} {349}},\ \bibinfo {pages} {1510--1513} (\bibinfo {year}
  {2015})}\BibitemShut {NoStop}%
\bibitem [{\citenamefont {{Stuhl}}\ \emph {et~al.}(2015)\citenamefont
  {{Stuhl}}, \citenamefont {{Lu}}, \citenamefont {{Aycock}}, \citenamefont
  {{Genkina}},\ and\ \citenamefont {{Spielman}}}]{Stuhl:2015}%
  \BibitemOpen
  \bibfield  {author} {\bibinfo {author} {\bibfnamefont {B.~K.}\ \bibnamefont
  {{Stuhl}}}, \bibinfo {author} {\bibfnamefont {H.~I.}\ \bibnamefont {{Lu}}},
  \bibinfo {author} {\bibfnamefont {L.~M.}\ \bibnamefont {{Aycock}}}, \bibinfo
  {author} {\bibfnamefont {D.}~\bibnamefont {{Genkina}}}, \ and\ \bibinfo
  {author} {\bibfnamefont {I.~B.}\ \bibnamefont {{Spielman}}},\ }\bibfield
  {title} {\enquote {\bibinfo {title} {{Visualizing edge states with an atomic
  Bose gas in the quantum Hall regime}},}\ }\href {\doibase
  10.1126/science.aaa8515} {\bibfield  {journal} {\bibinfo  {journal}
  {Science}\ }\textbf {\bibinfo {volume} {349}},\ \bibinfo {pages} {1514--1518}
  (\bibinfo {year} {2015})}\BibitemShut {NoStop}%
\bibitem [{\citenamefont {{Chalopin}}\ \emph {et~al.}(2020)\citenamefont
  {{Chalopin}}, \citenamefont {{Satoor}}, \citenamefont {{Evrard}},
  \citenamefont {{Makhalov}}, \citenamefont {{Dalibard}}, \citenamefont
  {{Lopes}},\ and\ \citenamefont {{Nascimbene}}}]{Chalopin:2020}%
  \BibitemOpen
  \bibfield  {author} {\bibinfo {author} {\bibfnamefont {T.}~\bibnamefont
  {{Chalopin}}}, \bibinfo {author} {\bibfnamefont {T.}~\bibnamefont
  {{Satoor}}}, \bibinfo {author} {\bibfnamefont {A.}~\bibnamefont {{Evrard}}},
  \bibinfo {author} {\bibfnamefont {V.}~\bibnamefont {{Makhalov}}}, \bibinfo
  {author} {\bibfnamefont {J.}~\bibnamefont {{Dalibard}}}, \bibinfo {author}
  {\bibfnamefont {R.}~\bibnamefont {{Lopes}}}, \ and\ \bibinfo {author}
  {\bibfnamefont {S.}~\bibnamefont {{Nascimbene}}},\ }\bibfield  {title}
  {\enquote {\bibinfo {title} {{Probing chiral edge dynamics and bulk topology
  of a synthetic Hall system}},}\ }\href {\doibase 10.1038/s41567-020-0942-5}
  {\bibfield  {journal} {\bibinfo  {journal} {Nature Physics}\ }\textbf
  {\bibinfo {volume} {16}},\ \bibinfo {pages} {1017--1021} (\bibinfo {year}
  {2020})}\BibitemShut {NoStop}%
\bibitem [{\citenamefont {Roccuzzo}\ \emph {et~al.}(2021)\citenamefont
  {Roccuzzo}, \citenamefont {Stringari},\ and\ \citenamefont
  {Recati}}]{Roccuzzo:2021}%
  \BibitemOpen
  \bibfield  {author} {\bibinfo {author} {\bibfnamefont {S.~M.}\ \bibnamefont
  {Roccuzzo}}, \bibinfo {author} {\bibfnamefont {S.}~\bibnamefont {Stringari}},
  \ and\ \bibinfo {author} {\bibfnamefont {A.}~\bibnamefont {Recati}},\
  }\href@noop {} {\enquote {\bibinfo {title} {Supersolid edge and bulk phases
  of a dipolar quantum gas in a box},}\ } (\bibinfo {year} {2021}),\ \Eprint
  {http://arxiv.org/abs/2104.01068} {arXiv:2104.01068} \BibitemShut {NoStop}%
\bibitem [{\citenamefont {B\"ottcher}\ \emph {et~al.}(2019)\citenamefont
  {B\"ottcher}, \citenamefont {Schmidt}, \citenamefont {Wenzel}, \citenamefont
  {Hertkorn}, \citenamefont {Guo}, \citenamefont {Langen},\ and\ \citenamefont
  {Pfau}}]{Bottcher:2019}%
  \BibitemOpen
  \bibfield  {author} {\bibinfo {author} {\bibfnamefont {F.}~\bibnamefont
  {B\"ottcher}}, \bibinfo {author} {\bibfnamefont {J.-N.}\ \bibnamefont
  {Schmidt}}, \bibinfo {author} {\bibfnamefont {M.}~\bibnamefont {Wenzel}},
  \bibinfo {author} {\bibfnamefont {J.}~\bibnamefont {Hertkorn}}, \bibinfo
  {author} {\bibfnamefont {M.}~\bibnamefont {Guo}}, \bibinfo {author}
  {\bibfnamefont {T.}~\bibnamefont {Langen}}, \ and\ \bibinfo {author}
  {\bibfnamefont {T.}~\bibnamefont {Pfau}},\ }\bibfield  {title} {\enquote
  {\bibinfo {title} {Transient supersolid properties in an array of dipolar
  quantum droplets},}\ }\href {\doibase 10.1103/PhysRevX.9.011051} {\bibfield
  {journal} {\bibinfo  {journal} {Phys. Rev. X}\ }\textbf {\bibinfo {volume}
  {9}},\ \bibinfo {pages} {011051} (\bibinfo {year} {2019})}\BibitemShut
  {NoStop}%
\bibitem [{\citenamefont {Tanzi}\ \emph {et~al.}(2019)\citenamefont {Tanzi},
  \citenamefont {Lucioni}, \citenamefont {Fam\`a}, \citenamefont {Catani},
  \citenamefont {Fioretti}, \citenamefont {Gabbanini}, \citenamefont {Bisset},
  \citenamefont {Santos},\ and\ \citenamefont {Modugno}}]{Tanzi:2019}%
  \BibitemOpen
  \bibfield  {author} {\bibinfo {author} {\bibfnamefont {L.}~\bibnamefont
  {Tanzi}}, \bibinfo {author} {\bibfnamefont {E.}~\bibnamefont {Lucioni}},
  \bibinfo {author} {\bibfnamefont {F.}~\bibnamefont {Fam\`a}}, \bibinfo
  {author} {\bibfnamefont {J.}~\bibnamefont {Catani}}, \bibinfo {author}
  {\bibfnamefont {A.}~\bibnamefont {Fioretti}}, \bibinfo {author}
  {\bibfnamefont {C.}~\bibnamefont {Gabbanini}}, \bibinfo {author}
  {\bibfnamefont {R.~N.}\ \bibnamefont {Bisset}}, \bibinfo {author}
  {\bibfnamefont {L.}~\bibnamefont {Santos}}, \ and\ \bibinfo {author}
  {\bibfnamefont {G.}~\bibnamefont {Modugno}},\ }\bibfield  {title} {\enquote
  {\bibinfo {title} {Observation of a dipolar quantum gas with metastable
  supersolid properties},}\ }\href {\doibase 10.1103/PhysRevLett.122.130405}
  {\bibfield  {journal} {\bibinfo  {journal} {Phys. Rev. Lett.}\ }\textbf
  {\bibinfo {volume} {122}},\ \bibinfo {pages} {130405} (\bibinfo {year}
  {2019})}\BibitemShut {NoStop}%
\bibitem [{\citenamefont {Chomaz}\ \emph {et~al.}(2019)\citenamefont {Chomaz},
  \citenamefont {Petter}, \citenamefont {Ilzh\"ofer}, \citenamefont {Natale},
  \citenamefont {Trautmann}, \citenamefont {Politi}, \citenamefont
  {Durastante}, \citenamefont {van Bijnen}, \citenamefont {Patscheider},
  \citenamefont {Sohmen}, \citenamefont {Mark},\ and\ \citenamefont
  {Ferlaino}}]{Chomaz:2019}%
  \BibitemOpen
  \bibfield  {author} {\bibinfo {author} {\bibfnamefont {L.}~\bibnamefont
  {Chomaz}}, \bibinfo {author} {\bibfnamefont {D.}~\bibnamefont {Petter}},
  \bibinfo {author} {\bibfnamefont {P.}~\bibnamefont {Ilzh\"ofer}}, \bibinfo
  {author} {\bibfnamefont {G.}~\bibnamefont {Natale}}, \bibinfo {author}
  {\bibfnamefont {A.}~\bibnamefont {Trautmann}}, \bibinfo {author}
  {\bibfnamefont {C.}~\bibnamefont {Politi}}, \bibinfo {author} {\bibfnamefont
  {G.}~\bibnamefont {Durastante}}, \bibinfo {author} {\bibfnamefont {R.~M.~W.}\
  \bibnamefont {van Bijnen}}, \bibinfo {author} {\bibfnamefont
  {A.}~\bibnamefont {Patscheider}}, \bibinfo {author} {\bibfnamefont
  {M.}~\bibnamefont {Sohmen}}, \bibinfo {author} {\bibfnamefont {M.~J.}\
  \bibnamefont {Mark}}, \ and\ \bibinfo {author} {\bibfnamefont
  {F.}~\bibnamefont {Ferlaino}},\ }\bibfield  {title} {\enquote {\bibinfo
  {title} {Long-lived and transient supersolid behaviors in dipolar quantum
  gases},}\ }\href {\doibase 10.1103/PhysRevX.9.021012} {\bibfield  {journal}
  {\bibinfo  {journal} {Phys. Rev. X}\ }\textbf {\bibinfo {volume} {9}},\
  \bibinfo {pages} {021012} (\bibinfo {year} {2019})}\BibitemShut {NoStop}%
\bibitem [{\citenamefont {Norcia}\ \emph {et~al.}(2021)\citenamefont {Norcia},
  \citenamefont {Politi}, \citenamefont {Klaus}, \citenamefont {Poli},
  \citenamefont {Sohmen}, \citenamefont {Mark}, \citenamefont {Bisset},
  \citenamefont {Santos},\ and\ \citenamefont {Ferlaino}}]{Norcia:2021}%
  \BibitemOpen
  \bibfield  {author} {\bibinfo {author} {\bibfnamefont {M.~A.}\ \bibnamefont
  {Norcia}}, \bibinfo {author} {\bibfnamefont {C.}~\bibnamefont {Politi}},
  \bibinfo {author} {\bibfnamefont {L.}~\bibnamefont {Klaus}}, \bibinfo
  {author} {\bibfnamefont {E.}~\bibnamefont {Poli}}, \bibinfo {author}
  {\bibfnamefont {M.}~\bibnamefont {Sohmen}}, \bibinfo {author} {\bibfnamefont
  {M.~J.}\ \bibnamefont {Mark}}, \bibinfo {author} {\bibfnamefont
  {R.}~\bibnamefont {Bisset}}, \bibinfo {author} {\bibfnamefont
  {L.}~\bibnamefont {Santos}}, \ and\ \bibinfo {author} {\bibfnamefont
  {F.}~\bibnamefont {Ferlaino}},\ }\href@noop {} {\enquote {\bibinfo {title}
  {Two-dimensional supersolidity in a dipolar quantum gas},}\ } (\bibinfo
  {year} {2021}),\ \Eprint {http://arxiv.org/abs/2102.05555} {arXiv:2102.05555}
  \BibitemShut {NoStop}%
\bibitem [{\citenamefont {{Hertkorn}}\ \emph {et~al.}(2021)\citenamefont
  {{Hertkorn}}, \citenamefont {{Schmidt}}, \citenamefont {{Guo}}, \citenamefont
  {{B{\"o}ttcher}}, \citenamefont {{Ng}}, \citenamefont {{Graham}},
  \citenamefont {{Uerlings}}, \citenamefont {{B{\"u}chler}}, \citenamefont
  {{Langen}}, \citenamefont {{Zwierlein}},\ and\ \citenamefont
  {{Pfau}}}]{Hertkorn:2021b}%
  \BibitemOpen
  \bibfield  {author} {\bibinfo {author} {\bibfnamefont {J.}~\bibnamefont
  {{Hertkorn}}}, \bibinfo {author} {\bibfnamefont {J.~N.}\ \bibnamefont
  {{Schmidt}}}, \bibinfo {author} {\bibfnamefont {M.}~\bibnamefont {{Guo}}},
  \bibinfo {author} {\bibfnamefont {F.}~\bibnamefont {{B{\"o}ttcher}}},
  \bibinfo {author} {\bibfnamefont {K.~S.~H.}\ \bibnamefont {{Ng}}}, \bibinfo
  {author} {\bibfnamefont {S.~D.}\ \bibnamefont {{Graham}}}, \bibinfo {author}
  {\bibfnamefont {P.}~\bibnamefont {{Uerlings}}}, \bibinfo {author}
  {\bibfnamefont {H.~P.}\ \bibnamefont {{B{\"u}chler}}}, \bibinfo {author}
  {\bibfnamefont {T.}~\bibnamefont {{Langen}}}, \bibinfo {author}
  {\bibfnamefont {M.}~\bibnamefont {{Zwierlein}}}, \ and\ \bibinfo {author}
  {\bibfnamefont {T.}~\bibnamefont {{Pfau}}},\ }\href@noop {} {\enquote
  {\bibinfo {title} {{Supersolidity in Two-Dimensional Trapped Dipolar Droplet
  Arrays}},}\ } (\bibinfo {year} {2021}),\ \Eprint
  {http://arxiv.org/abs/2103.09752} {arXiv:2103.09752} \BibitemShut {NoStop}%
\bibitem [{\citenamefont {{B{\"o}ttcher}}\ \emph {et~al.}(2021)\citenamefont
  {{B{\"o}ttcher}}, \citenamefont {{Schmidt}}, \citenamefont {{Hertkorn}},
  \citenamefont {{Ng}}, \citenamefont {{Graham}}, \citenamefont {{Guo}},
  \citenamefont {{Langen}},\ and\ \citenamefont {{Pfau}}}]{Bottcher:2021}%
  \BibitemOpen
  \bibfield  {author} {\bibinfo {author} {\bibfnamefont {F.}~\bibnamefont
  {{B{\"o}ttcher}}}, \bibinfo {author} {\bibfnamefont {J.-N.}\ \bibnamefont
  {{Schmidt}}}, \bibinfo {author} {\bibfnamefont {J.}~\bibnamefont
  {{Hertkorn}}}, \bibinfo {author} {\bibfnamefont {K.~S.~H.}\ \bibnamefont
  {{Ng}}}, \bibinfo {author} {\bibfnamefont {S.~D.}\ \bibnamefont {{Graham}}},
  \bibinfo {author} {\bibfnamefont {M.}~\bibnamefont {{Guo}}}, \bibinfo
  {author} {\bibfnamefont {T.}~\bibnamefont {{Langen}}}, \ and\ \bibinfo
  {author} {\bibfnamefont {T.}~\bibnamefont {{Pfau}}},\ }\bibfield  {title}
  {\enquote {\bibinfo {title} {{New states of matter with fine-tuned
  interactions: quantum droplets and dipolar supersolids}},}\ }\href {\doibase
  10.1088/1361-6633/abc9ab} {\bibfield  {journal} {\bibinfo  {journal} {Reports
  on Progress in Physics}\ }\textbf {\bibinfo {volume} {84}},\ \bibinfo {eid}
  {012403} (\bibinfo {year} {2021})}\BibitemShut {NoStop}%
\bibitem [{\citenamefont {{Mazurenko}}\ \emph {et~al.}(2017)\citenamefont
  {{Mazurenko}}, \citenamefont {{Chiu}}, \citenamefont {{Ji}}, \citenamefont
  {{Parsons}}, \citenamefont {{Kan{\'a}sz-Nagy}}, \citenamefont {{Schmidt}},
  \citenamefont {{Grusdt}}, \citenamefont {{Demler}}, \citenamefont {{Greif}},\
  and\ \citenamefont {{Greiner}}}]{Mazurenko:2017}%
  \BibitemOpen
  \bibfield  {author} {\bibinfo {author} {\bibfnamefont {A.}~\bibnamefont
  {{Mazurenko}}}, \bibinfo {author} {\bibfnamefont {C.~S.}\ \bibnamefont
  {{Chiu}}}, \bibinfo {author} {\bibfnamefont {G.}~\bibnamefont {{Ji}}},
  \bibinfo {author} {\bibfnamefont {M.~F.}\ \bibnamefont {{Parsons}}}, \bibinfo
  {author} {\bibfnamefont {M.}~\bibnamefont {{Kan{\'a}sz-Nagy}}}, \bibinfo
  {author} {\bibfnamefont {R.}~\bibnamefont {{Schmidt}}}, \bibinfo {author}
  {\bibfnamefont {F.}~\bibnamefont {{Grusdt}}}, \bibinfo {author}
  {\bibfnamefont {E.}~\bibnamefont {{Demler}}}, \bibinfo {author}
  {\bibfnamefont {D.}~\bibnamefont {{Greif}}}, \ and\ \bibinfo {author}
  {\bibfnamefont {M.}~\bibnamefont {{Greiner}}},\ }\bibfield  {title} {\enquote
  {\bibinfo {title} {{A cold-atom Fermi-Hubbard antiferromagnet}},}\ }\href
  {\doibase 10.1038/nature22362} {\bibfield  {journal} {\bibinfo  {journal}
  {Nature}\ }\textbf {\bibinfo {volume} {545}},\ \bibinfo {pages} {462--466}
  (\bibinfo {year} {2017})}\BibitemShut {NoStop}%
\bibitem [{\citenamefont {{Gall}}\ \emph {et~al.}(2021)\citenamefont {{Gall}},
  \citenamefont {{Wurz}}, \citenamefont {{Samland}}, \citenamefont {{Chan}},\
  and\ \citenamefont {{K{\"o}hl}}}]{Gall:2021}%
  \BibitemOpen
  \bibfield  {author} {\bibinfo {author} {\bibfnamefont {M.}~\bibnamefont
  {{Gall}}}, \bibinfo {author} {\bibfnamefont {N.}~\bibnamefont {{Wurz}}},
  \bibinfo {author} {\bibfnamefont {J.}~\bibnamefont {{Samland}}}, \bibinfo
  {author} {\bibfnamefont {C.~F.}\ \bibnamefont {{Chan}}}, \ and\ \bibinfo
  {author} {\bibfnamefont {M.}~\bibnamefont {{K{\"o}hl}}},\ }\bibfield  {title}
  {\enquote {\bibinfo {title} {{Competing magnetic orders in a bilayer Hubbard
  model with ultracold atoms}},}\ }\href {\doibase 10.1038/s41586-020-03058-x}
  {\bibfield  {journal} {\bibinfo  {journal} {Nature}\ }\textbf {\bibinfo
  {volume} {589}},\ \bibinfo {pages} {40--43} (\bibinfo {year}
  {2021})}\BibitemShut {NoStop}%
\bibitem [{\citenamefont {Fulde}\ and\ \citenamefont
  {Ferrell}(1964)}]{Fulde:1964}%
  \BibitemOpen
  \bibfield  {author} {\bibinfo {author} {\bibfnamefont {P.}~\bibnamefont
  {Fulde}}\ and\ \bibinfo {author} {\bibfnamefont {R.~A.}\ \bibnamefont
  {Ferrell}},\ }\bibfield  {title} {\enquote {\bibinfo {title}
  {Superconductivity in a strong spin-exchange field},}\ }\href {\doibase
  10.1103/PhysRev.135.A550} {\bibfield  {journal} {\bibinfo  {journal} {Phys.
  Rev.}\ }\textbf {\bibinfo {volume} {135}},\ \bibinfo {pages} {A550--A563}
  (\bibinfo {year} {1964})}\BibitemShut {NoStop}%
\bibitem [{\citenamefont {Larkin}\ and\ \citenamefont
  {Ovchinnikov}(1964)}]{Larkin:1964}%
  \BibitemOpen
  \bibfield  {author} {\bibinfo {author} {\bibfnamefont {A.~I.}\ \bibnamefont
  {Larkin}}\ and\ \bibinfo {author} {\bibfnamefont {Y.~N.}\ \bibnamefont
  {Ovchinnikov}},\ }\bibfield  {title} {\enquote {\bibinfo {title} {{Nonuniform
  state of superconductors}},}\ }\href@noop {} {\bibfield  {journal} {\bibinfo
  {journal} {Zh. Eksp. Teor. Fiz.}\ }\textbf {\bibinfo {volume} {47}},\
  \bibinfo {pages} {1136--1146} (\bibinfo {year} {1964})}\BibitemShut {NoStop}%
\bibitem [{\citenamefont {{Kinnunen}}\ \emph {et~al.}(2018)\citenamefont
  {{Kinnunen}}, \citenamefont {{Baarsma}}, \citenamefont {{Martikainen}},\ and\
  \citenamefont {{T{\"o}rm{\"a}}}}]{Kinnunen:2018}%
  \BibitemOpen
  \bibfield  {author} {\bibinfo {author} {\bibfnamefont {J.~J.}\ \bibnamefont
  {{Kinnunen}}}, \bibinfo {author} {\bibfnamefont {J.~E.}\ \bibnamefont
  {{Baarsma}}}, \bibinfo {author} {\bibfnamefont {J.-P.}\ \bibnamefont
  {{Martikainen}}}, \ and\ \bibinfo {author} {\bibfnamefont {P.}~\bibnamefont
  {{T{\"o}rm{\"a}}}},\ }\bibfield  {title} {\enquote {\bibinfo {title} {{The
  Fulde-Ferrell-Larkin-Ovchinnikov state for ultracold fermions in lattice and
  harmonic potentials: a review}},}\ }\href {\doibase 10.1088/1361-6633/aaa4ad}
  {\bibfield  {journal} {\bibinfo  {journal} {Reports on Progress in Physics}\
  }\textbf {\bibinfo {volume} {81}},\ \bibinfo {eid} {046401} (\bibinfo {year}
  {2018})}\BibitemShut {NoStop}%
\bibitem [{\citenamefont {Lee}\ and\ \citenamefont {Yang}(1957)}]{Lee:1957a}%
  \BibitemOpen
  \bibfield  {author} {\bibinfo {author} {\bibfnamefont {T.~D.}\ \bibnamefont
  {Lee}}\ and\ \bibinfo {author} {\bibfnamefont {C.~N.}\ \bibnamefont {Yang}},\
  }\bibfield  {title} {\enquote {\bibinfo {title} {Many-body problem in quantum
  mechanics and quantum statistical mechanics},}\ }\href {\doibase
  10.1103/PhysRev.105.1119} {\bibfield  {journal} {\bibinfo  {journal} {Phys.
  Rev.}\ }\textbf {\bibinfo {volume} {105}},\ \bibinfo {pages} {1119--1120}
  (\bibinfo {year} {1957})}\BibitemShut {NoStop}%
\bibitem [{\citenamefont {{Reppy}}\ \emph {et~al.}(2000)\citenamefont
  {{Reppy}}, \citenamefont {{Crooker}}, \citenamefont {{Hebral}}, \citenamefont
  {{Corwin}}, \citenamefont {{He}},\ and\ \citenamefont
  {{Zassenhaus}}}]{Reppy:2000}%
  \BibitemOpen
  \bibfield  {author} {\bibinfo {author} {\bibfnamefont {J.~D.}\ \bibnamefont
  {{Reppy}}}, \bibinfo {author} {\bibfnamefont {B.~C.}\ \bibnamefont
  {{Crooker}}}, \bibinfo {author} {\bibfnamefont {B.}~\bibnamefont {{Hebral}}},
  \bibinfo {author} {\bibfnamefont {A.~D.}\ \bibnamefont {{Corwin}}}, \bibinfo
  {author} {\bibfnamefont {J.}~\bibnamefont {{He}}}, \ and\ \bibinfo {author}
  {\bibfnamefont {G.~M.}\ \bibnamefont {{Zassenhaus}}},\ }\bibfield  {title}
  {\enquote {\bibinfo {title} {{Density Dependence of the Transition
  Temperature in a Homogeneous Bose-Einstein Condensate}},}\ }\href {\doibase
  10.1103/PhysRevLett.84.2060} {\bibfield  {journal} {\bibinfo  {journal}
  {\prl}\ }\textbf {\bibinfo {volume} {84}},\ \bibinfo {pages} {2060--2063}
  (\bibinfo {year} {2000})}\BibitemShut {NoStop}%
\bibitem [{\citenamefont {Arnold}\ and\ \citenamefont
  {Moore}(2001)}]{Arnold:2001}%
  \BibitemOpen
  \bibfield  {author} {\bibinfo {author} {\bibfnamefont {P.}~\bibnamefont
  {Arnold}}\ and\ \bibinfo {author} {\bibfnamefont {G.}~\bibnamefont {Moore}},\
  }\bibfield  {title} {\enquote {\bibinfo {title} {{BEC} transition temperature
  of a dilute homogeneous imperfect {B}ose gas},}\ }\href {\doibase
  10.1103/PhysRevLett.87.120401} {\bibfield  {journal} {\bibinfo  {journal}
  {Phys. Rev. Lett.}\ }\textbf {\bibinfo {volume} {87}},\ \bibinfo {pages}
  {120401} (\bibinfo {year} {2001})}\BibitemShut {NoStop}%
\bibitem [{\citenamefont {Kashurnikov}\ \emph {et~al.}(2001)\citenamefont
  {Kashurnikov}, \citenamefont {Prokof'ev},\ and\ \citenamefont
  {Svistunov}}]{Kashurnikov:2001}%
  \BibitemOpen
  \bibfield  {author} {\bibinfo {author} {\bibfnamefont {V.~A.}\ \bibnamefont
  {Kashurnikov}}, \bibinfo {author} {\bibfnamefont {N.~V.}\ \bibnamefont
  {Prokof'ev}}, \ and\ \bibinfo {author} {\bibfnamefont {B.~V.}\ \bibnamefont
  {Svistunov}},\ }\bibfield  {title} {\enquote {\bibinfo {title} {Critical
  temperature shift in weakly interacting {B}ose gas},}\ }\href {\doibase
  10.1103/PhysRevLett.87.120402} {\bibfield  {journal} {\bibinfo  {journal}
  {Phys. Rev. Lett.}\ }\textbf {\bibinfo {volume} {87}},\ \bibinfo {pages}
  {120402} (\bibinfo {year} {2001})}\BibitemShut {NoStop}%
\bibitem [{\citenamefont {Andersen}(2004)}]{Andersen:2004}%
  \BibitemOpen
  \bibfield  {author} {\bibinfo {author} {\bibfnamefont {J.~O.}\ \bibnamefont
  {Andersen}},\ }\bibfield  {title} {\enquote {\bibinfo {title} {Theory of the
  weakly interacting {B}ose gas},}\ }\href {\doibase 10.1103/RevModPhys.76.599}
  {\bibfield  {journal} {\bibinfo  {journal} {Rev. Mod. Phys.}\ }\textbf
  {\bibinfo {volume} {76}},\ \bibinfo {pages} {599--639} (\bibinfo {year}
  {2004})}\BibitemShut {NoStop}%
\bibitem [{\citenamefont {{Holzmann}}\ \emph {et~al.}(2004)\citenamefont
  {{Holzmann}}, \citenamefont {{Fuchs}}, \citenamefont {{Baym}}, \citenamefont
  {{Blaizot}},\ and\ \citenamefont {{Lalo{\"e}}}}]{Holzmann:2004}%
  \BibitemOpen
  \bibfield  {author} {\bibinfo {author} {\bibfnamefont {M.}~\bibnamefont
  {{Holzmann}}}, \bibinfo {author} {\bibfnamefont {J.-N.}\ \bibnamefont
  {{Fuchs}}}, \bibinfo {author} {\bibfnamefont {G.~A.}\ \bibnamefont {{Baym}}},
  \bibinfo {author} {\bibfnamefont {J.-P.}\ \bibnamefont {{Blaizot}}}, \ and\
  \bibinfo {author} {\bibfnamefont {F.}~\bibnamefont {{Lalo{\"e}}}},\
  }\bibfield  {title} {\enquote {\bibinfo {title} {{Bose Einstein transition
  temperature in a dilute repulsive gas}},}\ }\href {\doibase
  10.1016/j.crhy.2004.01.003} {\bibfield  {journal} {\bibinfo  {journal}
  {Comptes Rendus Physique}\ }\textbf {\bibinfo {volume} {5}},\ \bibinfo
  {pages} {21--37} (\bibinfo {year} {2004})}\BibitemShut {NoStop}%
\bibitem [{\citenamefont {Ensher}\ \emph {et~al.}(1996)\citenamefont {Ensher},
  \citenamefont {Jin}, \citenamefont {Matthews}, \citenamefont {Wieman},\ and\
  \citenamefont {Cornell}}]{Ensher:1996}%
  \BibitemOpen
  \bibfield  {author} {\bibinfo {author} {\bibfnamefont {J.~R.}\ \bibnamefont
  {Ensher}}, \bibinfo {author} {\bibfnamefont {D.~S.}\ \bibnamefont {Jin}},
  \bibinfo {author} {\bibfnamefont {M.~R.}\ \bibnamefont {Matthews}}, \bibinfo
  {author} {\bibfnamefont {C.~E.}\ \bibnamefont {Wieman}}, \ and\ \bibinfo
  {author} {\bibfnamefont {E.~A.}\ \bibnamefont {Cornell}},\ }\bibfield
  {title} {\enquote {\bibinfo {title} {{B}ose--{E}instein condensation in a
  dilute gas: Measurement of energy and ground-state occupation},}\ }\href
  {\doibase 10.1103/PhysRevLett.77.4984} {\bibfield  {journal} {\bibinfo
  {journal} {Phys. Rev. Lett.}\ }\textbf {\bibinfo {volume} {77}},\ \bibinfo
  {pages} {4984--4987} (\bibinfo {year} {1996})}\BibitemShut {NoStop}%
\bibitem [{\citenamefont {Gerbier}\ \emph {et~al.}(2004)\citenamefont
  {Gerbier}, \citenamefont {Thywissen}, \citenamefont {Richard}, \citenamefont
  {Hugbart}, \citenamefont {Bouyer},\ and\ \citenamefont
  {Aspect}}]{Gerbier:2004}%
  \BibitemOpen
  \bibfield  {author} {\bibinfo {author} {\bibfnamefont {F.}~\bibnamefont
  {Gerbier}}, \bibinfo {author} {\bibfnamefont {J.~H.}\ \bibnamefont
  {Thywissen}}, \bibinfo {author} {\bibfnamefont {S.}~\bibnamefont {Richard}},
  \bibinfo {author} {\bibfnamefont {M.}~\bibnamefont {Hugbart}}, \bibinfo
  {author} {\bibfnamefont {P.}~\bibnamefont {Bouyer}}, \ and\ \bibinfo {author}
  {\bibfnamefont {A.}~\bibnamefont {Aspect}},\ }\bibfield  {title} {\enquote
  {\bibinfo {title} {Critical temperature of a trapped, weakly interacting
  {B}ose gas},}\ }\href {\doibase 10.1103/PhysRevLett.92.030405} {\bibfield
  {journal} {\bibinfo  {journal} {Phys. Rev. Lett.}\ }\textbf {\bibinfo
  {volume} {92}},\ \bibinfo {eid} {030405} (\bibinfo {year}
  {2004})}\BibitemShut {NoStop}%
\bibitem [{\citenamefont {Meppelink}\ \emph {et~al.}(2010)\citenamefont
  {Meppelink}, \citenamefont {Rozendaal}, \citenamefont {Koller}, \citenamefont
  {Vogels},\ and\ \citenamefont {van~der Straten}}]{Meppelink:2010}%
  \BibitemOpen
  \bibfield  {author} {\bibinfo {author} {\bibfnamefont {R.}~\bibnamefont
  {Meppelink}}, \bibinfo {author} {\bibfnamefont {R.~A.}\ \bibnamefont
  {Rozendaal}}, \bibinfo {author} {\bibfnamefont {S.~B.}\ \bibnamefont
  {Koller}}, \bibinfo {author} {\bibfnamefont {J.~M.}\ \bibnamefont {Vogels}},
  \ and\ \bibinfo {author} {\bibfnamefont {P.}~\bibnamefont {van~der
  Straten}},\ }\bibfield  {title} {\enquote {\bibinfo {title} {Thermodynamics
  of {B}ose--{E}instein-condensed clouds using phase-contrast imaging},}\
  }\href {\doibase 10.1103/PhysRevA.81.053632} {\bibfield  {journal} {\bibinfo
  {journal} {Phys. Rev. A}\ }\textbf {\bibinfo {volume} {81}},\ \bibinfo
  {pages} {053632} (\bibinfo {year} {2010})}\BibitemShut {NoStop}%
\bibitem [{\citenamefont {Smith}\ \emph
  {et~al.}(2011{\natexlab{a}})\citenamefont {Smith}, \citenamefont {Campbell},
  \citenamefont {Tammuz},\ and\ \citenamefont {Hadzibabic}}]{Smith:2011}%
  \BibitemOpen
  \bibfield  {author} {\bibinfo {author} {\bibfnamefont {R.~P.}\ \bibnamefont
  {Smith}}, \bibinfo {author} {\bibfnamefont {R.~L.~D.}\ \bibnamefont
  {Campbell}}, \bibinfo {author} {\bibfnamefont {N.}~\bibnamefont {Tammuz}}, \
  and\ \bibinfo {author} {\bibfnamefont {Z.}~\bibnamefont {Hadzibabic}},\
  }\bibfield  {title} {\enquote {\bibinfo {title} {Effects of interactions on
  the critical temperature of a trapped {B}ose gas},}\ }\href {\doibase
  10.1103/PhysRevLett.106.250403} {\bibfield  {journal} {\bibinfo  {journal}
  {Phys. Rev. Lett.}\ }\textbf {\bibinfo {volume} {106}},\ \bibinfo {pages}
  {250403} (\bibinfo {year} {2011}{\natexlab{a}})}\BibitemShut {NoStop}%
\bibitem [{\citenamefont {Smith}\ \emph
  {et~al.}(2011{\natexlab{b}})\citenamefont {Smith}, \citenamefont {Tammuz},
  \citenamefont {Campbell}, \citenamefont {Holzmann},\ and\ \citenamefont
  {Hadzibabic}}]{Smith:2011b}%
  \BibitemOpen
  \bibfield  {author} {\bibinfo {author} {\bibfnamefont {R.~P.}\ \bibnamefont
  {Smith}}, \bibinfo {author} {\bibfnamefont {N.}~\bibnamefont {Tammuz}},
  \bibinfo {author} {\bibfnamefont {R.~L.~D.}\ \bibnamefont {Campbell}},
  \bibinfo {author} {\bibfnamefont {M.}~\bibnamefont {Holzmann}}, \ and\
  \bibinfo {author} {\bibfnamefont {Z.}~\bibnamefont {Hadzibabic}},\ }\bibfield
   {title} {\enquote {\bibinfo {title} {Condensed fraction of an atomic {B}ose
  gas induced by critical correlations},}\ }\href {\doibase
  10.1103/PhysRevLett.107.190403} {\bibfield  {journal} {\bibinfo  {journal}
  {Phys. Rev. Lett.}\ }\textbf {\bibinfo {volume} {107}},\ \bibinfo {pages}
  {190403} (\bibinfo {year} {2011}{\natexlab{b}})}\BibitemShut {NoStop}%
\bibitem [{\citenamefont {Giorgini}\ \emph {et~al.}(1996)\citenamefont
  {Giorgini}, \citenamefont {Pitaevskii},\ and\ \citenamefont
  {Stringari}}]{Giorgini:1996}%
  \BibitemOpen
  \bibfield  {author} {\bibinfo {author} {\bibfnamefont {S.}~\bibnamefont
  {Giorgini}}, \bibinfo {author} {\bibfnamefont {L.~P.}\ \bibnamefont
  {Pitaevskii}}, \ and\ \bibinfo {author} {\bibfnamefont {S.}~\bibnamefont
  {Stringari}},\ }\bibfield  {title} {\enquote {\bibinfo {title} {Condensate
  fraction and critical temperature of a trapped interacting {B}ose gas},}\
  }\href {\doibase 10.1103/PhysRevA.54.R4633} {\bibfield  {journal} {\bibinfo
  {journal} {Phys. Rev. A}\ }\textbf {\bibinfo {volume} {54}},\ \bibinfo
  {pages} {R4633--R4636} (\bibinfo {year} {1996})}\BibitemShut {NoStop}%
\bibitem [{\citenamefont {{Shkedrov}}\ \emph {et~al.}(2021)\citenamefont
  {{Shkedrov}}, \citenamefont {{Menashes}}, \citenamefont {{Ness}},
  \citenamefont {{Vainbaum}},\ and\ \citenamefont
  {{Sagi}}}]{shkedrov2021absence}%
  \BibitemOpen
  \bibfield  {author} {\bibinfo {author} {\bibfnamefont {C.}~\bibnamefont
  {{Shkedrov}}}, \bibinfo {author} {\bibfnamefont {M.}~\bibnamefont
  {{Menashes}}}, \bibinfo {author} {\bibfnamefont {G.}~\bibnamefont {{Ness}}},
  \bibinfo {author} {\bibfnamefont {A.}~\bibnamefont {{Vainbaum}}}, \ and\
  \bibinfo {author} {\bibfnamefont {Y.}~\bibnamefont {{Sagi}}},\ }\href@noop {}
  {\enquote {\bibinfo {title} {{Absence of heating in a uniform Fermi gas
  created by periodic driving}},}\ } (\bibinfo {year} {2021}),\ \Eprint
  {http://arxiv.org/abs/2102.09506} {arXiv:2102.09506} \BibitemShut {NoStop}%
\bibitem [{\citenamefont {Becker}\ \emph {et~al.}(2018)\citenamefont {Becker},
  \citenamefont {Lachmann}, \citenamefont {Seidel}, \citenamefont {Ahlers},
  \citenamefont {Dinkelaker}, \citenamefont {Grosse}, \citenamefont {Hellmig},
  \citenamefont {M{\"u}ntinga}, \citenamefont {Schkolnik}, \citenamefont
  {Wendrich}, \citenamefont {Wenzlawski}, \citenamefont {Weps}, \citenamefont
  {Corgier}, \citenamefont {Franz}, \citenamefont {Gaaloul}, \citenamefont
  {Herr}, \citenamefont {L{\"u}dtke}, \citenamefont {Popp}, \citenamefont
  {Amri}, \citenamefont {Duncker}, \citenamefont {Erbe}, \citenamefont
  {Kohfeldt}, \citenamefont {Kubelka-Lange}, \citenamefont {Braxmaier},
  \citenamefont {Charron}, \citenamefont {Ertmer}, \citenamefont {Krutzik},
  \citenamefont {L{\"a}mmerzahl}, \citenamefont {Peters}, \citenamefont
  {Schleich}, \citenamefont {Sengstock}, \citenamefont {Walser}, \citenamefont
  {Wicht}, \citenamefont {Windpassinger},\ and\ \citenamefont
  {Rasel}}]{Becker:2018}%
  \BibitemOpen
  \bibfield  {author} {\bibinfo {author} {\bibfnamefont {D.}~\bibnamefont
  {Becker}}, \bibinfo {author} {\bibfnamefont {M.~D.}\ \bibnamefont
  {Lachmann}}, \bibinfo {author} {\bibfnamefont {S.~T.}\ \bibnamefont
  {Seidel}}, \bibinfo {author} {\bibfnamefont {H.}~\bibnamefont {Ahlers}},
  \bibinfo {author} {\bibfnamefont {A.~N.}\ \bibnamefont {Dinkelaker}},
  \bibinfo {author} {\bibfnamefont {J.}~\bibnamefont {Grosse}}, \bibinfo
  {author} {\bibfnamefont {O.}~\bibnamefont {Hellmig}}, \bibinfo {author}
  {\bibfnamefont {H.}~\bibnamefont {M{\"u}ntinga}}, \bibinfo {author}
  {\bibfnamefont {V.}~\bibnamefont {Schkolnik}}, \bibinfo {author}
  {\bibfnamefont {T.}~\bibnamefont {Wendrich}}, \bibinfo {author}
  {\bibfnamefont {A.}~\bibnamefont {Wenzlawski}}, \bibinfo {author}
  {\bibfnamefont {B.}~\bibnamefont {Weps}}, \bibinfo {author} {\bibfnamefont
  {R.}~\bibnamefont {Corgier}}, \bibinfo {author} {\bibfnamefont
  {T.}~\bibnamefont {Franz}}, \bibinfo {author} {\bibfnamefont
  {N.}~\bibnamefont {Gaaloul}}, \bibinfo {author} {\bibfnamefont
  {W.}~\bibnamefont {Herr}}, \bibinfo {author} {\bibfnamefont {D.}~\bibnamefont
  {L{\"u}dtke}}, \bibinfo {author} {\bibfnamefont {M.}~\bibnamefont {Popp}},
  \bibinfo {author} {\bibfnamefont {S.}~\bibnamefont {Amri}}, \bibinfo {author}
  {\bibfnamefont {H.}~\bibnamefont {Duncker}}, \bibinfo {author} {\bibfnamefont
  {M.}~\bibnamefont {Erbe}}, \bibinfo {author} {\bibfnamefont {A.}~\bibnamefont
  {Kohfeldt}}, \bibinfo {author} {\bibfnamefont {A.}~\bibnamefont
  {Kubelka-Lange}}, \bibinfo {author} {\bibfnamefont {C.}~\bibnamefont
  {Braxmaier}}, \bibinfo {author} {\bibfnamefont {E.}~\bibnamefont {Charron}},
  \bibinfo {author} {\bibfnamefont {W.}~\bibnamefont {Ertmer}}, \bibinfo
  {author} {\bibfnamefont {M.}~\bibnamefont {Krutzik}}, \bibinfo {author}
  {\bibfnamefont {C.}~\bibnamefont {L{\"a}mmerzahl}}, \bibinfo {author}
  {\bibfnamefont {A.}~\bibnamefont {Peters}}, \bibinfo {author} {\bibfnamefont
  {W.~P.}\ \bibnamefont {Schleich}}, \bibinfo {author} {\bibfnamefont
  {K.}~\bibnamefont {Sengstock}}, \bibinfo {author} {\bibfnamefont
  {R.}~\bibnamefont {Walser}}, \bibinfo {author} {\bibfnamefont
  {A.}~\bibnamefont {Wicht}}, \bibinfo {author} {\bibfnamefont
  {P.}~\bibnamefont {Windpassinger}}, \ and\ \bibinfo {author} {\bibfnamefont
  {E.~M.}\ \bibnamefont {Rasel}},\ }\bibfield  {title} {\enquote {\bibinfo
  {title} {Space-borne {B}ose--{E}instein condensation for precision
  interferometry},}\ }\href {\doibase 10.1038/s41586-018-0605-1} {\bibfield
  {journal} {\bibinfo  {journal} {Nature}\ }\textbf {\bibinfo {volume} {562}},\
  \bibinfo {pages} {391--395} (\bibinfo {year} {2018})}\BibitemShut {NoStop}%
\bibitem [{\citenamefont {{Aveline}}\ \emph {et~al.}(2020)\citenamefont
  {{Aveline}}, \citenamefont {{Williams}}, \citenamefont {{Elliott}},
  \citenamefont {{Dutenhoffer}}, \citenamefont {{Kellogg}}, \citenamefont
  {{Kohel}}, \citenamefont {{Lay}}, \citenamefont {{Oudrhiri}}, \citenamefont
  {{Shotwell}}, \citenamefont {{Yu}},\ and\ \citenamefont
  {{Thompson}}}]{Aveline:2020}%
  \BibitemOpen
  \bibfield  {author} {\bibinfo {author} {\bibfnamefont {D.~C.}\ \bibnamefont
  {{Aveline}}}, \bibinfo {author} {\bibfnamefont {J.~R.}\ \bibnamefont
  {{Williams}}}, \bibinfo {author} {\bibfnamefont {E.~R.}\ \bibnamefont
  {{Elliott}}}, \bibinfo {author} {\bibfnamefont {C.}~\bibnamefont
  {{Dutenhoffer}}}, \bibinfo {author} {\bibfnamefont {J.~R.}\ \bibnamefont
  {{Kellogg}}}, \bibinfo {author} {\bibfnamefont {J.~M.}\ \bibnamefont
  {{Kohel}}}, \bibinfo {author} {\bibfnamefont {N.~E.}\ \bibnamefont {{Lay}}},
  \bibinfo {author} {\bibfnamefont {K.}~\bibnamefont {{Oudrhiri}}}, \bibinfo
  {author} {\bibfnamefont {R.~F.}\ \bibnamefont {{Shotwell}}}, \bibinfo
  {author} {\bibfnamefont {N.}~\bibnamefont {{Yu}}}, \ and\ \bibinfo {author}
  {\bibfnamefont {R.~J.}\ \bibnamefont {{Thompson}}},\ }\bibfield  {title}
  {\enquote {\bibinfo {title} {{Observation of Bose-Einstein condensates in an
  Earth-orbiting research lab}},}\ }\href {\doibase 10.1038/s41586-020-2346-1}
  {\bibfield  {journal} {\bibinfo  {journal} {Nature}\ }\textbf {\bibinfo
  {volume} {582}},\ \bibinfo {pages} {193--197} (\bibinfo {year}
  {2020})}\BibitemShut {NoStop}%
\bibitem [{\citenamefont {Frye}\ \emph {et~al.}(2021)\citenamefont {Frye},
  \citenamefont {Abend}, \citenamefont {Bartosch}, \citenamefont {Bawamia},
  \citenamefont {Becker}, \citenamefont {Blume}, \citenamefont {Braxmaier},
  \citenamefont {Chiow}, \citenamefont {Efremov}, \citenamefont {Ertmer} \emph
  {et~al.}}]{Frye:2021bose}%
  \BibitemOpen
  \bibfield  {author} {\bibinfo {author} {\bibfnamefont {K.}~\bibnamefont
  {Frye}}, \bibinfo {author} {\bibfnamefont {S.}~\bibnamefont {Abend}},
  \bibinfo {author} {\bibfnamefont {W.}~\bibnamefont {Bartosch}}, \bibinfo
  {author} {\bibfnamefont {A.}~\bibnamefont {Bawamia}}, \bibinfo {author}
  {\bibfnamefont {D.}~\bibnamefont {Becker}}, \bibinfo {author} {\bibfnamefont
  {H.}~\bibnamefont {Blume}}, \bibinfo {author} {\bibfnamefont
  {C.}~\bibnamefont {Braxmaier}}, \bibinfo {author} {\bibfnamefont {S.-W.}\
  \bibnamefont {Chiow}}, \bibinfo {author} {\bibfnamefont {M.~A.}\ \bibnamefont
  {Efremov}}, \bibinfo {author} {\bibfnamefont {W.}~\bibnamefont {Ertmer}},
  \emph {et~al.},\ }\bibfield  {title} {\enquote {\bibinfo {title} {The
  {B}ose--{E}instein condensate and cold atom laboratory},}\ }\href {\doibase
  https://doi.org/10.1140/epjqt/s40507-020-00090-8} {\bibfield  {journal}
  {\bibinfo  {journal} {EPJ Quantum Technology}\ }\textbf {\bibinfo {volume}
  {8}},\ \bibinfo {pages} {1--38} (\bibinfo {year} {2021})}\BibitemShut
  {NoStop}%
\end{thebibliography}

\end{document}